\documentclass[aps,showpacs,prl,twocolumn,superscriptaddress,apsrev]{revtex4-1}
\usepackage{multirow}
\usepackage{color}
\usepackage{amsfonts}
\usepackage{amsmath}
\usepackage{mathrsfs}
\usepackage{amssymb}
\usepackage{wasysym}
\usepackage{xfrac}
\usepackage{graphicx}
\usepackage{CJK}

\begin{document}
\title{Floquet-Induced Superfluidity with Periodically Modulated Interactions of 
Two-Species Hardcore Bosons in a One-dimensional Optical Lattice}
%
\author{Tao Wang}
\affiliation{Hubei Key Laboratory of Optical Information and Pattern Recognition, Wuhan Institute of Technology, Wuhan, 438000, China}
\affiliation{Physics Department and Research Center OPTIMAS,
Technische Universit\"{a}t Kaiserslautern, 67663 Kaiserslautern, Germany}
\affiliation{Department of Physics, Chongqing University, Chongqing, 401331, China}

\author{Shijie Hu}
\email{Corresponding author: shijiehu201@gmail.com}
\affiliation{Physics Department and Research Center OPTIMAS,
Technische Universit\"{a}t Kaiserslautern, 67663 Kaiserslautern, Germany}
\author{Sebastian Eggert}
\affiliation{Physics Department and Research Center OPTIMAS,
Technische Universit\"{a}t Kaiserslautern, 67663 Kaiserslautern, Germany}
\author{Michael Fleischhauer}
\affiliation{Physics Department and Research Center OPTIMAS,
Technische Universit\"{a}t Kaiserslautern, 67663 Kaiserslautern, Germany}
\author{Axel Pelster}
\affiliation{Physics Department and Research Center OPTIMAS,
Technische Universit\"{a}t Kaiserslautern, 67663 Kaiserslautern, Germany}
\author{Xue-Feng Zhang}
\affiliation{Department of Physics, Chongqing University, Chongqing, 401331, China}
\begin{abstract}
We consider two species of hard-core bosons with density dependent hopping in a 
one-dimensional optical lattice, for which we propose experimental realizations using
time-periodic driving.  The 
quantum phase diagram for half-integer filling is determined by 
combining 
different advanced numerical simulations with analytic calculations. 
We find that a {\it reduction} of
 the density-dependent hopping induces 
a Mott-insulator to superfluid transition.  For negative hopping 
a previously unknown state is found, where one species induces 
a gauge phase of the other species, which leads to a 
superfluid phase of gauge-paired particles.
The corresponding experimental signatures 
are discussed. 
\end{abstract}
\maketitle

%
Recent developments for ultra-cold atomic systems provide  useful platforms 
for quantum simulations in a wide window of tunable 
parameters~\cite{Anderson_1995, Davis_1995}. 
Interacting bosons in an optical lattice show a quantum phase transition 
from a superfluid (SF) to a Mott-insulator (MI)~\cite{Fisher_1989,Jaksch_1998}, 
which has been experimentally shown by ``time-of-flight"  
measurements~\cite{Greiner_2002} of the momentum distribution~\cite{Hoffmann_2009}. 
In a mixture of different species, 
the interaction strengths for both inter- and intra-species scattering  
can be tuned via Feshbach resonances~\cite{Chin_2010}. 
As a result, a large variety of interesting new phases were predicted for spinor bosons~\cite{Gross_2002, Demler_2002,Mobarak,Duan_2003}, interacting multi-species bosons or fermions~\cite{Hofstetter_2002, Kuklov_2003,Altman_2003, Kuno_2013, ring} and Bose-Fermi mixtures~\cite{M_lmer_1998,  Viverit_2000, Bhaseen_2009}. 
Recently, time-dependent and driven
optical lattices have opened an era of exploring 
exotic dynamical quantum states~\cite{Bretin_2004, Schweikhard_2004, Lin_2009, Lin_2011, Vidanovic, Liberto_2014, Greschner_2015, Tang_2015, Str_ter_2016, Keilmann_2011, Ramos_2008, Pollack_2010, Rapp_2012, Wang_2014,Greschner_2014,Meinert_2016,Eckardt_2017,Arimondo_2012,Struck_2011, Struck_2012, Hauke_2012, Struck_2013, Aidelsburger_2011, Lignier_2007, exp2, exp3}.
For instance, assisted Raman tunneling and shaking 
were proposed to induce a density-dependent complex phase 
in the hopping elements, which may allow the experimental realization 
of anyonic physics~\cite{Keilmann_2011, Greschner_2015, Tang_2015, Str_ter_2016}.
On the other hand, 
a fast time-periodic modulation of the interaction \cite{Ramos_2008,Pollack_2010} will lead to an effective 
hopping matrix element depending on the density 
difference~\cite{Rapp_2012, Wang_2014,Greschner_2014, Meinert_2016},
which gives rise to pair superfluidity in one dimension (1D)~\cite{Rapp_2012}, while superfluidity is
suppressed in higher dimensions~\cite{Wang_2014}.
Experimental realizations of time-periodic driving \cite{Zenesini_2009, Struck_2011, Struck_2012, Hauke_2012, Struck_2013, Aidelsburger_2011, Lignier_2007,exp2,exp3, Meinert_2016} demonstrate that
signatures of interesting effective models can be observed before 
heating or decoherence destroys the so-called Floquet states.

In this Letter we propose a realization of a density-dependent hopping model
of two interacting boson species in 1D via time-periodic driving. 
The corresponding
quantum phase diagram is determined using a combination of 
advanced numerical methods. 
We find that a reduction of the density-dependent hopping by driving, counter-intuitively, causes a MI to SF quantum phase transition.
 For larger driving we obtain negative effective hopping, 
which gives rise to an exotic SF phase of gauge-dressed composite particles.

\begin{figure}[t]
\includegraphics[width=0.99 \columnwidth]{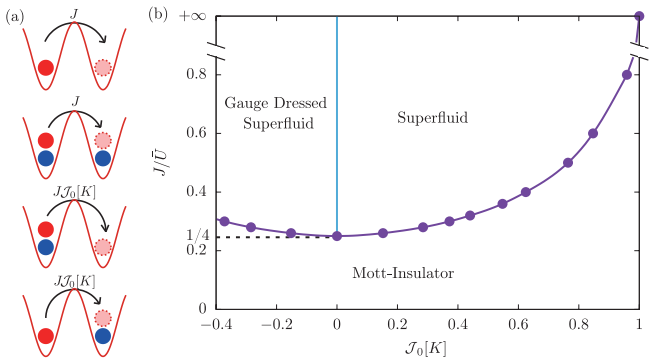}
\caption{(a)
Hopping processes of one species ``$a$" (red filled circle) in the effective model in Eq.~(\ref{He}). The other species ``$b$" is denoted by blue filled circles. Hopping between 
two neighboring single occupied or double/empty sites are suppressed
by  ${\cal J}_{0}[K]$.  
(b) Ground-state quantum phase diagram of the effective model in Eq.~(\ref{He}) at 
half-filling.}\label{fig1}
\end{figure}

The model for hard-core bosons with two hyperfine states (marked by ``$a$" and ``$b$")
in a 1D optical lattice is given in terms of 
corresponding creation and annihilation operators 
$\hat a_{l}^{\dagger}$, ${\hat a^{\phantom{\dagger}}}_{l}$, 
$\hat b_{l}^{\dagger}$, ${\hat b^{\phantom{\dagger}}}_{l}$ at each {site}
\begin{eqnarray}
\nonumber
\hat{H}(t)&=&-J \sum_{l} \left({\hat a}^{\dagger}_{l} {\hat a}_{l+1} + {\hat b}^{\dagger}_{l} {\hat b}_{l+1} + \rm{h.c.}\right) \\
&&+ U \sum_{l} {\hat n}^{a}_{l}{\hat n}^{b}_{l} + J_{\Omega} \sum_{l}\left(\hat{a}_{l}^{\dagger}\hat{b}_{l}+\rm{h.c.}\right),\label{Ht}
\end{eqnarray}
where we have included a possible Rabi coupling $J_\Omega$.  Here
$U$ is the repulsive interaction between densities of opposite species 
${\hat n}_{l}^{a} = \hat a_{l}^{\dagger}{\hat a^{\phantom{\dagger}}}_{l}$
and ${\hat n}_{l}^{b} = \hat b_{l}^{\dagger}{\hat b^{\phantom{\dagger}}}_{l}$, 
 which can be achieved by a magnetic field just below 
the inter-species Feshbach resonance in a deep lattice potential 
(see Supplemental Materials for details \cite{SM}). 
We consider a time-periodic modulation of the Feshbach resonance 
$U=\bar{U}+\delta U\cos(\omega t)$ without Rabi transitions $J_{\Omega}=0$, or
-- {\it alternatively} -- an oscillating Rabi 
amplitude $J_{\Omega}=J_{\Omega}^{0}\cos(\omega t)$ for constant $U$.  The concrete 
details  for the  experimental
realization of the two  alternative setups are described in the Supplemental 
Materials \cite{SM}, which require a lattice depth of $s\sim 20$ 
to obey the hardcore constraint, corresponding to $J/h\sim 20\rm Hz$ for Rubidium-87 \cite{SM}.
Choosing driving frequency and amplitude around 1 kHz, we have
$\hbar \omega \gg J, \bar U$, while transitions to higher bands are still suppressed.
Using 
Floquet theory in this limit~\cite{Rapp_2012,Wang_2014,Greschner_2014,Meinert_2016, Eckardt_2017,Arimondo_2012}, in both cases 
a time-independent effective Hamiltonian with density-dependent hopping can
be reached by adiabatically increasing the driving amplitude 

%
\begin{eqnarray}
\hat{H}_{e}&=&\sum^{L}_{l=1}\left(-\hat J_l^{a}{\hat a}_{l}^{\dagger}{\hat a^{\phantom{\dagger}}}_{l+1}-\hat J_l^{b}{\hat b}_{l}^{\dagger}{\hat b^{\phantom{\dagger}}}_{l+1}+\bar U{\hat n}_{l}^{a}{\hat n}_{l}^{b}\right),\label{He}
\end{eqnarray}
where the hoppings $\hat J_l^{a/b}$ are now operators depending on the 
local densities of the opposite species 
\begin{equation}
\hat J_l^{b} = J \, {\cal J}_{0}\left[K({\hat n}_{l}^{a}-{\hat n}_{l+1}^{a})\right] 
\end{equation}
with matrix elements 
\begin{equation}
\left\{
\begin{array}{ll}
   J &  {\  \rm for\ }  n_l^a-n_{l+1}^a=0 \\
   J  {\cal J}_0[K]  & {\ \rm for\ }  | n_l^a- n_{l+1}^a| =1 
\end{array}
\right.
\label{hatJ}
\end{equation}
and analogously for $\hat J_l^{a}$.
Here ${\cal J}_{0}[K]$ denotes the zeroth-order Bessel function of the first kind
and the dimensionless {driving amplitude} $K=\delta U/\hbar \omega$ or
-- {\it alternatively} -- $K=2J_{\Omega}^{0}/\hbar \omega$ gives 
the modulation strength in units of $\hbar \omega$ \cite{SM}.  As illustrated in 
Fig.~\ref{fig1}(a) the effect of driving is therefore to suppress hopping 
of hard-core type-$a$ bosons by ${\cal J}_0[K]$ if the 
occupation of type-$b$ bosons is different and vice versa.  
The suppression decreases with increasing driving 
$K$ from ${\cal J}_0[0]=1$ to the negative minimum value of ${\cal J}_0[3.8717] \approx -0.4024$.  
Further tuning parameters of the model are possible, e.g.~by an asymmetry in the
pulse sequence \cite{Struck_2012}, which makes  
this setup an interesting general platform.  In this Letter we will 
focus on the phase diagram of the model (\ref{He}) at half-filling $\langle \hat{n}_l^a \rangle
= \langle \hat{n}_l^b \rangle = \sfrac{1}{2}$.
In this case, the undriven system \mbox{${\cal J}_0[0]=1$} 
is known to be in the Mott state 
 for any $\bar U>0$ without a
quantum phase transition \cite{Lieb_1968}.
However, as we will see below, the selective {\it reduction} of
hopping elements by driving will destroy the MI state.

An interesting point is reached at the zeros of the Bessel function since
for ${\cal J}_0[K] = 0$ the hopping between neighboring double 
occupied and empty sites is not possible in this case
as shown in Fig.~\ref{fig1}~(a). 
Because the Hamiltonian no longer distinguishes
between double occupied and empty sites, we can denote both of them with
pseudo-spin up $|\!\!\uparrow\rangle$ (for $n^a_l=n^b_l$).  
Likewise hopping between neighboring single occupied sites is forbidden 
regardless if they are type-$a$ or $b$, so both can be denoted with pseudo-spin  down 
$|\!\!\downarrow\rangle$ (for $n^a_l\neq n^b_l$).
The corresponding total occupation numbers 
for the four different possible local states ($a$, $b$, double, empty) are all conserved  
and the resulting Hamiltonian for half-filling is expressed exactly as 
\begin{equation}
\hat{H}_{e}=\sum^{L}_{l=1}[-J({\hat S}_{l}^{+} {\hat S}_{l+1}^{-}+{\rm h.c.})+\frac{\bar U}{2}({\hat S}_{l}^{z}+1/2)], \label{Hxx}
\end{equation}
where ${\hat S}_{l}^{+/-}$ and ${\hat S}_{l}^{z}$ represent the respective 
pseudo-spin-$1/2$ operators.  There is a macroscopic degeneracy $2^L$ increasing with
the number of sites $L$, since each pseudospin state represent two different
but equivalent local states for each site.  
The $xy-$model in Eq.~(\ref{Hxx}) is exactly solvable, where
 ${\bar U}$ provides a Zeemann splitting between $|\!\!\uparrow\rangle$ and $|\!\!\downarrow\rangle$. 
For  ${\bar U}>4J$, 
the system is saturated with only single occupied sites and a finite charge gap
corresponding to the MI phase.
When ${\bar U}\leq 4J$, the ground state is in a gapless $xy$ phase {\it without} SF response
indicated by a blue vertical line
in Fig.~\ref{fig1}(b). Details of the solution and correlations at the degenerate line
are discussed in the Supplemental Materials \cite{SM}.

To obtain 
the full quantum phase diagram at half-filling, 
we now use a combination of three independent advanced 
numerical simulation methods.  The density matrix renormalization group (DMRG) 
method~\cite{White_1992, White_1993, Peschel_1999, Schollwoeck_2005} is used to 
measure properties of finite-size chains, such as the charge gap $\Delta_c$, the superfluid 
density $\rho_s$,  
and correlation functions using
up to $M=4096$ states.
With the further development of the DMRG to infinite systems 
(iDMRG)~\cite{McCulloch_2008, Hu_2011, Hu_2014}, we can moreover determine the fidelity 
susceptibility $\chi_F$ and the entanglement entropy $\cal S$ directly 
in the thermodynamic limit.
Last but not least the stochastic series expansion algorithm of the quantum Monte Carlo (QMC) method with parallel tempering~\cite{sse1,sse2,sse3}
is used to calculate the compressibility $\kappa$ close to the zero temperature limit.

\begin{figure}[t]
\includegraphics[width=0.99 \columnwidth]{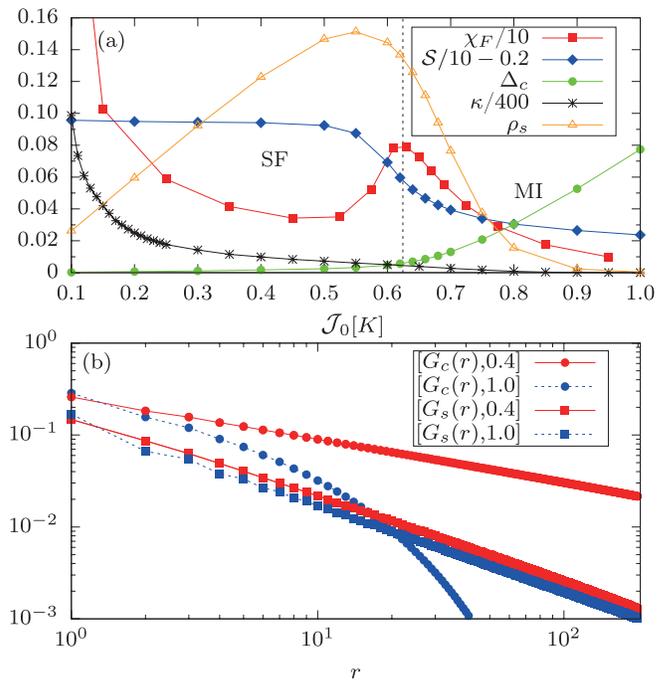}
\caption{Different observables 
at $J/{\bar U}=0.4$.  
(a) Fidelity susceptibility $\chi_{F}$ and entanglement entropy $\cal S$ 
from iDMRG ($L=\infty$); 
charge gap $\Delta_{c}$ 
and superfluid density $\rho_s$
from DMRG ($L=100$); 
compressibility $\kappa$ from QMC ($L=100$).
(b) Single-particle correlation 
$G_c(r)=\langle {\hat a}^{\dagger}_{0} {\hat a}_{r} \rangle$ ($\bigcirc$)
and density-hole-pair 
correlation $G_s(r)=\langle {\hat a}^{\dagger}_{0} {\hat b}_{0} {\hat a}_{r} {\hat b}^{\dagger}_{r} \rangle$ ($\square$) as a function of distance $r$ 
relative to $L/4$ for ${\cal J}_{0}[K]=0.4$ (solid line) and $1$ (dashed line)
 obtained by DMRG ($L=100$).
}\label{fig2}
\end{figure}

As shown in Fig.~\ref{fig2}(a) for $J=0.4 \bar U$
we now observe signatures of a quantum phase transition at half-filling
as a function of the effective hopping ${\cal J}_0[K]$, which is reduced 
by the driving amplitude $K$.
Because the phase transition
is of Berezinskii-Kosterlitz-Thouless (BKT) type~\cite{Kuhner}, 
finite-size effects are only logarithmically small.
Therefore, measuring the transition point
numerically by physical observables 
is very tricky and inaccurate, so we employ a combination of methods.
Only in the full thermodynamic limit, the charge gap increases from zero, the global compressibility goes to zero, the entanglement entropy drops from infinity to a finite value and the 
fidelity susceptibility becomes extremely sharp at the transition point.
The superfluid density $\rho_s$ can be obtained {using} DMRG
from the 
second-order response $[E_{0} (\theta)-E_{0}(0)]/\theta^2$ of the ground-state energy $E_{0}$ to a twist-angle $\theta$~\cite{Roth_2003}.
The response $\rho_s$
is finite and increasing for small ${\cal J}_0[K]$, which shows that the system is indeed in 
a superfluid phase for this part of the phase diagram.  The increase of $\rho_s$ with 
effective hopping ${\cal J}_0[K]$ Fig.~\ref{fig2}(a) is not surprising,
since for smaller ${\cal J}_0[K]$ the hopping
of type-$a$ bosons is blocked by a changing occupation of type-$b$ and vice versa.
However, for larger ${\cal J}_0[K]$ a maximum and sudden drop
to $\rho_s \to 0$ as ${\cal J}_0[K]\to 1$ signals a quantum phase transition to the well-established
Mott state in the undriven system~\cite{book,Lieb_1968}.
To pinpoint the transition point, we consider 
the fidelity susceptibility 
$\chi_F ({\bar x}) = -2\ln F(x_{1}, x_{2})/\delta^2$, which  is defined via the 
overlap 
of ground states
 $F(x_{1}, x_{2})=\langle \psi_{0} (x_{1}) | \psi_{0} (x_{2})\rangle$ 
with $\delta=|x_{1}-x_{2}|$
and ${\bar x}=(x_{1}+x_{2})/2$ for two close values   
$x_{1}$ and $x_{2}$ of the parameter ${\cal J}_0[K]$.
A  peak in $\chi_F$ 
is a clear signal of a quantum phase transition~\cite{You_2007,Campos_Venuti_2007}, which 
occurs at ${\cal J}_{0}[K]_{c}=0.624(6)$.
In addition, 
the entanglement entropy ${\cal S}=-\textrm{Tr} \rho_{r} \ln \rho_{r}$ is obtained
from the partial trace of the reduced density matrix for half the 
system~\cite{Osterloh_2002,Wu_2004,Laflorencie_2016}, which shows a distinct drop 
in the vicinity of the transition point.  
Using QMC we find the compressibility 
$\kappa=\langle {\hat {\cal N}}^{2} \rangle - \langle {\hat {\cal N}} \rangle^2$
for $L=100$ sites {at low temperatures}
which vanishes in the deep Mott phase. The charge gap 
$\Delta_{c}=E_{p}+E_{h}-2E_{0}$ is found by DMRG from the energies of systems
with one additional particle $E_p$ and one additional 
 hole $E_h$ relative to the ground state and becomes
finite in the MI.
After finite-size scaling analysis on ${\cal J}_{0}[K]_{c}$ by level-spectroscopic technique~\cite{SM,fss}, we find it matches well with the maxima in $\chi_F$ within errorbars, so we use the latter to obtain the full phase diagram in Fig.~\ref{fig1}(b). 

At first sight it is strange that the reduction in hopping  ${\cal J}_0[K]$
can induce a SF state, since normally  weaker hopping makes the MI more stable.
However, in this case the density-dependent processes in Fig.~\ref{fig1}
are responsible for a virtual exchange, which reduces the energy of an
alternating density order $ababab...$ to 
second order $4 J^2 {\cal J}_0^2[K]/\bar U$ \cite{Altman_2003}. Therefore, 
by selectively tuning away those processes via periodic driving,
the alternating order
and the corresponding MI are actually {\it destabilized}, which in turn enables
a SF for finite $\bar U$.
For ${\cal J}_0[K]=0$ the system has no
$a$-$b$-density correlations, which leads to the degeneracy discussed above.

It is instructive to
analyze the characteristic correlation functions for the different phases as 
shown in Fig.~\ref{fig2}(b) for $J=0.4 \bar U$.
The single-particle correlation $G_c(r)=\langle {\hat a}^{\dagger}_{0} {\hat a}_{r}\rangle$
shows a typical power-law decay in the SF phase ${\cal J}_{0}[K]=0.4<{\cal J}_{0}[K]_{c}$, while 
an exponential decay is a signature of a MI for ${\cal J}_{0}[K]=1>{\cal J}_{0}[K]_{c}$.
The particle-hole-pair correlation $G_s(r)=\langle {\hat a}^{\dagger}_{0}
{\hat b}_{0} {\hat a}_{r} {\hat b}^{\dagger}_{r}\rangle$ on the other hand shows
a slow power-law decay in either phase.

\begin{figure}[t]
	\includegraphics[width=0.99 \columnwidth]{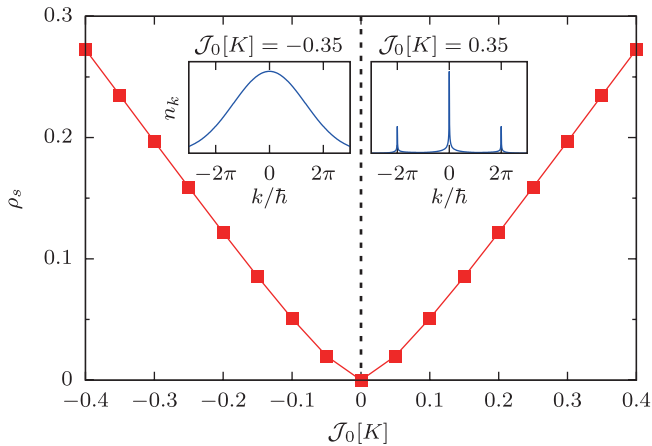}
	\caption{Superfluid density $\rho_{s}$ per site calculated by DMRG with $L=100$ at
$J= \bar U$.    
Inset: Momentum distribution $n^b_{k}$ at ${\cal J}_{0}[K]=\pm 0.35$, 
which is normalized by its maximal value.}\label{fig3}
\end{figure}

We now turn to negative effective hopping ${\cal J}_0[K] < 0$.
The corresponding phase diagram and the superfluid density are shown in
Fig.~\ref{fig1}(b) and Fig.~\ref{fig3}, respectively.  At first sight the
results look perfectly symmetric around ${\cal J}_0[K]=0$, which would suggest that 
negative hopping has the same effect as positive hopping.  However, the 
underlying states for positive and negative values
are quite different, which becomes clear by looking at the
	signature of the momentum distribution (MD) $n^{b}_{k}$ defined by 
\begin{equation}
n^{b}_{k}=\left|w(k)\right|^2 \sum^{L}_{l,l'=1} \exp[ik(l-l')/\hbar]\langle 
{\hat b}_{l}^{\dagger} {\hat b}_{l'}^{\phantom{\dagger}}\rangle,
\end{equation}
as a function of momentum $k$, where 
$w(k)$ stands for the Fourier transformation of the Wannier function in a 1D optical lattice
with lattice spacing equal to one~\cite{wannier}.
As shown in the inset of Fig.~\ref{fig3} the MD shows 
an interference pattern with sharp peaks at $k=0$ (modulo $2\pi$) for 
positive values ${\cal J}_{0}[K]=0.35$, which originates from the phase coherence of bosons 
in the normal SF. However in the region ${\cal J}_{0}[K]<0$, no sharp interference pattern is
observed. 

Both the symmetry in $\rho_s$ and the 
difference in the MD interference pattern can be explained by a
gauge transformation which defines new quasi-particles of type-$\beta$
${\hat \beta}_{l}={\hat b}_{l} \exp(i\pi{\hat n}_{l}^{a})$ and analogous for type-$\alpha$.
We see that the hopping terms in Eqs.~(\ref{He})--(\ref{hatJ}) can then be written 
as
\begin{eqnarray}
\hspace*{-0.3cm}\hat J_l^b \hat b_l^\dagger \hat b_{l+1}^{\phantom{\dagger}} & = & 
 J\left(\tfrac{1+{\cal J}_0[K]}{2} + \tfrac{1-{\cal J}_0[K]}{2}e^{i\pi({\hat n}_{l}^{a}+{\hat n}_{l+1}^{a})}\right) \hat b_l^\dagger \hat b_{l+1}^{\phantom{\dagger}} \nonumber \\
 &= & J\left(\tfrac{1+{\cal J}_0[K]}{2}e^{i\pi({\hat n}_{l}^{a}+{\hat n}_{l+1}^{a})} + \tfrac{1-{\cal J}_0[K]}{2} \right)  \hat \beta_l^\dagger \hat \beta_{l+1}^{\phantom{\dagger}} \label{hop}
\end{eqnarray}
and likewise for $\hat J_l^a \hat a_l^\dagger \hat a_{l+1}^{\phantom{\dagger}}$.
Since the densities are not affected $\hat n_l^{a/b} = \hat n_l^{\alpha/\beta}$, 
a change of sign ${\cal J}_0[K]\to -{\cal J}_0[K]$ is therefore equivalent to a transformation
$\hat b_l \to \hat \beta_l$ and $\hat a_l \to \hat \alpha_l$ in Eq.~(\ref{hop}).  Accordingly, the energies and
phase transition lines are identical for positive and negative ${\cal J}_0[K]$, but the 
superfluid density for negative sign corresponds to a response of gauge-paired 
particles $\hat \alpha, \ \hat \beta$ and is therefore called a {\it gauge-dressed SF}
with a different MD shown in the inset of Fig.~\ref{fig3}. The transition to
such an exotic condensed density can also be captured by a 
Gutzwiller mean-field argument which is discussed in the Supplemental Materials \cite{SM}.
Note, that the symmetry transformation to new gauge-paired particles in Eq.~(\ref{hop}) is
independent of the dimensionality and geometry of the lattice.

Thus, the gauge dressed SF is characterized by a lattice gauge $\exp(i \pi {\hat n}^{a}_{l})$ provided by one species (type-$a$) which couples to the hopping 
of the other species (type-$b$) and vice versa. As can be seen from Eq.~(\ref{hop})
the gauge dressed hopping becomes dominant in the strongly driven region ${\cal J}_{0}[K]<0$, 
resulting in a superfluid response from gauge dressed particles.
The quantum phase transition to a MI is analogous to an ordinary SF and happens at exactly the
same critical value of $J/\bar U$ {in Fig.~\ref{fig1}(b)
as for corresponding positive ${\cal J}_{0}[K]>0$ since the gauge does not change the
energy response to a twist-angle $\theta$.} 
The gauge dressed SF is therefore different from pair superfluidity, where
correlated hopping is observed due to a strong coupling of the 
hopping directly to the density \cite{Rapp_2012,Schmidt_2006}.
The so-called counterflow SF is another type of correlated hopping~\cite{Altman_2003, Kuklov_2003, Kuno_2013}, where 
hopping of particles of one species is facilitated by 
holes of the opposite species.  In contrast, in the new gauge dressed SF the 
hopping is facilitated 
by gauges $\exp(i \pi {\hat n}^{a}_{l})$, which can also be viewed as particles that 
are their own anti-particles, analogous to a Majorana description.

For the experimental realization of these phases several critical questions must be 
solved.  First of all 
accessing the steady state 
by adiabatic ramping of the driving amplitude from the ground
state 
is only possible, when no dense avoided
level crossing of the Floquet quasi-energy take place.  
Our analysis of the quasi-energy spectrum
in the Supplemental Materials \cite{SM} 
ensures that there are no critical avoided level crossings in the relevant parameter
range. 
Secondly, a measurement can be affected by the unitary transformation into the effective
Floquet basis, if the operators do not commute with the Kick operator \cite{SM}.
For stroboscopic measurements at times of integer multiples of period $T=2 \pi/\omega$ we 
show in the Supplemental Materials \cite{SM} that this effect
is reduced by $J/\hbar \omega$ in the high frequency limit and calculate the
corrections from higher order terms, in order to predict the 
experimental mapping of the phase diagram by time-of-flight and compressibility 
measurements.  For a separate check of the predictions we also performed
real-time simulations {for a small lattice $L=6$} \cite{SM} which clearly show the stability of the
effective Hamiltonian and the feasibility of real-time dynamic measurements on finite 
time- and length-scales. 

In conclusion, we proposed a setup of a 1D lattice with two species of 
hard-core bosons and time-periodically modulated fields, which can be 
described by  density-dependent tunneling with an interesting quantum phase diagram.  
By controlling the driving amplitude, density dependent hopping processes are
selectively tuned away, which are 
responsible for an alternating density $a$-$b$ order.  
This in turn leads to a transition from the MI to a SF at half-filling 
in contrast to the undriven case. 
By tuning away these terms completely at ${\cal J}_0[K]=0$, a
highly degenerate state is obtained corresponding to 
an exactly solvable model without $a$-$b$ correlations. 
For many-body systems
the study of nearly degenerate points is a very active research area, e.g.~in the context of 
frustrated models, spin ice, and spin liquids.  Much theoretical activity is
devoted to studying novel quantum states, which are dominated by
the quantum fluctuations near degenerate points, but we are not aware of any such studies
for driving-induced degeneracy.  In this case, dynamical effects will likely dominate the 
quantum correlations, which opens an interesting research field beyond our current abilities.
For even larger driving amplitudes, negative hopping parameters ${\cal J}_0[K]<0$ 
lead to a new {\it gauge dressed SF} with a novel type of pairing mechanism,  where
an atom of one species and a gauge phase of the other are bound to contribute to a
nonzero superfluidity.
This gauge dressed SF has different correlations from an ordinary
SF, as shown in Fig.~\ref{fig3} for the momentum distribution.  Nonetheless, 
an exact hidden transformation to the positive hopping case
can be found.

\begin{acknowledgments}
We thank Youjin Deng, Shaon Sahoo, Oliver Thomas, and Zhensheng Yuan for useful discussion. This research was supported
by the Special Foundation from NSFC for theoretical physics Research Program of China (No 11647165),
by the Nachwuchsring of the TU Kaiserslautern,
by the German Research Foundation (DFG) via the Collaborative Research Center
and SFB/TR185 (OSCAR).
Especially, we gratefully acknowledge the computing time granted by the John von Neumann Institute for Computing (NIC)
and provided on the supercomputer JURECA at J\"ulich Supercomputing Centre (JSC). X.-F. Z. acknowledges funding from Project No. 2018CDQYWL0047 and 2019CDJDWL0005 supported by the Fundamental Research Funds for the Central Universities, Grant No. cstc2018jcyjAX0399 by Chongqing Natural Science Foundation, and from the National Science Foundation of China under Grants No. 11804034, No. 11874094 and No. 11847301.
\end{acknowledgments} 

\begin{widetext}

\section{APPENDIX}

Here, we discuss two proposals of the experimental realization, higher orders in the effective Hamiltonian, the role of the Kick operator, avoided level crossing of the quasi-energy spectrum, real-time dynamics of the original time-dependent Hamiltonian, finite size scaling, and the Gutzwiller mean field method.

\section{Experimental realization} 
Initially we assume to have ultra-cold atoms in one hyperfine state confined in an optical dipole trap.\cite{Lin_2011} The trap consists of a pair of counter-propagating laser beams with wave length $\lambda_{L}$ along
the $x$-axis, so the potential function reads
\begin{eqnarray}
V(x) = V_{0} \sin^{2} \left(k_{r} x\right)
\end{eqnarray}
with the lattice depth $V_{0}$ and the wave vector $k_{r} = 2 \pi/\lambda_{L}$. Using an initially off-resonant radio-frequency magnetic field, we suggest to adiabatically ramp the Zeeman fields to zero and decrease
the radio-frequency coupling strength to a certain value for a while. Then we propose to suddenly turn off the coupling strength projecting the BEC into an {\it equal} superposition of two hyperfine states. In the single-band
approximation, the movement of the atoms appears in form of hoppings between two neighboring sites, e.g.~the minima of the lattice potential with spacing $\lambda_{L}/2$, and the related Hamiltonian reads
\begin{equation}
{\hat H}_{T}=-\sum_{l} \left(J_{a} {\hat a}^{\dagger}_{l} {\hat a}_{l+1} + J_{b} {\hat b}^{\dagger}_{l} {\hat b}_{l+1} + \rm{h.c.}\right)\,,
\end{equation}
where ${\hat a}_{l}$ (${\hat b}_{l}$) and ${\hat a}^{\dagger}_{l}$ (${\hat b}^{\dagger}_{l}$) are the annihilation and creation operators,  $J_{a}$ ($J_{b}$) the hopping coefficient of atoms with hyperfine level
``$a$" (``$b$") and the site index $l$ runs over the whole lattice. Obviously, $J_{a} = J_{b} = J$ because the hopping processes are independent of the hyperfine internal states 
of atoms.\cite{Demler_2002} %

The depth of the optical lattice potential can affect the on-site repulsive interaction between the ultra-cold atoms,\cite{Demler_2002} which is independent of the hyperfine states unless we are close to a Feshbach resonance.
Therefore, increasing the lattice depth $V_{0}$ will generate 
large intra- and inter-species repulsive interactions independent of the hyperfine states 
\begin{equation}
\hat{H}_{L} = \frac{U_{L}}{2} \sum_{l} \left({\hat n}^{a}_{l} + {\hat n}^{b}_{l}\right) \left({\hat n}^{a}_{l} + {\hat n}^{b}_{l} - 1\right)\, ,
\end{equation}
where ${\hat n}^{a}_{l}={\hat a}^{\dagger}_{l} {\hat a}_{l}$ (${\hat n}^{b}_{l}={\hat b}^{\dagger}_{l} {\hat b}_{l}$) denote the particle number operator of the species ``$a$" (``$b$") and $U_{L} \gg J$.

Furthermore, we suggest to add a static magnetic field and tune its amplitude $B$ to be very close to the Feshbach resonance point $B_{0}$, where two-species atoms form s-wave bound states,
while it is far away from intra-species Fesh\-bach resonance points for both species.\cite{Chin_2010,Verhaar_2002} On the side of the negative scattering length, an attractive inter-species interaction emerges, namely
\begin{equation}
\hat{H}_{F} = -U_{F} \sum_{l} {\hat n}^{a}_{l}{\hat n}^{b}_{l}
\end{equation}
with $U_{F} \gg J$. It can compensate the inter-species repulsion $U_{L}$ and results in a total finite inter-species repulsion $U=U_{L} - U_{F}$, which is assumed to be
of the order of $J$. Meanwhile a large intra-species repulsion $U_{L}$ still
leads to a {\it hard-core} constraint, which means more than one atom from the same species is forbidden at the same lattice site.  Taking $^{87}$Rb atoms for example, we can choose the magnetic field a little
smaller than the inter-species Feshbach resonance point $1259.96$~G and far away from the intra-species ones, which amounts to be $685.43$~G for $|F=1, m_{F}=1\rangle$ (``a") and $661.43$~G
for $|F=1, m_{F}=0\rangle$ (``b").\cite{Verhaar_2002}

To fulfill the hardcore constraint, the lattice depth $V_0$ must be
choosen significantly larger than the recoil energy $E_r=h^2/2m\lambda^2_L$, i.e.
we need 
$s = V_0/E_r \agt  20$ or larger,\cite{Jaksch_1998} 
while the hopping is approximately
 $J \sim 4 E_r s^{3/4} e^{-2 \sqrt{s}}$.\cite{zwerger} 
For Rubidium-87 we therefore have $E_r/h \sim 3.5  \rm kHz$ in a
400nm lattice, which gives $J/h\sim 17  \rm Hz$.  The rotating 
frequency of the time-periodic driving
must be much larger than $J$ and $U$, but not too large to avoid ``photon-assisted hopping"
between different energy bands of the optical lattice.
In the following discussion of possible realizations
we therefore assume a rotating frequency $\omega$ of the order of 1 kHz, 
which will also be the order of magnitude of the driving amplitude.

\subsection{Periodically modulated inter-species interaction}

A straight-forward, but technologically challenging time-periodic driving can be 
realized by an oscillating magnetic field
$B(t)={\bar B}+\delta B\cos \omega t$ near $B_{0}$, where $\bar B$ denotes the time-average strength of the magnetic field, $\delta B$ represents the oscillation amplitude
of the magnetic field and $\omega$ stands for the oscillating frequency as shown in Fig.~\ref{figs1} (a). Thus the relevant s-wave scattering length can be written as
\begin{equation}
a_{s}(t)=a_{\rm bg} \left(1-\frac{\Delta}{{\bar B}+\delta B \cos (\omega t)-B_{0}}\right),
\end{equation}
where $\Delta$ is the width of the Feshbach resonance and $a_{\rm bg}$ represents the background scattering length, which is determined by the lattice depth. If we choose $\delta B \ll |{\bar B}-B_{0}|$, we can further perform
a Taylor series expansion of $a_{s}(t)$  with respect to a small value of $\delta B/({\bar B}-B_{0})$ and get
\begin{figure}[t]
	\includegraphics[width=0.31 \columnwidth]{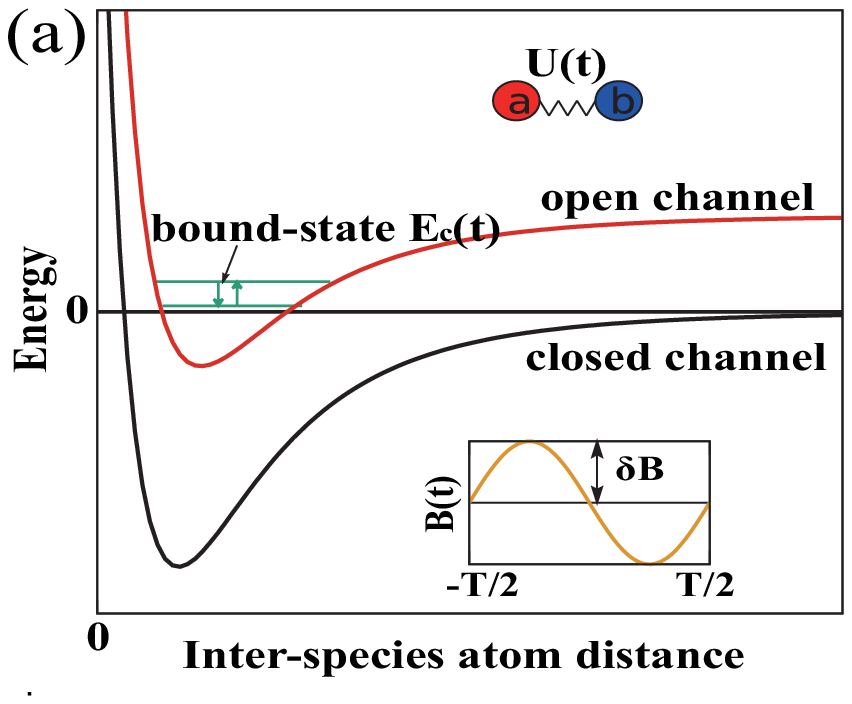}\includegraphics[width=0.31 \columnwidth]{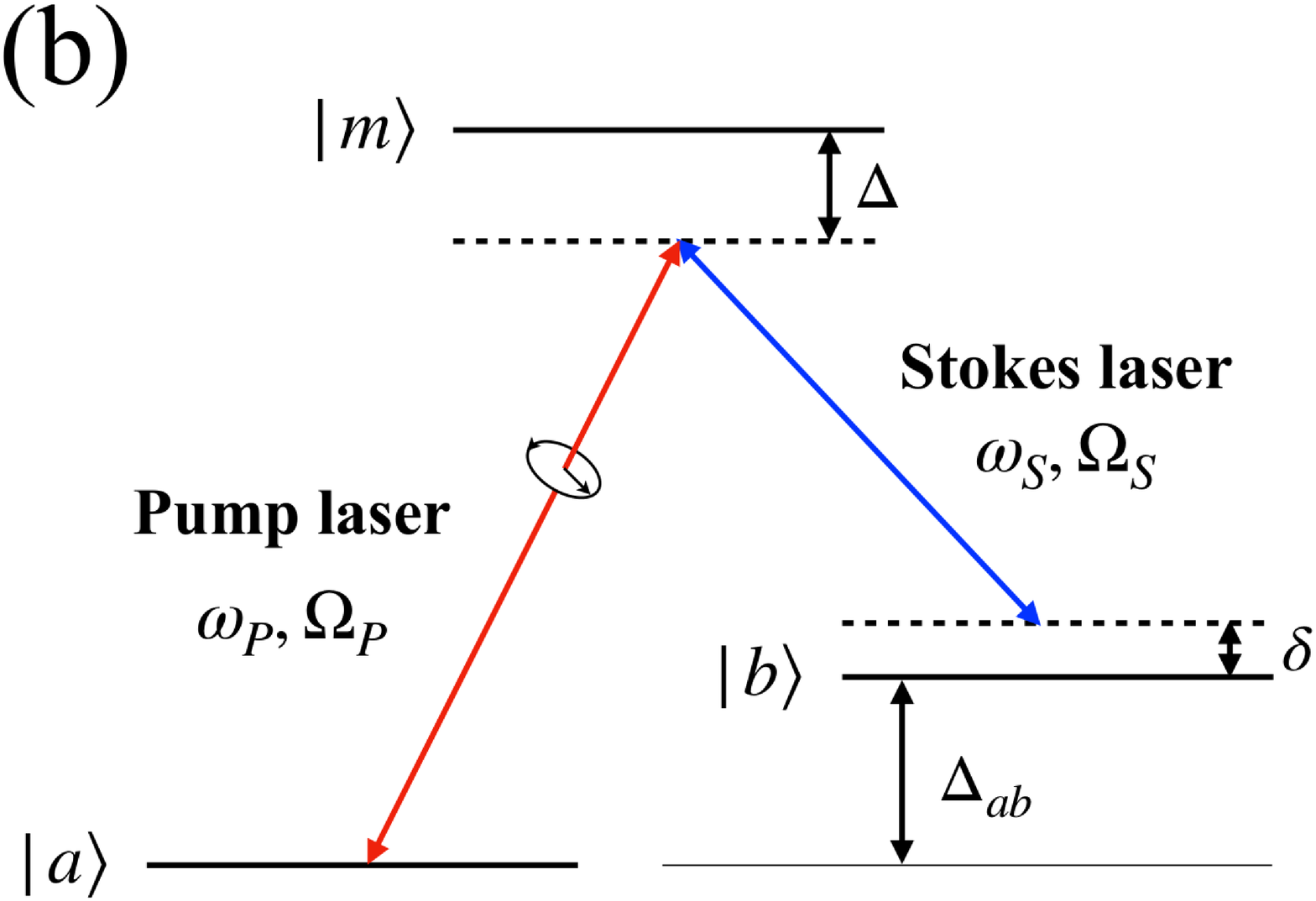}\includegraphics[width=0.31 \columnwidth]{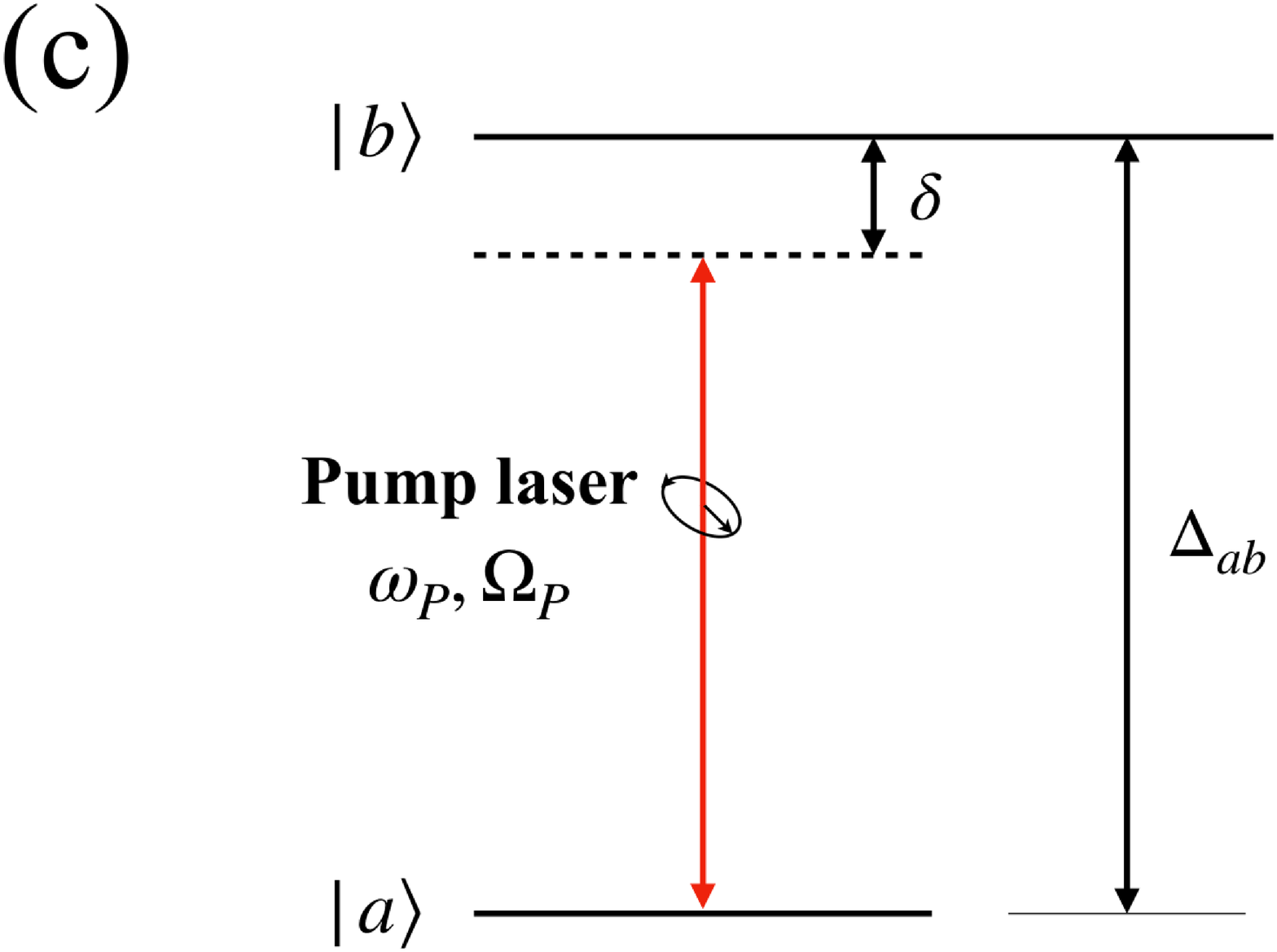}
	\caption{(a) The realization of time-periodically modulated inter-species interaction potential energy by imposing a small cosine type periodically modulated magnetic field (see inset) near Feshbach resonance.
          (b) Schematic picture of the standard simulated Raman transition. An atom jumps from the hyperfine state ``$a$" to the intermediate state ``$m$" by absorbing a photon from a pump laser (red line,
          frequency $\omega_{P}$, {{fast rotating linearly polarized}}, coupling strength $\Omega_{P}$). Similarly an atom jumps from ``$m$" to ``$b$" by emitting a photon from a Stokes laser (blue line,
          frequency $\omega_{S}$, {\it linearly polarized}, coupling strength $\Omega_{S}$).\cite{Naber_2016} (c) Schematic picture of a two-level system directly coupled via a pump laser. An atom jumps from
          the hyperfine state ``$a$" to ``$b$" by absorbing a photon from a pump laser (red line, frequency $\omega_{P}$, {{fast rotating linearly polarized}}, coupling strength $\Omega_{P}$).
          {One realization of the pump laser is the output of a circularly polarized laser passing a $1/4$ wave plate which is connected to a fast rotating mechanical motor (frequency $\omega$ sets several kHz).}
	}\label{figs1}
\end{figure}

\begin{equation}
a_{s}(t) = a^{(0)}_{s}+a^{(1)}_{s} \cos(\omega t) + {\cal O}\left( \left(\frac{\delta B}{{\bar B}-B_{0}}\right)^2\right),
\end{equation}
where the coefficients of the leading orders are $a^{(0)}_{s}=a_{\rm bg}\left[1-\Delta/(\bar{B}-B_{0})\right]$ and $a^{(1)}_{s}=a_{\rm bg} \Delta \delta B/({\bar B}-B_{0})$. Note that a pure cosine oscillation of $a_s$ is in principle also possible for larger amplitude $\delta B$
if the waveform of the magnetic field is adjusted correspondingly.
The inter-species interaction energy $U(t)$ is proportional to the related scattering length, which means $U(t)={\bar U}+\delta U\cos(\omega t)$, where the time-average energy ${\bar U}$ is proportional to $a^{(0)}_{s}$ and
the oscillation amplitude $\delta U$ is proportional to $a^{(1)}_{s}$ if we neglect higher-order terms. In this case, the system can be described by the following Hamiltonian:
\begin{eqnarray}
{\hat H}_{1} (t) =-J \sum_{l} \left({\hat a}^{\dagger}_{l} {\hat a}_{l+1} + {\hat b}^{\dagger}_{l} {\hat b}_{l+1} + \rm{h.c.}\right) + U(t) \sum_{l} {\hat n}^{a}_{l}{\hat n}^{b}_{l}\,.
\label{H1}
\end{eqnarray}

\subsection{Periodically modulated Rabi oscillation}

Fast oscillating fields are possible but challenging, so 
an alternative experimental realization in a static magnetic field is useful.
To this end we propose to gradually switch on a pair of Raman laser beams, which are coupled to the atomic cloud with the frequency difference $\delta \omega_{R}$.
An atomic transition from the hyperfine state
``$a$" to ``$b$" by a two-photon emission-absorption has the standard $\Lambda$-form\cite{Naber_2016} shown in Fig.~\ref{figs1} (b). One pump laser initiates that atoms jump from the hyperfine state ``$a$" to the
intermediate state ``$m$" by absorbing a photon with frequency $\omega_{P}$ and coupling strength $\Omega_{P}$, while the other Stokes laser triggers atoms to jump from ``$m$" to ``$b$" by emitting a
photon with frequency $\omega_{S}$ and coupling strength $\Omega_{S}$.\cite{Naber_2016}
Both coupling strengths depend on the projection of a transition dipole moment onto the polarization of the Raman laser beams,\cite{Naber_2016} namely
\begin{eqnarray}
\Omega_{P} = \langle m | {\bf d}_{P} \cdot {\bf E}_{P}  | a \rangle,\quad \Omega_{S} = \langle b | {\bf d}_{S} \cdot {\bf E}_{S}| m \rangle\, ,
\end{eqnarray}
where ${\bf d}_{P/S}$ denote the transition dipole momenta and ${\bf E}_{P/S}$ represent the electric field of the respective Raman laser beams. The
transition dipole momenta only depend on the initial and final states and are unchanged
during the two-photon transition processes.
%

%
At the stimulated Raman transition, the frequency difference $\delta \omega_{R}$ exactly coincides with the hyperfine splitting between two levels $\Delta_{ab}$.
In order to avoid a resonant excitation of the intermediate state,
the detuning $\Delta<0$ of the Raman beam from the one-photon transition has to be much larger than the linewidth of the excited level and it should also be much larger than the Raman detuning $\delta$ from the
two-photon resonance. In this way, the system can be considered as an effective two-level systems ``$a$" and ``$b$". Its Hamiltonian reads 
\begin{eqnarray}
{\hat H}_{\Omega} = \left(
\begin{array}{cc}
E_{P} & J_{\Omega} \\
J_{\Omega} & E_{S} + \delta \\
\end{array}
\right)
\end{eqnarray}
in the basis of the two hyperfine levels with the energies $E_{P/S}=\Omega^2_{P/S}/\Delta$ and the so-called ``Rabi frequency'' $J_{\Omega}=\Omega_{P}\Omega_{S}/\Delta$.
After the second quantization, we obtain the Hamiltonian for the Rabi oscillation
\begin{eqnarray}
{\hat H}_{\Omega}=J_{\Omega} \sum_{l}\left(\hat{a}_{l}^{\dagger}\hat{b}_{l}+\rm{h.c.}\right) \, ,
\end{eqnarray}
where we neglect the small shift of the chemical potential $\delta$ between two hyperfine levels.

In order to produce a time-periodic oscillating Rabi coupling strength, we let the polarization direction of the pump laser circulate in time, e.g.~$E_{x}=A \cos(\omega t)$ and $E_{z}=A \sin(\omega t)$ with the amplitude
of the polarization $A$.
For realization, a circularly-polarized pump laser may pass through $1/4$ wave plate to get a 
linearly-polarized laser beam as output, the polarization direction of which is 45 degree shifted to the optical axis of the wave plate. Sequentially, the wave plate is connected to a mechanical motor with rotating frequency $\omega$ in the kHz range, 
which is much lower than the laser frequency of several hundreds of THz ($10^{12}$Hz). Therefore the  polarization of the pump laser is also rotating and effectively provides a time-modulated Rabi
coupling.
In order to avoid
coupling to other magnetic sublevels when the linear
polarization is rotated, we assume a sufficiently strong Zeeman splitting.
Therefore assuming $d_{z} = 0$, we get
\begin{eqnarray}
\Omega_{P} = \langle m | A d_{x} \cos(\omega t) | a \rangle = \Omega^{0}_{P} \cos(\omega t)
\end{eqnarray}
with $\Omega^{0}_{P} = A d_{x}$. As a result, the effective Rabi-frequency turns out to be time-periodic
\begin{eqnarray}
J_{\Omega}(t)=J^{0}_{\Omega} \cos(\omega t)
\end{eqnarray}
with the amplitude $J^{0}_{\Omega} = \Omega^{0}_{P}\Omega_{S}/\Delta$. We can use, for instance, an acousto-optic modulator (AOM) to change both the amplitude and the polarization of the pump laser.\cite{Sapriel_1979,Eklund_1975}

In an extra scheme, we suggest to prepare a cloud of ultra-cold atoms with equally-weighted species but with different total angular momenta {and long life time}. They directly couple to a {fast rotating linearly polarized pump laser beam with rotating 
frequency in the kHz range}. In this way, we can generate the same time-periodic Rabi oscillation in the Eq.~(9), see Fig.~{\ref{figs1}} (c). 

In general, the full Hamiltonian reads then
\begin{eqnarray}
{\hat H} (t)
=-J \sum_{l} \left({\hat a}^{\dagger}_{l} {\hat a}_{l+1} + {\hat b}^{\dagger}_{l} {\hat b}_{l+1} + \rm{h.c.}\right) +\bar{U}\sum_{l} {\hat n}^{a}_{l}{\hat n}^{b}_{l}
+ J_{\Omega} (t) \sum_{l}\left(\hat{a}_{l}^{\dagger}\hat{b}_{l}+\rm{h.c.}\right). \label{H2}
\end{eqnarray}

\section{Effective Hamiltonian and kick operators}
For the wave function of a system, which is described by a time-periodically driven Hamiltonian $\hat{H}(t)=\hat{H}(t+T)$ with the period $T=2 \pi / \omega$ being determined by the driving frequency $\omega$,
the time-dependent Schr\"{o}dinger equation holds
\begin{eqnarray}
i\hbar \frac{\partial}{\partial t} |\psi(t)\rangle = {\hat H}(t) |\psi(t)\rangle\,,\label{TDSE}
\end{eqnarray}
where the steady-state after long times obeys the Floquet theory.\cite{Floquet} As a result, the wave function is of the form $|\psi(t)\rangle = e^{-i\epsilon t/\hbar} |\phi(t)\rangle$
with time-periodic Floquet modes $|\phi(t)\rangle = |\phi(t+T)\rangle$, where $\epsilon$ represents the corresponding quasi-energy. With this
the time-dependent Schr\"{o}dinger equation (\ref{TDSE}) becomes

\begin{eqnarray}
{\hat {\cal H}}(t) |\phi(t)\rangle = \epsilon |\phi(t)\rangle
\label{eigen}
\end{eqnarray}
with the Floquet Hamiltonian ${\hat {\cal H}}(t)={\hat H} (t)-i\hbar \partial/\partial t$. In the extended Hilbert space including the time dimension, Eq.~(\ref{eigen}) is the eigenvalue equation
of the Floquet Hamiltonian ${\hat {\cal H}}(t)$.  In the high frequency regime, where $\hbar \omega$ is large, it is natural to choose the eigenfunctions of the operator
$-i\hbar \partial/\partial t$ as the basis and to treat the whole Hamiltonian ${\hat H}(t)$ perturbatively.
Then, we can  use a nearly-degenerate perturbative method to solve Eq.~(\ref{eigen}).\cite{Eckardt_2015} Although the full solution of Eq.~(\ref{eigen}) is still
difficult because of the intricate quasi-energy spectrum, we can calculate the time-independent effective Hamiltonian and the kick operator order by order. Thus, the whole dynamics of the system can be
described by the effective Hamiltonian and the corresponding kick operator.

However in our cases we can not perform directly perturbative calculations because of the extra energy scales $\delta U$ and $J^{0}_{\Omega} \sim \hbar \omega$.
Therefore, we need to apply a rotation ${\hat V}$ at the preliminary step in order
to eliminate these extra terms by moving them to the phases of the respective hopping terms.\cite{Bukov_2015}
In the rotating frame after the unitary transformation, the wave function turns out to obey the transformed time-dependent Schr\"{o}dinger equation
\begin{eqnarray}
i\hbar \frac{\partial}{\partial t} |\psi_{r}(t)\rangle = {\hat H}_{r}(t) |\psi_{r}(t)\rangle\,,
\end{eqnarray}
with $|\psi_{r}(t)\rangle = {\hat V}^{\dagger} |\psi(t)\rangle$ and ${\hat H}_{r} (t) = {\hat V}^{\dagger} [ {\hat H} (t) - i\hbar\partial /\partial t ] {\hat V}$, i.e.~the quickly 
rotating
phases have been absorbed in the transformation.

\subsection{Periodically modulated inter-species interaction}
For the periodically modulated inter-species Hamiltonian (\ref{H1}) the rotation operator is given by ${\hat V}(t)=\exp[-i {\tilde K}_{U} \sum_{l}{\hat n}^{a}_{l}{\hat n}^{b}_{l}]$ with dimensionless modulation strengths
$K_{U} = \delta U/\hbar \omega$ and ${\tilde K}_{U} = K_{U} \sin(\omega t)$. With this the Hamiltonian after the rotation results in 
\begin{equation}
\hat{H}_{r}(t)=-J \sum_{l} \left[ {\hat a}^{\dagger}_{l}e^{i {\tilde K}_{U} ({\hat n}^{b}_{l}-{\hat n}^{b}_{l+1})} {\hat a}_{l+1}
+ {\hat b}^{\dagger}_{l}e^{i{\tilde K}_{U} ({\hat n}^{a}_{l}-{\hat n}^{a}_{l+1})} {\hat b}_{l+1} + \rm{h.c.}\right] + {\hat H}_{\bar U}\,.
\end{equation}
Here $\hat{H}_{r}$ is again periodic and can thus be expanded into a Fourier series 
$\hat{H}_{r}=\sum_{n=-\infty}^{\infty}\hat{H}_r^{(n)}e^{i\omega t}$ with
\begin{equation}
\hat{H}_r^{(n)}=-J \sum_{l} \left\{ {\hat a}^{\dagger}_{l} {\cal J}_{n}\left[K_{U}({\hat n}^{b}_{l}-{\hat n}^{b}_{l+1})\right]{\hat a}_{l+1} + {\hat b}^{\dagger}_{l}{\cal J}_{n}\left[K_{U}({\hat n}^{a}_{l}-{\hat n}^{a}_{l+1})\right] {\hat b}_{l+1}
+ \rm{h.c.}\right\} +\delta_{n,0}{\hat H}_{\bar U}\,,
\end{equation}
where ${\cal J}_{n}$ denotes the $n$th order Bessel function of first kind.

Now we can calculate the effective Hamiltonian order by order.\cite{Eckardt_2015,Bukov_2015} The zeroth-order effective Hamiltonian is
\begin{equation}
\hat{H}_{e}=\hat{H}_r^{(0)}=-J \sum_{l} \left({\hat a}^{\dagger}_{l} {\cal J}_{0}\left[K_{U}({\hat n}^{b}_{l}-{\hat n}^{b}_{l+1})\right]{\hat a}_{l+1}
+ {\hat b}^{\dagger}_{l}{\cal J}_{0}\left[K_{U}({\hat n}^{a}_{l}-{\hat n}^{a}_{l+1})\right] {\hat b}_{l+1} + \rm{h.c.}\right)+\bar{U} \sum_{l} {\hat n}^{a}_{l}{\hat n}^{b}_{l}\,.
\end{equation}
Furthermore, the first-order effective Hamiltonian vanishes
\begin{equation}
\hat{H}_{e}^{(1)}=\sum^{+\infty}_{n=1} \frac{1}{n\hbar\omega} \left[{\hat H}_r^{(n)}, {\hat H}_r^{(-n)}\right] = 0,
\end{equation}
where we use the property ${\hat H}_r^{(n)}=(-1)^{n}{\hat H}_r^{(-n)}$. All second-order corrections consist of many terms, which are accompanied by the prefactor $(J/\hbar\omega)^2$ and are not listed here.
Because of $J/\hbar\omega \ll 1$ we conclude that all higher-order corrections to the zeroth-order effective Hamiltonian are small.

In order to describe the whole dynamics of the system, we also calculated the kick operator up to first order:
\begin{eqnarray}
{\hat {\cal K}}^{(0)}(t) &=& 0\,, \\
{\hat {\cal K}}^{(1)}(t) &=& \frac{1}{i\hbar\omega} \sum_{n=\pm 1, \pm2, ...} \frac{e^{in\omega t}}{n} {\hat H}_r^{(n)} \approx \frac{2{\hat H}_r^{(1)}\cos(\omega t)}{i\hbar\omega}\,. 
\end{eqnarray}
The effective Hamiltonian and the kick operator calculated above are defined in the rotating frame, but we are interested in observables in the lab frame. The link between both frames
is provided by the fact that the time evolving operators $\hat{U}(t_2,t_1)$
in the laboratory frame and $\hat{U}_{r}(t_2,t_1)$ in the rotating frame are connected by a rotation transformation, namely  $\hat{U}(t_2,t_1)=\hat{V}(t_2)\hat{U}_{r}(t_2,t_1)\hat{V}^{\dagger}(t_1)$. In this paper,
we are only interested in the stroboscopic dynamics at time $t=nT$, so we conclude $\hat{U}(t_2,t_1)=\hat{U}_{r}(t_2,t_1)$ and any observable turns out to be the same in both frames.
Furthermore, if one prepares the system in the ground state of the non-driven Hamiltonian, then adiabatically turning on the driving has the consequence that
the ground state of the system will follow the instantaneous stroboscopic Floquet
Hamiltonian.\cite{Bukov_2015} Thus the time-evolving wave-function consists of 
the ground state of the effective Hamiltonian and a phase factor from the kick operator
\begin{equation}
|\psi (t) \rangle=e^{-i {\hat {\cal K}}(t)}|\psi_{e} \rangle\,.
\end{equation}
Up to the first order, we get ${\hat {\cal K}}(nT)={\hat {\cal K}}^{(1)}(nT)=2{\hat H}_r^{(1)}/i\hbar\omega$. The expectation value of an observable $\hat {\cal O}$ results then in
\begin{equation}
\langle \psi(nT) | \hat {\cal O} |\psi(nT) \rangle=\langle  e^{\frac{2{\hat H}_r^{(1)}}{\hbar\omega}} \hat {\cal O}  e^{-\frac{2{\hat H}_r^{(1)}}{\hbar\omega}}  \rangle_e\,,
\end{equation}
where $\langle \hat A \rangle_e \equiv \langle \psi_{e} |  \hat A | \psi_{e} \rangle$.
Thus, the expectation value of an observable in the lab coincides with that of the dressed observable $e^{\frac{2{\hat H}_r^{(1)}}{\hbar\omega}} \hat {\cal O}  e^{-\frac{2{\hat H}_r^{(1)}}{\hbar\omega}}$, which is
determined by the effective Hamiltonian. As  $J/\hbar \omega$ is small, we only need to keep the two lowest orders
\begin{equation}
  \langle \hat {\cal O} \rangle =\langle \hat {\cal O} \rangle_{e}-\frac{2J}{\hbar \omega} \left\langle\left[\sum_{l} \left({\hat a}^{\dagger}_{l} {\cal J}_{1}\left[K_{U}({\hat n}^{b}_{l}-{\hat n}^{b}_{l+1})\right]{\hat a}_{l+1}
    + {\hat b}^{\dagger}_{l}{\cal J}_{1}\left[K_{U}({\hat n}^{a}_{l}-{\hat n}^{a}_{l+1})\right] {\hat b}_{l+1} + \rm{h.c.}\right),\hat {\cal O}\right] \right\rangle_{e}.
\end{equation}
As a concrete example we take the expectation value of $\hat{a}_{k}\hat{a}_{q}$, which has been used for calculating the density distribution in momentum space
$\langle \hat{a}_{k}\hat{a}_{q}\rangle=\langle \hat{a}_{k}\hat{a}_{q}\rangle_{e} + 2{\cal J}_{1}[K]\langle \hat{A} \rangle_{e} J /\hbar \omega$ with
\begin{eqnarray}
\hat{A}&=& \left(\hat{n}_{k-1}^{b}-\hat{n}_{k}^{b}\right)\left(1-2\hat{n}_{k}^{a}\right)\hat{a}_{k-1}^{\dagger}\hat{a}_{q}-
\left(\hat{b}_{k-1}^{\dagger}\hat{b}_{k}+\hat{b}_{k}^{\dagger}\hat{b}_{k-1}\right)\hat{a}_{k}^{\dagger}\hat{a}_{q} \nonumber \\
&&+\left(\hat{n}_{k}^{b}-\hat{n}_{k+1}^{b}\right)\left(1-2\hat{n}_{k}^{a}\right)\hat{a}_{k+1}^{\dagger}\hat{a}_{q}+
\left(\hat{b}_{k}^{\dagger}\hat{b}_{k+1}+\hat{b}_{k+1}^{\dagger}\hat{b}_{k}\right)\hat{a}_{k}^{\dagger}\hat{a}_{q} \nonumber \\
&&+\left(\hat{n}_{q-1}^{b}-\hat{n}_{q}^{b}\right)\left(2\hat{n}_{q}^{a}-1\right)\hat{a}_{k}^{\dagger}\hat{a}_{q-1}+
\left(\hat{b}_{q-1}^{\dagger}\hat{b}_{q}+\hat{b}_{q}^{\dagger}\hat{b}_{q-1}\right)\hat{a}_{k}^{\dagger}\hat{a}_{q} \nonumber \\
&&+\left(\hat{n}_{q}^{b}-\hat{n}_{q+1}^{b}\right)\left(2\hat{n}_{q}^{a}-1\right)\hat{a}_{k}^{\dagger}\hat{a}_{q+1}
-\left(\hat{b}_{q}^{\dagger}\hat{b}_{q+1}+\hat{b}_{q+1}^{\dagger}\hat{b}_{q}\right)\hat{a}_{k}^{\dagger}\hat{a}_{q}\,.
\end{eqnarray}
We read off that the correction operator $\hat{A}$ includes finite local terms. Note that one can always reduce the effect of the correction by tuning the value of $J/\hbar \omega$.

\subsection{Periodically modulated Rabi oscillation}
We now deal with the Hamiltonian (\ref{H2}), which describes a periodically modulated Rabi oscillation. In case of $\delta U=0$ and $U={\bar U}$, the rotation transformation is given by
$\hat{V}=\exp[-i {\tilde K}_{\Omega} (\hat{a}_{l}^{\dagger}\hat{b}_{l}+\rm{h.c.})]$ with dimensionless modulation strengths $K_{\Omega} = J^{0}_{\Omega}/\hbar \omega$ and ${\tilde K}_{\Omega} = K_{\Omega} \sin(\omega t)$, so we get
\begin{eqnarray}
{\hat V}^{\dagger} {\hat a}_{l}^{\dagger} {\hat V}&=&\cos({\tilde K}_{\Omega}){\hat a}_{l}^{\dagger}+i\sin({\tilde K}_{\Omega}){\hat b}^{\dagger}_{l}(1-2{\hat n}^{a}_{l}),\nonumber\\
{\hat V}^{\dagger} {\hat a}_{l}{\hat V}&=&\cos({\tilde K}_{\Omega}){\hat a}_{l}-i\sin({\tilde K}_{\Omega}){\hat b}_{l}(1-2{\hat n}^{a}_{l})\,.
\end{eqnarray}
From this we conclude 
\begin{eqnarray}
  {\hat V}^{\dagger} ( {\hat a}_{l}^{\dagger} {\hat a}_{l+1} + {\hat b}_{l}^{\dagger} {\hat b}_{l+1}) {\hat V} &=&
  \cos\left[2{\tilde K}_{\Omega} \left({\hat n}^{b}_{l} - {\hat n}^{b}_{l+1}\right)\right] {\hat a}_{l}^{\dagger} {\hat a}_{l+1}
  + \cos\left[2{\tilde K}_{\Omega} \left({\hat n}^{a}_{l} - {\hat n}^{a}_{l+1}\right)\right] {\hat b}_{l}^{\dagger} {\hat b}_{l+1} \nonumber\\
  && -i \sin\left[2{\tilde K}_{\Omega} \left({\hat n}^{a}_{l} - {\hat n}^{b}_{l+1}\right)\right] {\hat b}_{l}^{\dagger} {\hat a}_{l+1}
  - i \sin\left[2{\tilde K}_{\Omega} \left({\hat n}^{b}_{l} - {\hat n}^{a}_{l+1}\right)\right] {\hat a}_{l}^{\dagger} {\hat b}_{l+1}\,.
\end{eqnarray}
Thus, the rotated Hamiltonian results in
\begin{eqnarray}
{\hat H}_{r} (t)=\hat{V}^{\dagger} \left({\hat H}-i\hbar \partial /\partial t \right) \hat{V} = {\hat H}^{T}_{r} (t) + {\hat H}^{U}_{r} (t)\, ,
\end{eqnarray}
where we have
\begin{eqnarray}
{\hat H}_{r}^{T} (t) &=& -J \sum_{l} \left\{ \cos\left[{2{\tilde K}_{\Omega}\left( {\hat n}^{b}_{l} - {\hat n}^{b}_{l+1} \right)}\right] {\hat a}^{\dagger}_{l} {\hat a}_{l+1}
+ \cos\left[{2{\tilde K}_{\Omega} \left( {\hat n}^{a}_{l} - {\hat n}^{a}_{l+1} \right)}\right] {\hat b}^{\dagger}_{l} {\hat b}_{l+1}\right.\nonumber\\
& & \left. - i \sin\left[2{\tilde K}_{\Omega} \left({\hat n}^{a}_{l} - {\hat n}^{b}_{l+1}\right)\right] {\hat b}_{l}^{\dagger} {\hat a}_{l+1}
- i \sin\left[2{\tilde K}_{\Omega} \left({\hat n}^{b}_{l} - {\hat n}^{a}_{l+1}\right)\right] {\hat a}_{l}^{\dagger} {\hat b}_{l+1} +  {\rm h.c.} \right\}
\end{eqnarray}
and the interaction term remains unchanged, e.g.~${\hat H}_{r}^{U} (t)=\bar{U} \sum_{l} {\hat n}^{a}_{l}{\hat n}^{b}_{l}$. Also here ${\hat H}_{r} (t)$ is
time-periodic and can be expanded into a Fourier series, namely ${\hat H}_{r} (t) = \sum^{+\infty}_{n=-\infty}  {\hat H}_r^{(l)} e^{in\omega t}$ with
\begin{eqnarray}
{\hat H}_r^{(2m)} &=& -J \sum_{l} \left\{ {\cal J}_{2m}\left[{2 K_{\Omega} \left( {\hat n}^{b}_{l} - {\hat n}^{b}_{l+1} \right)}\right] {\hat a}^{\dagger}_{l} {\hat a}_{l+1}
+ {\cal J}_{2m}\left[{2 K_{\Omega} \left( {\hat n}^{a}_{l} - {\hat n}^{a}_{l+1} \right)}\right] {\hat b}^{\dagger}_{l} {\hat b}_{l+1} + {\rm h.c.} \right\} + \delta_{2m,0} {\hat H}_{\bar U},\nonumber\\
{\hat H}_r^{(2m+1)} &=& J \sum_{l} \left\{ {\cal J}_{2m+1}\left[{2 K_{\Omega} \left( {\hat n}^{a}_{l} - {\hat n}^{b}_{l+1} \right)}\right] {\hat b}^{\dagger}_{l} {\hat a}_{l+1}
+ {\cal J}_{2m+1}\left[{2 K_{\Omega} \left( {\hat n}^{b}_{l} - {\hat n}^{a}_{l+1} \right)}\right] {\hat a}^{\dagger}_{l} {\hat b}_{l+1} + {\rm h.c.} \right\}
\end{eqnarray}
for even and odd orders, respectively.

Now we can use the formalism of the high-frequency expansion to calculate order by order the effective Hamiltonian\cite{Eckardt_2015,Bukov_2015}
in the rotating frame, namely ${\hat H}_{e} = \sum^{+\infty}_{n=0} {\hat H}^{(n)}_{e}$. The zeroth order effective Hamiltonian turns out to be
\begin{eqnarray}
{\hat H}^{(0)}_{e} = {\hat H}_r^{(0)} = -J \sum_{l} \left( {\cal J}_{0}\left[{2 K_{\Omega} \left( {\hat n}^{b}_{l} - {\hat n}^{b}_{l+1} \right)}\right] {\hat a}^{\dagger}_{l} {\hat a}_{l+1}
+ {\cal J}_{0}\left[{2 K_{\Omega}\left( {\hat n}^{a}_{l} - {\hat n}^{a}_{l+1} \right)}\right] {\hat b}^{\dagger}_{l} {\hat b}_{l+1} + {\rm h.c.} \right) + {\hat H}_{\bar U}
\end{eqnarray}
and the first order vanishes
\begin{eqnarray}
{\hat H}^{(1)}_{e} = \sum^{+\infty}_{n=1} \frac{1}{n\hbar\omega} \left[{\hat H}_r^{(n)}, {\hat H}_r^{(-n)}\right] = 0
\end{eqnarray}
because of the property ${\hat H}_r^{(n)}=(\pm 1)^{n} {\hat H}_r^{(-n)}$. Similarly the second correction consists of many terms proportional to $(J/\hbar\omega)^2$.

We also calculate the kick operator\cite{Eckardt_2015,Bukov_2015} ${\hat {\cal K}}_{e} = \sum^{+\infty}_{n=0} {\hat {\cal K}}^{(n)}_{e}$ up to first order:
\begin{eqnarray}
{\hat {\cal K}}^{(0)}_{e} (t) &=& 0\,,\\
{\hat {\cal K}}^{(1)}_{e} (t) &=& \frac{1}{i\hbar\omega} \sum_{n=\pm 1, \pm2, ...} \frac{e^{in\omega t}}{m} {\hat H}_r^{(n)} \approx \frac{2{\hat H}_r^{(1)}\cos(\omega t)}{i\hbar\omega}\,. 
\end{eqnarray}
With the same reasoning as described in the last subsection we only check the dynamics of the system stroboscopically at $t=nT$, so the expectation value of an observable $\hat {\cal O}$ is given by
\begin{equation}
\langle \psi(nT)| \hat {\cal O} |\psi(nT) \rangle=\langle \psi_{e}| e^{\frac{2{\hat H}_r^{(1)}}{\hbar\omega}} \hat {\cal O}  e^{-\frac{2{\hat H}_r^{(1)}}{\hbar\omega}} |\psi_{e} \rangle\,.
\end{equation}
The expectation value of an observable in the lab frame coincides with $\exp(2{\hat H}_r^{(1)}/\hbar\omega) \hat {\cal O}  \exp(-2{\hat H}_r^{(1)}/\hbar\omega)$, which is the one of the dressed observable being
determined by effective Hamiltonian. As $J/\hbar \omega$ is small, we only need to keep the two lowest orders
\begin{equation}
\langle \hat {\cal O} \rangle =\langle \hat {\cal O} \rangle_{e}-\frac{2J}{\hbar \omega} \left\langle\left[\sum_{l} \left({\hat a}^{\dagger}_{l} {\cal J}_{1}
\left[2K_{\Omega}({\hat n}^{b}_{l}-{\hat n}^{a}_{l+1})\right]{\hat b}_{l+1} + {\hat b}^{\dagger}_{l}{\cal J}_{1}\left[2K_{\Omega}({\hat n}^{a}_{l}-{\hat n}^{b}_{l+1})\right] {\hat a}_{l+1}
+ \rm{h.c.}\right),\hat {\cal O}\right] \right\rangle_{e}\,.
\end{equation}
As an example we also consider the expectation value of $\hat{a}_{k}\hat{a}_{q}$,
which is used for calculating the density distribution in momentum space $\langle \hat{a}_{k}\hat{a}_{q}\rangle=\langle \hat{a}_{k}\hat{a}_{q}\rangle_{e} + 2{\cal J}_{1}[K]\langle \hat{A} \rangle_{e} J /\hbar \omega$ with
\begin{eqnarray}
\hat{A}&=& \left(\hat{n}_{k-1}^{a}-\hat{n}_{k}^{b}\right)\left(1-2\hat{n}_{k}^{b}\right)\hat{b}_{k-1}^{\dagger}\hat{a}_{q}-
\left(\hat{a}_{k-1}^{\dagger}\hat{b}_{k}+\hat{b}_{k}^{\dagger}\hat{a}_{k-1}\right)\hat{a}_{k}^{\dagger}\hat{a}_{q} \nonumber \\
&&+\left(\hat{n}_{k}^{b}-\hat{n}_{k+1}^{a}\right)\left(1-2\hat{n}_{k}^{a}\right)\hat{b}_{k+1}^{\dagger}\hat{a}_{q}+
\left(\hat{b}_{k}^{\dagger}\hat{a}_{k+1}+\hat{a}_{k+1}^{\dagger}\hat{b}_{k}\right)\hat{a}_{k}^{\dagger}\hat{a}_{q} \nonumber \\
&&+\hat{a}_{k}^{\dagger}\left(\hat{n}_{q-1}^{a}-\hat{n}_{q}^{b}\right)\left(2\hat{n}_{q}^{a}-1\right)\hat{b}_{q-1}+
\hat{a}_{k}^{\dagger}\left(\hat{a}_{q-1}^{\dagger}\hat{b}_{q}+\hat{b}_{q}^{\dagger}\hat{a}_{q-1}\right)\hat{a}_{q} \nonumber \\
&&+\hat{a}_{k}^{\dagger}\left(\hat{n}_{q}^{b}-\hat{n}_{q+1}^{a}\right)\left(2\hat{n}_{q}^{a}-1\right)\hat{a}_{q+1}
-\hat{a}_{k}^{\dagger}\left(\hat{b}_{q}^{\dagger}\hat{a}_{q+1}+\hat{a}_{q+1}^{\dagger}\hat{b}_{q}\right)\hat{a}_{q}.
\end{eqnarray}
We read off that the correction operator $\hat{A}$ includes finite local terms. Again one can always reduce the effect of the correction 
by tuning the value of $J/\hbar \omega$ to lower values.

In general we conclude for both experimental realizations that the zeroth-order effective Hamiltonian reads
\begin{eqnarray}
{\hat H}^{(0)}_{e} = -J \sum_{l} \left\{ {\cal J}_{0}\left[{K \left( {\hat n}^{b}_{l} - {\hat n}^{b}_{l+1} \right)}\right] {\hat a}^{\dagger}_{l} {\hat a}_{l+1}
+ {\cal J}_{0}\left[{K\left( {\hat n}^{a}_{l} - {\hat n}^{a}_{l+1} \right)}\right] {\hat b}^{\dagger}_{l} {\hat b}_{l+1} + {\rm h.c.} \right\} + \bar{U} \sum_{l} {\hat n}^{a}_{l}{\hat n}^{b}_{l}
\end{eqnarray}
with $K=K_{U}=2K_{\Omega}$. As the first-order effective Hamiltonian vanishes, only this zeroth-order effective Hamiltonian is relevant.
Although the value of any observable in the lab differs from the ground state expectation value of the effective Hamiltonian, the difference between them is
always accompanied by a prefactor $J/\hbar \omega$, which can be decreased by increasing the driven frequency $\omega$.

\section{Real-time dynamics}

In the following we check the real-time dynamics for a small system and answer the question if it is feasible to complete the preparation of the sample and the measurement before the
thermalization sets in. To this end we consider two steps for switching on $K$, i.e. $K_{U}$ or $K_{\Omega}$, and ${\bar U}$, respectively.
At the first step, we initialize the system staying at the ground state of the non-driven model with a small on-site repulsion ${\bar U}^{i}$. Then for $t > 0$ the amplitude of the
time-periodic circularly-polarized Raman laser beams is gradually switched on following a linear function of time $t$
\begin{eqnarray}
K (t) = \left\{
\begin{array}{cc}
v_{K} t & 0 < t \le t_{1} \\
K^{f} & t > t_{1}\, ,
\end{array}
\right.
\end{eqnarray}
\begin{figure}[t]
\includegraphics[width=0.24 \columnwidth]{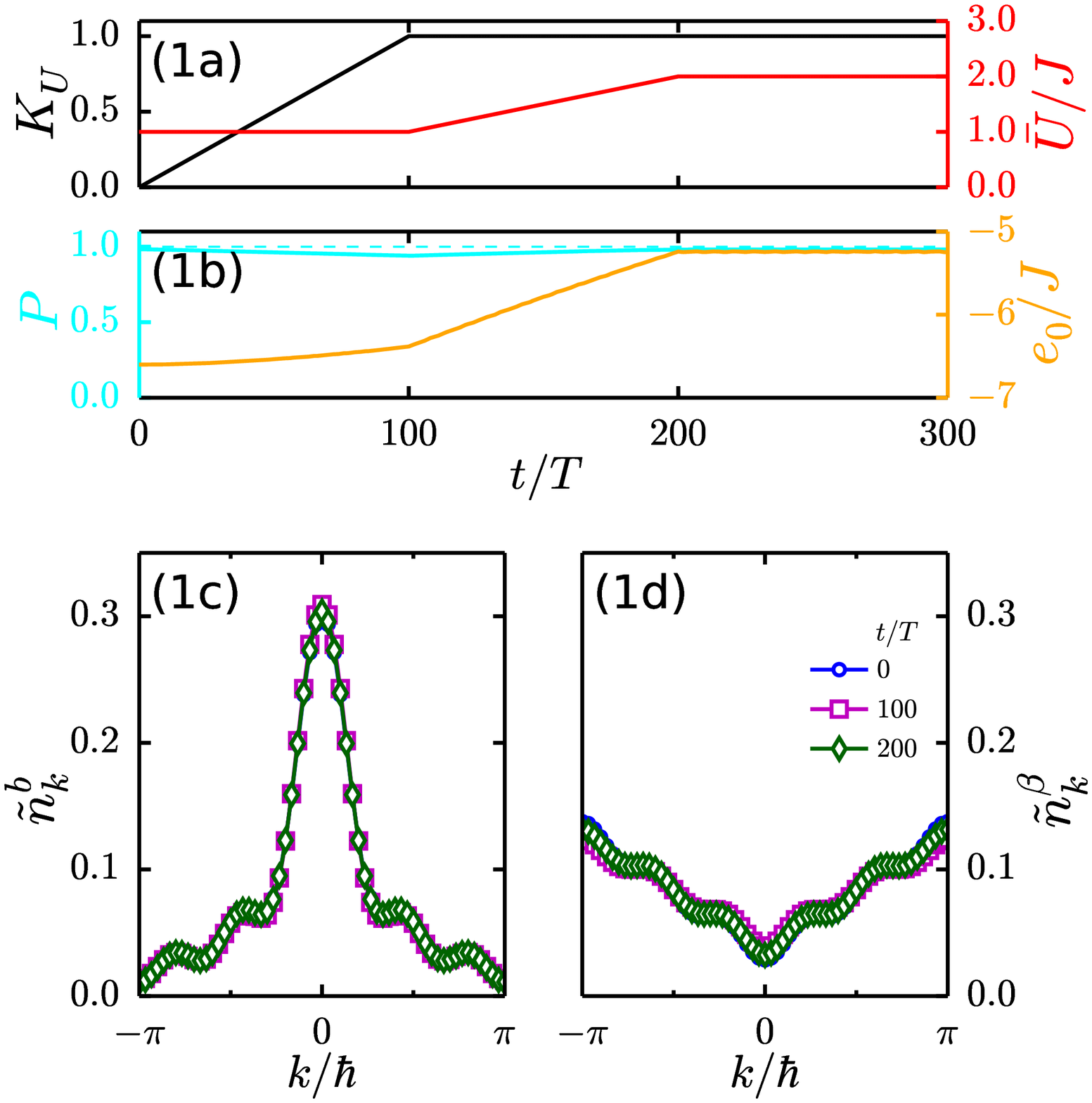}\includegraphics[width=0.24 \columnwidth]{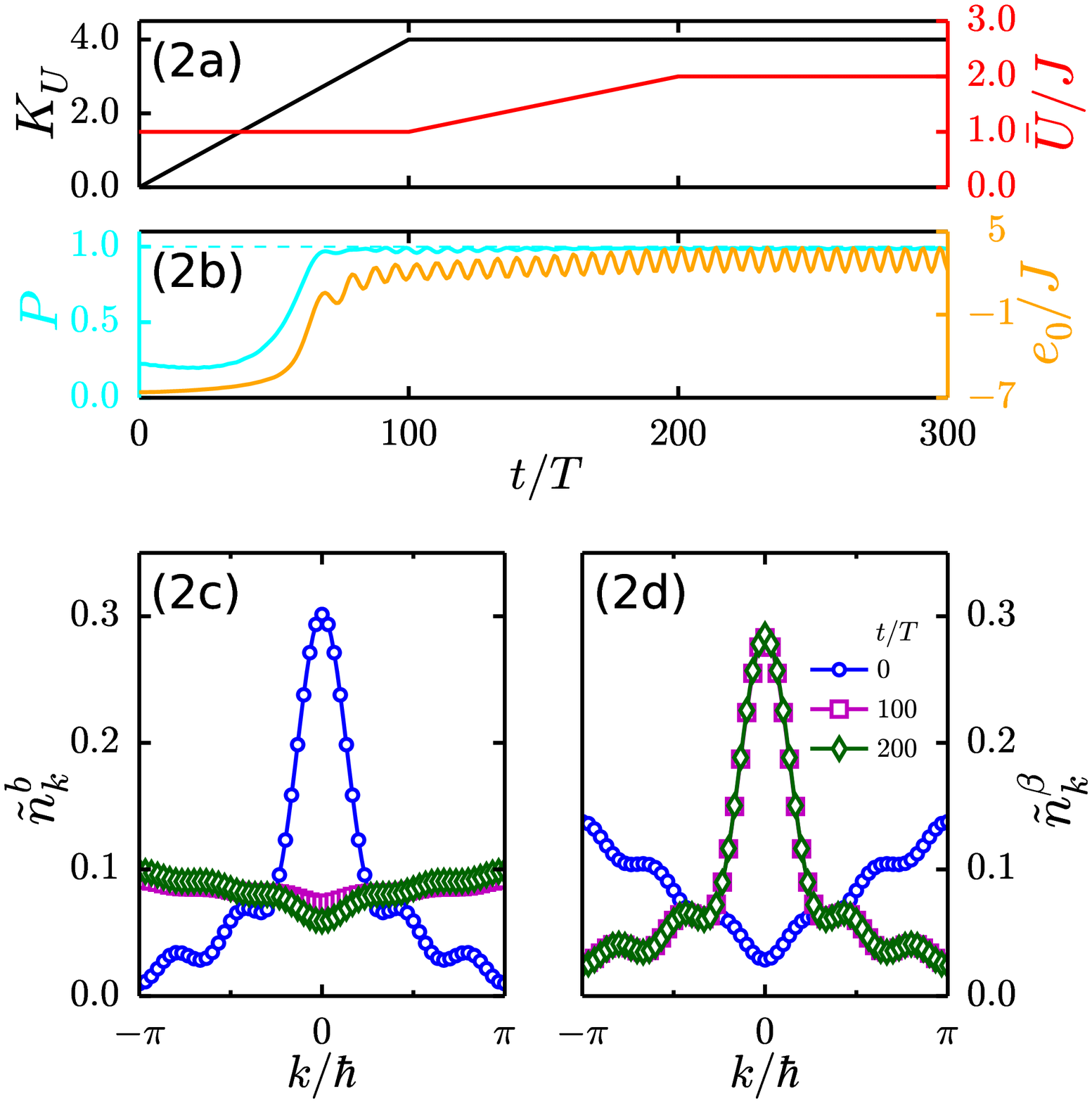}\includegraphics[width=0.24 \columnwidth]{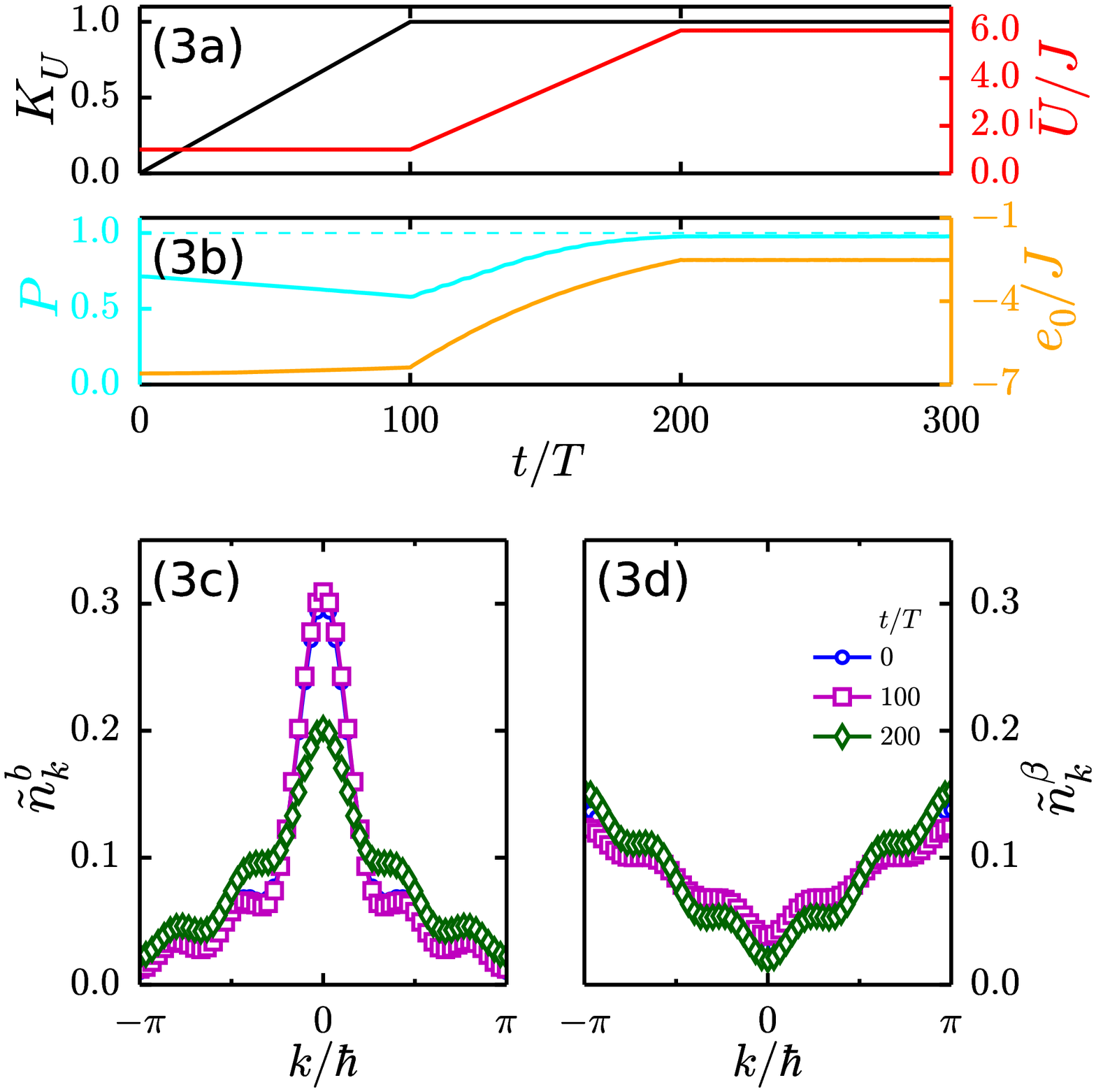}\includegraphics[width=0.24 \columnwidth]{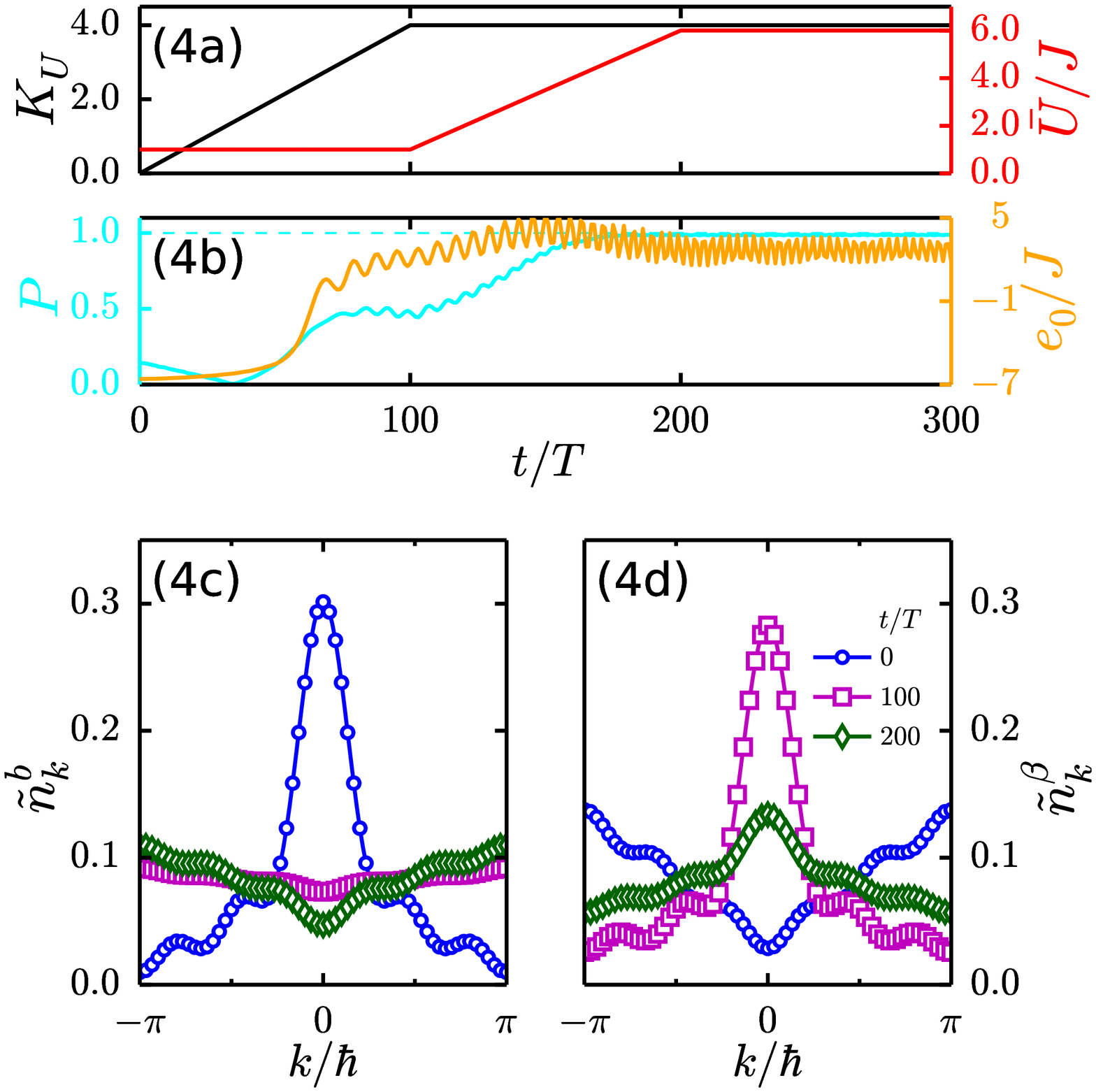}
\includegraphics[width=0.24 \columnwidth]{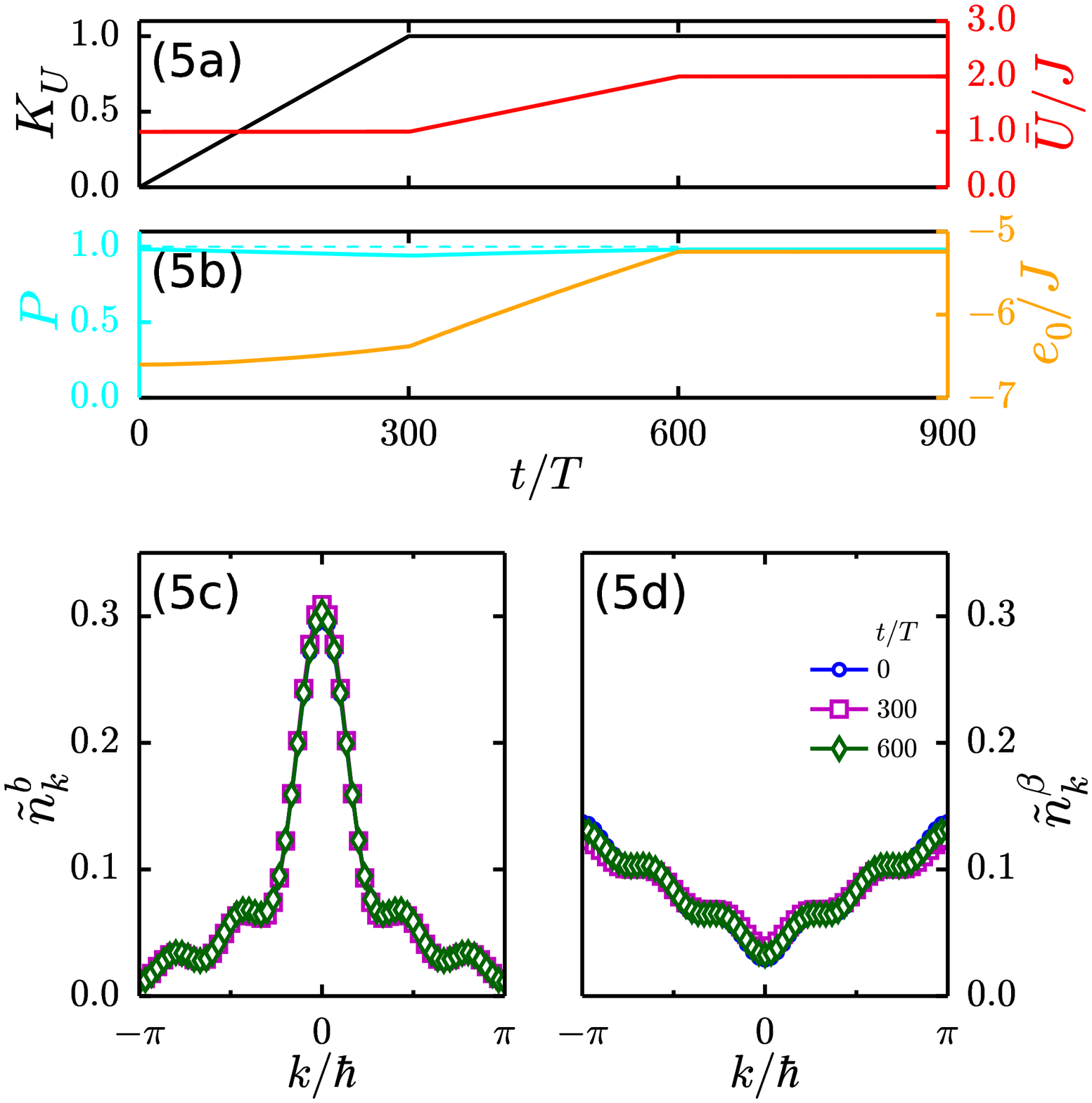}\includegraphics[width=0.24 \columnwidth]{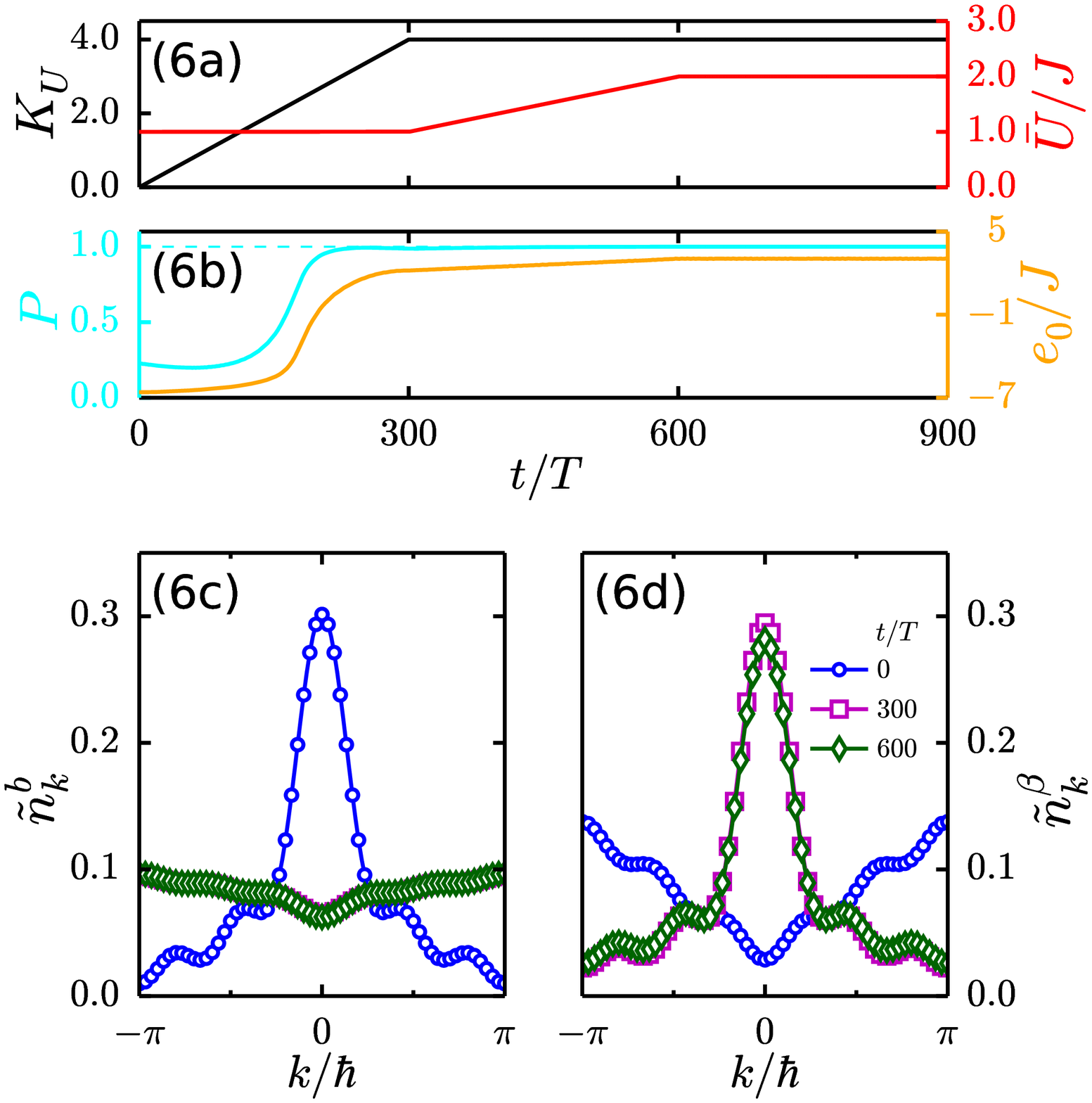}\includegraphics[width=0.24 \columnwidth]{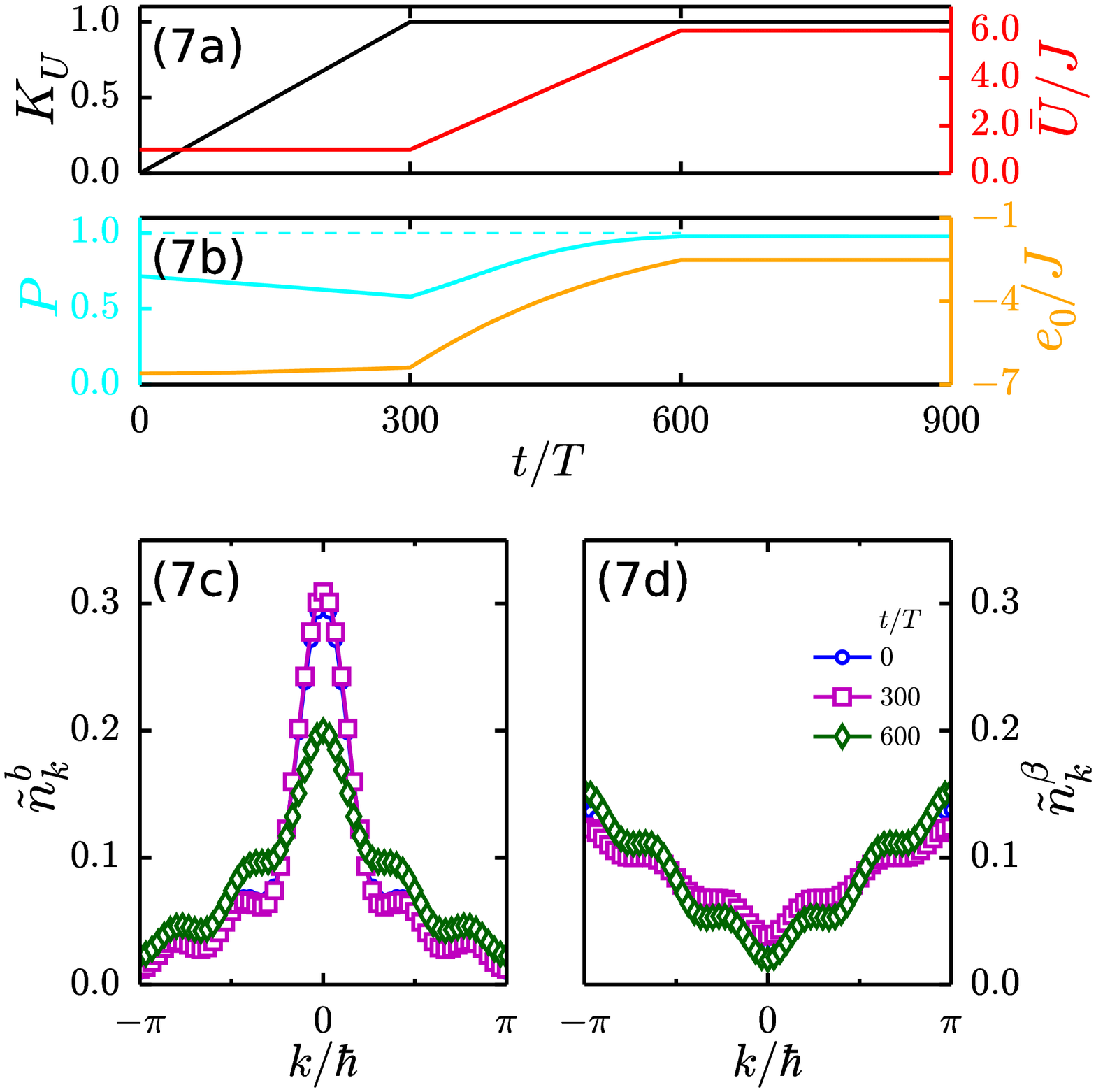}\includegraphics[width=0.24 \columnwidth]{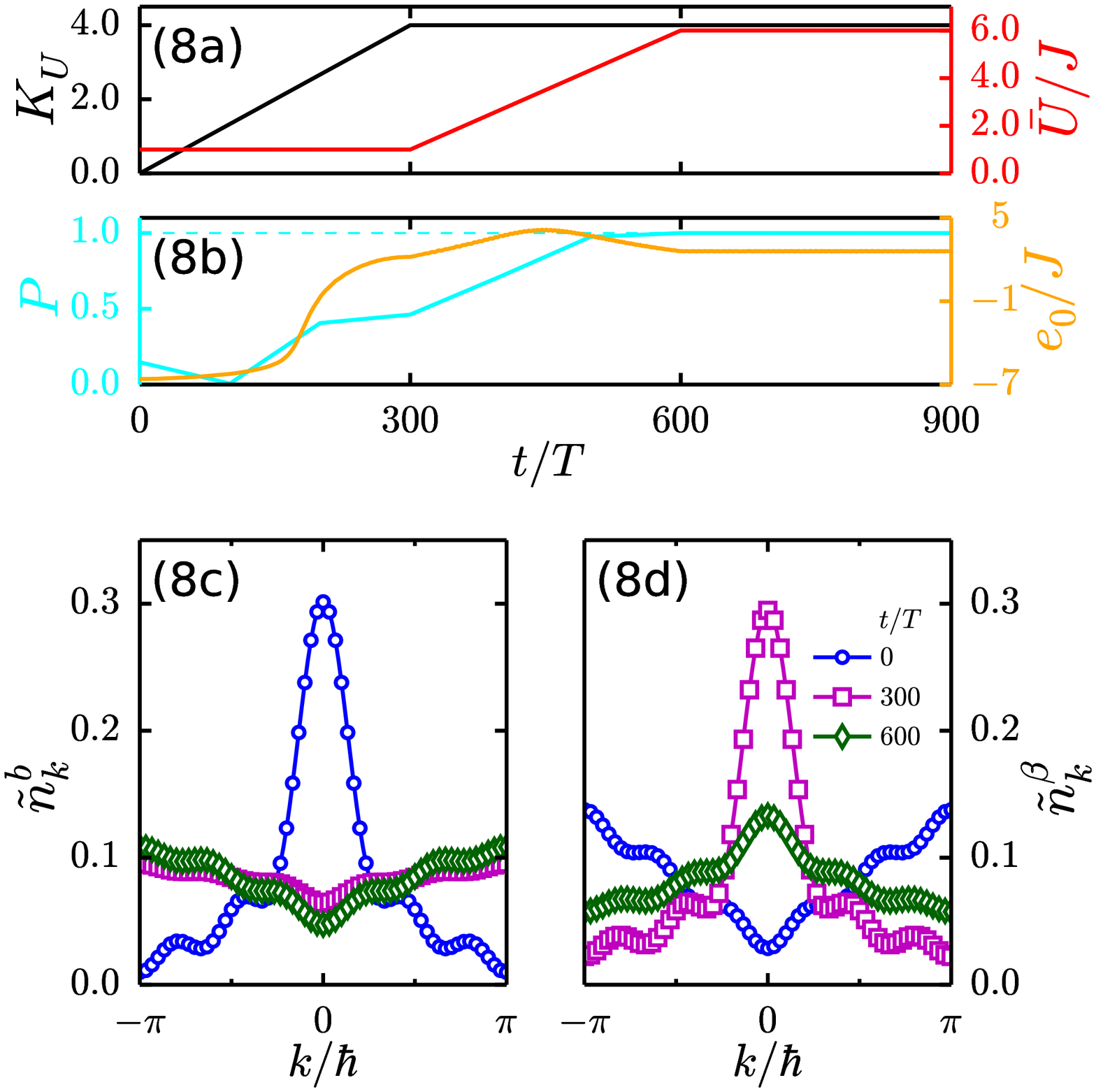}
\caption{
  Real-time dynamics for a chain of $6$ sites determined from the time-dependent Schr\"{o}dinger equation (\ref{TDSE}). Here we choose $J=1$ as an energy unit, a relatively high frequency $\omega=20$, $K_{\Omega}=0$, and
  balanced filling $N^{a}=N^{b}=3$. Initially, we let the system stay at the ground state of the Hamiltonian with a small on-site repulsion strength ${\bar U}^{i}=1$. Then $K_{U}$ linearly grows until $t_{1}$
  and persists as a working Rabi oscillation. At the moment $t_{1}$, we start to linearly increase ${\bar U}$ to a desired value ${\bar U}^{f}$ at $t_{2}$, which persists until the measurement. We consider
  two cases: $t_{1}=100~T$ and $t_{2}=200~T$ for the fast switching-on (upper four clusters of panels) while $t_{1}=300~T$ and $t_{2}=600~T$ for the slow switching-on (lower four clusters of panels). For each
  case we show the time-evolving behavior related to four different parameter sets: (1) and (5) for $K_{U}^{f}=1$ and ${\bar U}^{f}=2$; (2) and (6) for $K_{U}^{f}=4$ and ${\bar U}^{f}=2$; (3) and (7)
  for $K_{U}^{f}=1$ and ${\bar U}^{f}=6$; (4) and (8) for $K_{U}^{f}=4$ and ${\bar U}^{f}=6$. For each scheme we plot the modulated parameters $K_{U}$ (black lines) and $\bar U$ (red lines) as a function of time $t/T$
  in panel (a). And we plot the time-evolving overlap $P=|\langle \psi(t) |\psi_{e}\rangle|$ and the energy $e_{0}(t) = \langle \psi(t) | {\hat H}_{T} + {\hat H}_{U} | \psi(t) \rangle$ in panel (b).
  In panel (c) and (d), we exhibit the structure factors of ${\tilde n}^{b}_{k}$ and ${\tilde n}^{\beta}_{k}$ at the moments $t=0$ (blue circles), $t=t_{1}$ (magenta squares), and $t=t_{2}$ (green diamonds), respectively.
}\label{figs3}
\end{figure}
where the modulation lasts for the duration of $t_{1} = n_{01} T$ until $K$ reaches a desired value $K^{f}$, so the effective speed amounts to $v_{K} = K^{f} / t_{1}$. It persists as a working Rabi oscillation before the measurement.
In the second step, we turn on gradually the on-site interaction as a linear function of time $t$
\begin{eqnarray}
{\bar U} (t) = \left\{
\begin{array}{cc}
{\bar U}^{i} & 0 < t \le t_{1} \\
{\bar U}^{i} + v_{\bar U} t & t_{1} < t \le t_{2} \\
{\bar U}^{f} & t > t_{2} \, ,
\end{array}
\right.
\end{eqnarray}
where the modulation lasts for the duration $t_{2} - t_{1} = n_{12} T$ until the on-site interaction reaches a desired value ${\bar U}^{f}$, so the effective speed reads $v_{\bar U} = ({\bar U}^{f} - {\bar U}^{i}) / (t_{2} - t_{1})$.

After the modulation duration we expect that the low-energy behavior of the system can be described by the effective Hamiltonian with the desired parameter values ${\bar U}^{f}$ and $K^{f}$. And then the state should persist for
a while in order to complete the measurement. Here we exploit the fourth-order Runge-Kutta method to solve the time-dependent Schr\"{o}dinger equation (\ref{TDSE}) and to determine the time-evolving wave
function $|\psi(t)\rangle$. In order to understand how close it is to the desired one $|\psi_{e}\rangle$,
we measure the time-dependent overlap $P=|\langle \psi(t) |\psi_{e}\rangle|$. Furthermore, we measure the time-dependent energy $e_{0}(t) = \langle \psi(t) | {\hat H}_{T} + {\hat H}_{U} | \psi(t) \rangle$ and the structure factors
\begin{eqnarray}
{\tilde n}^{b}_{k} &=& \frac{1}{L^{2}} \sum^{L}_{l,l'=1} \exp\left[ik(l - l')/\hbar\right] \langle{\hat b}^{\dagger}_{l} {\hat b}_{l'}\rangle\nonumber\\
{\tilde n}^{\beta}_{k} &=& \frac{1}{L^{2}} \sum^{L}_{l,l'=1} \exp\left[ik(l - l')/\hbar\right] \langle{\hat \beta}^{\dagger}_{l} {\hat \beta}_{l'}\rangle\, ,
\end{eqnarray}
where ${\hat \beta}_{l}={\hat b}_{l} \exp(i\pi {\hat n}^{a}_{l})$ and ${\hat \beta}^{\dagger}_{l}={\hat b}^{\dagger}_{l} \exp(-i\pi {\hat n}^{a}_{l})$ are the annihilation and creation operators for the gauge-dressed particles.
\begin{figure}[t]
\includegraphics[width=0.24 \columnwidth]{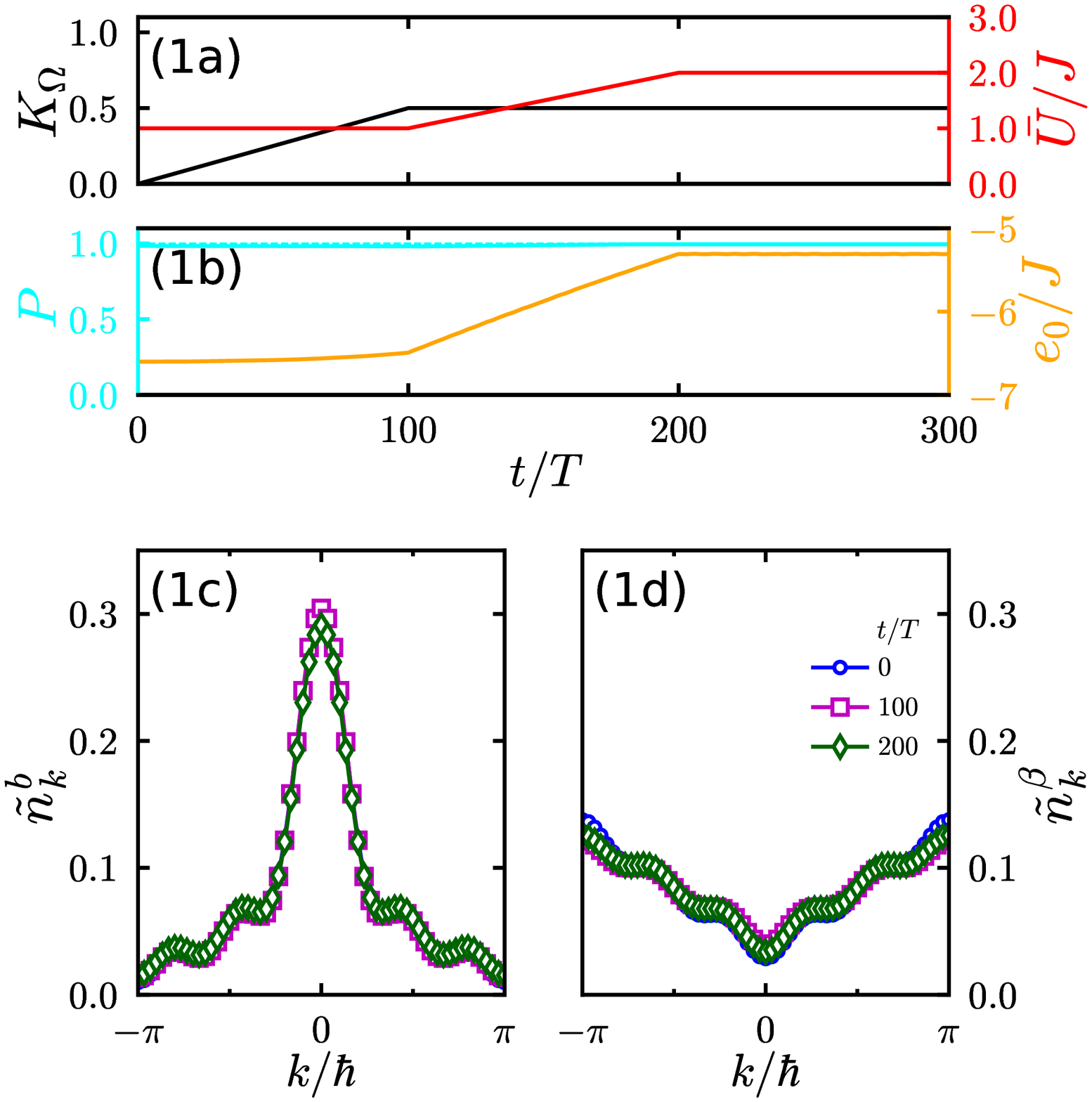}\includegraphics[width=0.24 \columnwidth]{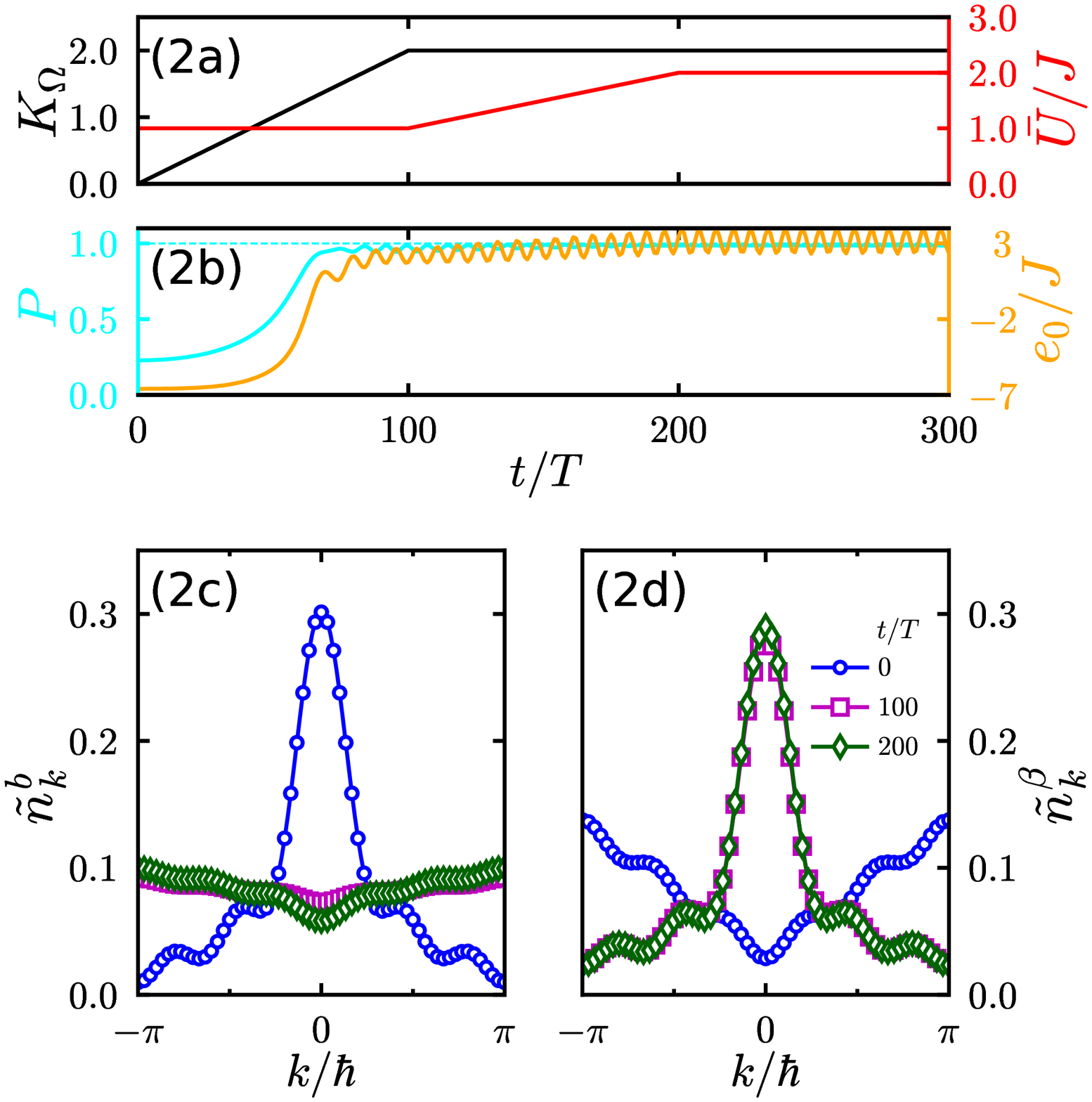}\includegraphics[width=0.24 \columnwidth]{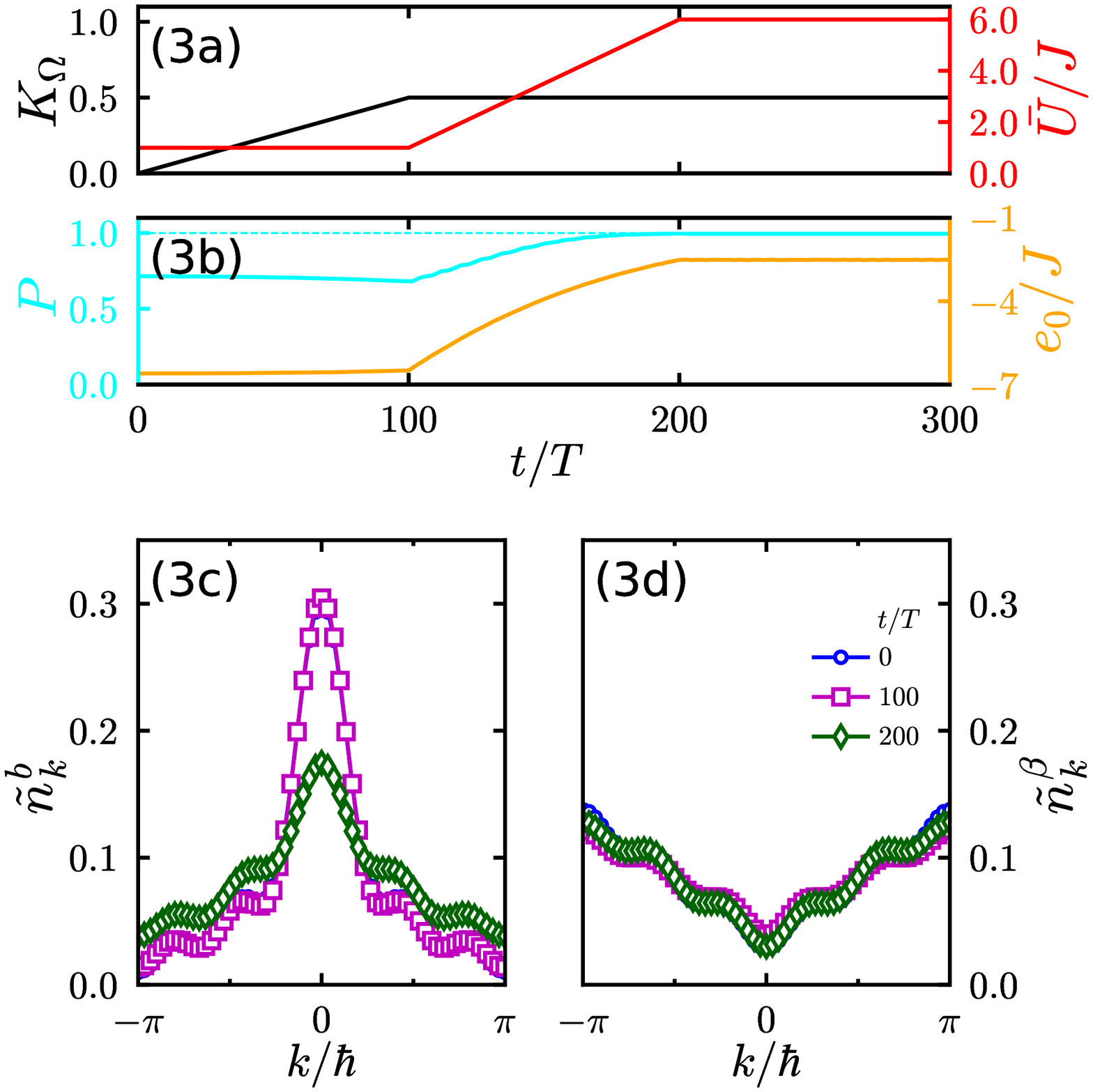}\includegraphics[width=0.24 \columnwidth]{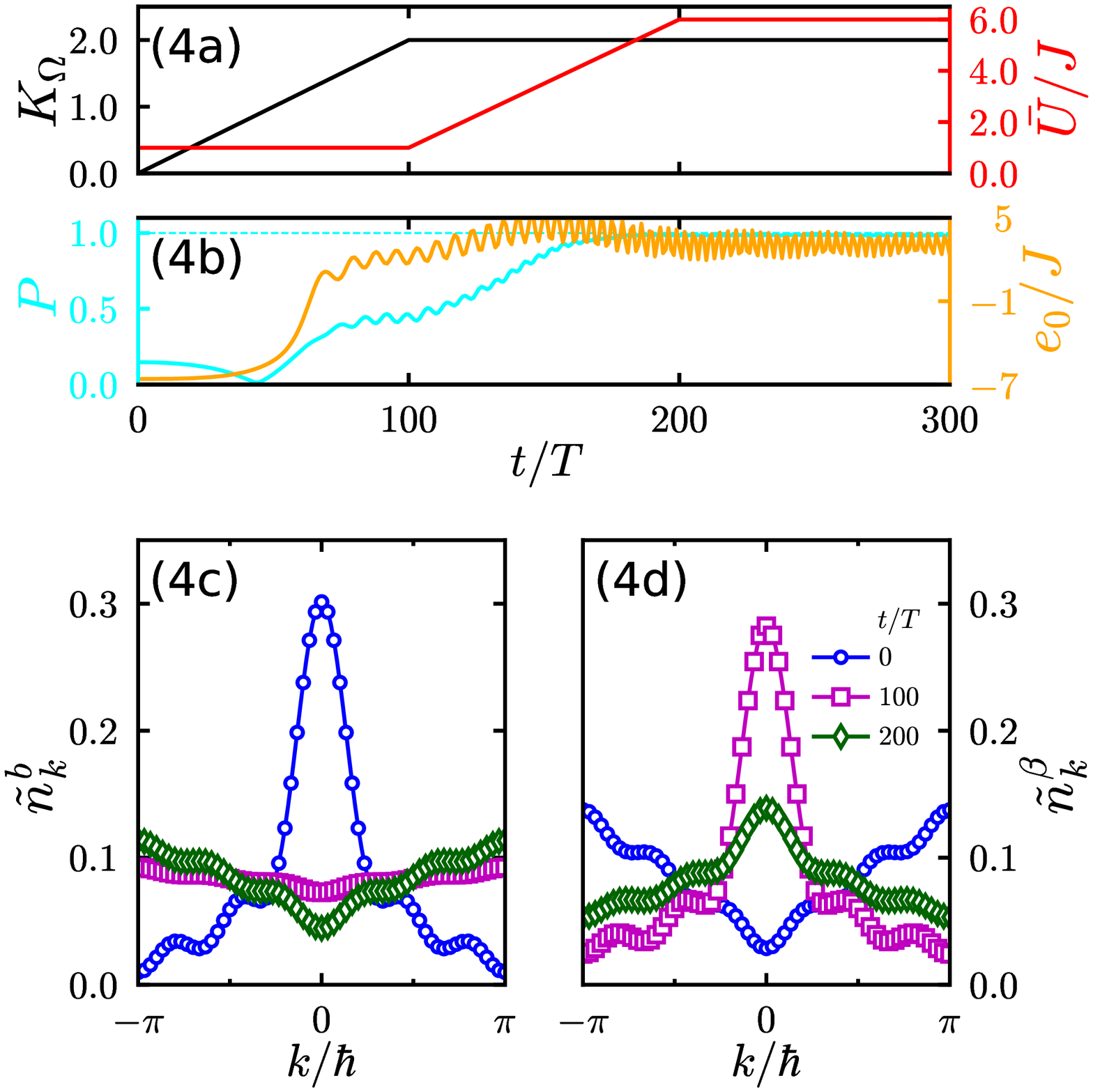}
\includegraphics[width=0.24 \columnwidth]{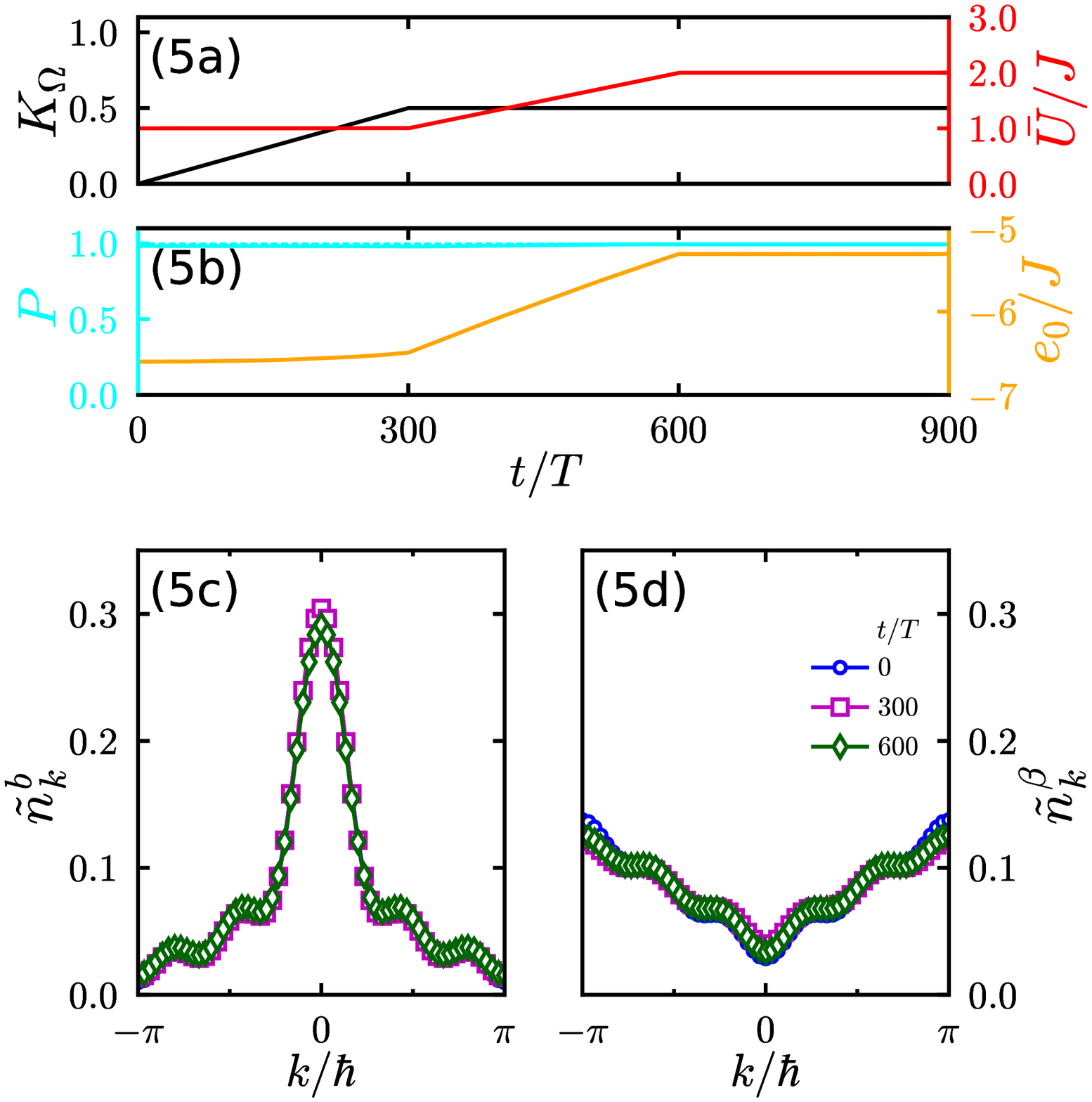}\includegraphics[width=0.24 \columnwidth]{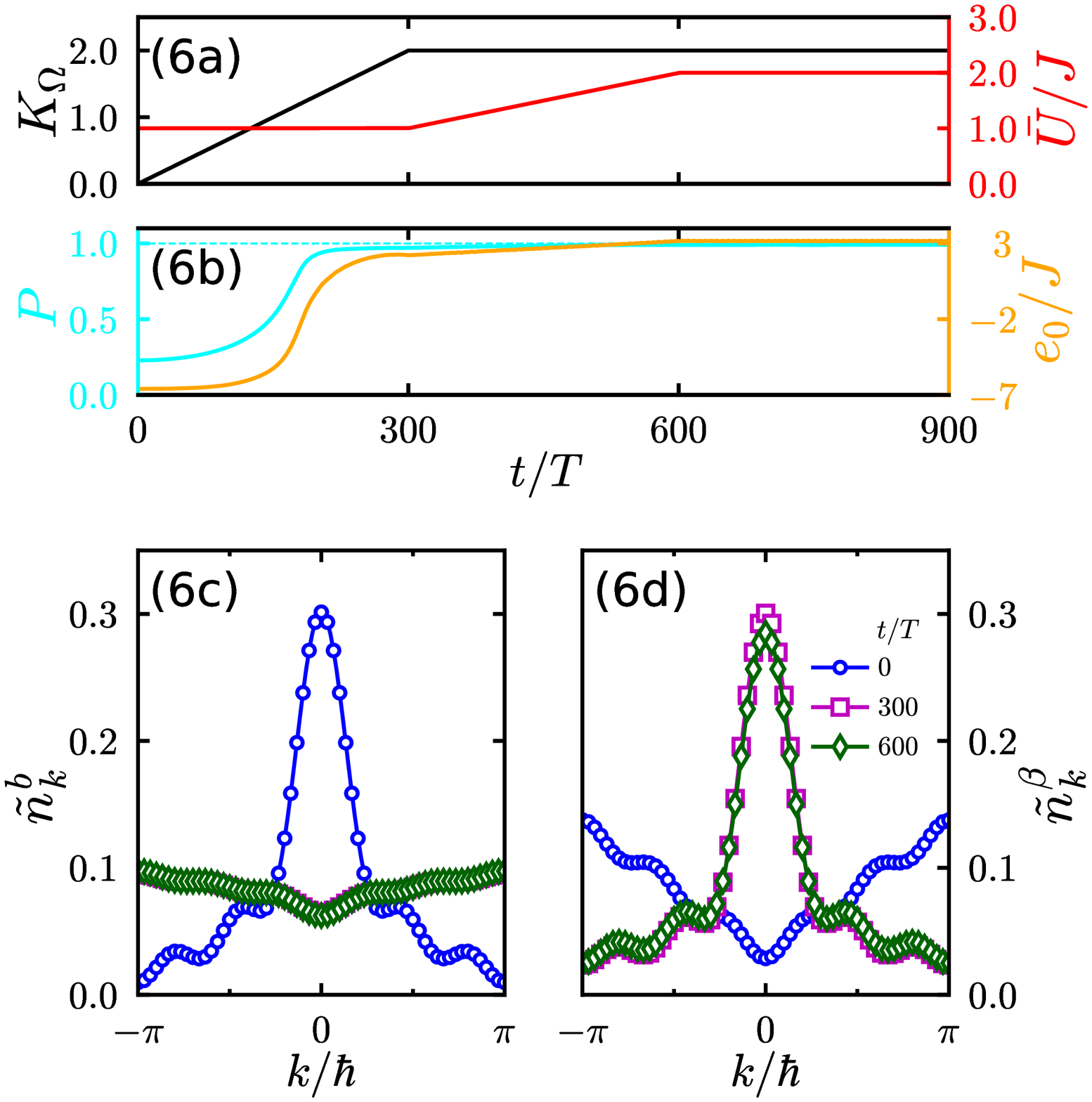}\includegraphics[width=0.24 \columnwidth]{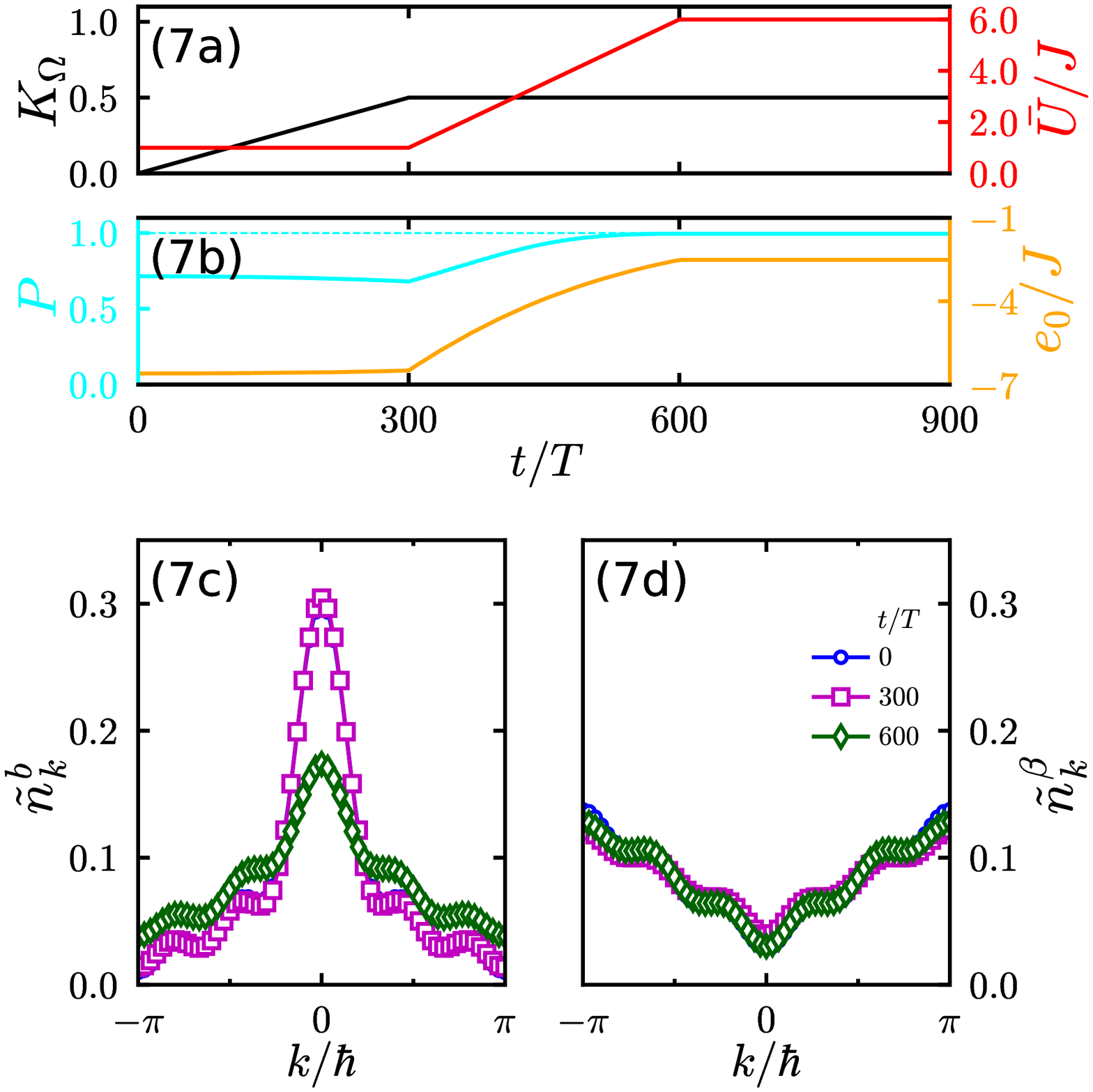}\includegraphics[width=0.24 \columnwidth]{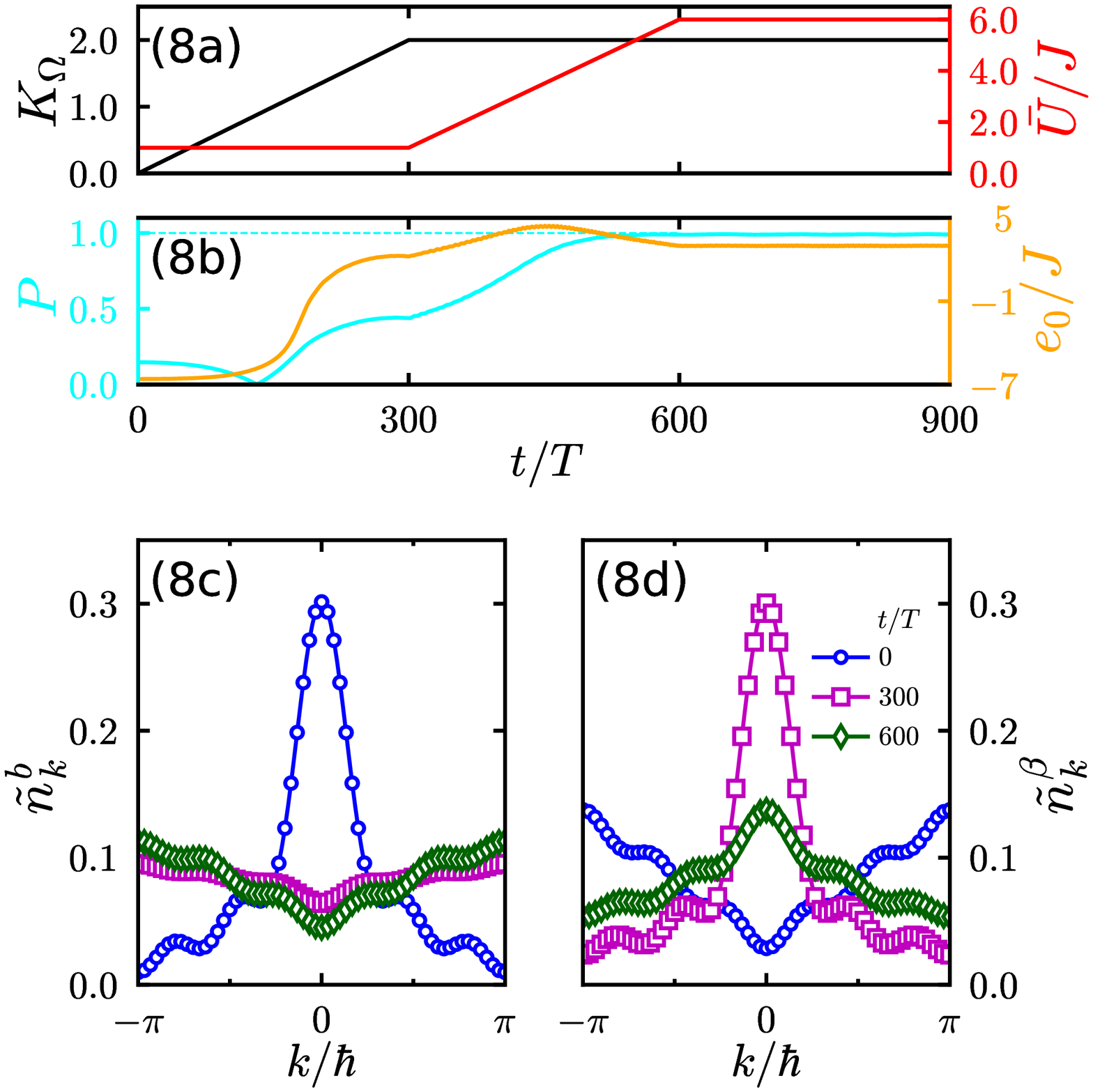}
\caption{
  Real-time dynamics for a chain of $6$ sites determined from the time-dependent Schr\"{o}dinger equation (\ref{TDSE}). Here we choose $J=1$ as an energy unit, a relatively high frequency $\omega=20$, $K_{U}=0$ and the
  integer-$1$ filling $N^{a}+N^{b}=6$. Initially, we let the system stay at the ground state of the Hamiltonian with a small on-site repulsion strength ${\bar U}^{i}=1$. Then $K_{\Omega}$ linearly grows until
  $t_{1}$ and persists as a working Rabi oscillation. At the moment $t_{1}$, we start to linearly increase ${\bar U}$ to a desired value ${\bar U}^{f}$ at $t_{2}$, which persists until the measurement. We consider
  two cases: $t_{1}=100~T$ and $t_{2}=200~T$ for the fast switching-on (upper four clusters of panels) while $t_{1}=300~T$ and $t_{2}=600~T$ for the slow switching-on (lower four clusters of panels). For each case
  we show the time-evolving behavior related to four different parameter sets: (1) and (5) for $K_{\Omega}^{f}=0.5$ and ${\bar U}^{f}=2$; (2) and (6) for $K_{\Omega}^{f}=2$ and ${\bar U}^{f}=2$; (3) and (7) for
  $K_{\Omega}^{f}=0.5$ and ${\bar U}^{f}=6$; (4) and (8) for $K_{\Omega}^{f}=2$ and ${\bar U}^{f}=6$. For each scheme, we plot the modulated $K_{\Omega}$ (black lines) and $\bar U$ (red lines) as a function of time $t/T$
  in panel (a). And we plot the time-evolving overlap $P=|\langle \psi(t) |\psi_{e}\rangle|$ and the energy $e_{0}(t) = \langle \psi(t) | {\hat H}_{T} + {\hat H}_{U} | \psi(t) \rangle$ in panel (b). In panel
  (c) and (d), we exhibit the structure factors of ${\tilde n}^{b}_{k}$ and ${\tilde n}^{\beta}_{k}$ at the moments $t=0$ (blue circles), $t=t_{1}$ (magenta squares), and $t=t_{2}$ (green diamonds), respectively.
}\label{figs4}
\end{figure}
In the Figs.~\ref{figs3} and \ref{figs4}, we systematically study the real-time dynamics for $6$ sites for a fast and a slow switching-on. Although the middle process is complicated, the
time-evolving overlap is close to unity and the energy $e_{0}$ is almost constant at the end of the modulations. This means that we can obtain the ground state of the effective Hamiltonian with the desired
physical parameters following our scheme of the sample preparation. Besides we note that a slow switching-on always works better than the fast one, so thus we suggest that experimentalists need to tune the parameters as
slow as possible before the thermalization happens.
%

\section{Avoided level-crossings}
After a cloud of ultracold atoms is confined in the optical lattice and has reached equilibrium in the ground-state, the relevant parameters $K_{U}$, $K_{\Omega}$, and $\bar U$ are adiabatically switched on in order to
obtain the ground state of the effective Hamiltonian in the specified parameter regime. In previous studies it was found that the request to the adiabatic modulation is not achievable if avoided
level-crossings between different Floquet bands occur.\cite{Eckardt_2008}

In this section, we investigate avoided level-crossings in the quasi-energy spectrum for a small system as a function of $K$ and $U$, respectively. Besides the perturbative treatment discussed above, for a
small system size we can exactly diagonalize the general Floquet Hamiltonian ${\hat {\cal H}} (t) = {\hat H} (t) - i\hbar \partial/\partial t$ which obeys the eigenvalue equation
\begin{eqnarray}
{\hat {\cal H}}(t) |\phi^{\alpha}(t)\rangle = \epsilon_{\alpha} |\phi^{\alpha}(t)\rangle\,,
\label{GFH}
\end{eqnarray}
where the Floquet mode $|\phi^{\alpha}(t)\rangle$ is a many-body state instead of the local Fock basis. The Floquet modes live in the Hilbert space of real dimensions $D_{\cal P}$. Because we have
$|\phi^{\alpha}(t)\rangle=|\phi^{\alpha}(t+T)\rangle$, each Floquet mode can be expanded by Fourier modes
\begin{eqnarray}
\hspace*{-0.7cm}|\phi^{\alpha}(t)\rangle = \sum^{+\infty}_{m=-\infty} \exp(im\omega t) |\phi^{\alpha}_{m}
\rangle = \sum^{+\infty}_{m=-\infty} \sum_{\{n^{a}_{l} n^{b}_{l}\}} \Lambda^{\alpha, m}_{\{n^{a}_{l} n^{b}_{l}\}} \exp(im\omega t) |\{n^{a}_{l} n^{b}_{l}\}\rangle=\sum^{+\infty}_{m=-\infty} \sum_{\{n^{a}_{l} n^{b}_{l}\}}
\Lambda^{\alpha, m}_{\{n^{a}_{l} n^{b}_{l}\}} |m, \{n^{a}_{l} n^{b}_{l}\}\rangle\, .
\end{eqnarray}
The new bases $|m, \{n^{a}_{l} n^{b}_{l}\}\rangle$ satisfies the relation of super-orthogonalization
\begin{eqnarray}
\langle\langle m, \{n^{a}_{l} n^{b}_{l}\}|m', \{n^{a}_{l} n^{b}_{l}\}'\rangle\rangle=\frac{1}{T}\int^{T}_{0} \langle \{n^{a}_{l} n^{b}_{l}\}|\{n^{a}_{l} n^{b}_{l}\}'\rangle e^{-i(m-m')\omega t} dt = \delta_{m,m'}
\delta_{\{n^{a}_{l} n^{b}_{l}\}, \{n^{a}_{l} n^{b}_{l}\}'}\, .
\end{eqnarray}
With this Eq.~(\ref{GFH}) can be interpreted as an eigenvalue problem, which is defined in the enlarged Hilbert space $D_{\cal P}\otimes D_{\cal T}$ with an infinite larger number of frequencies $D_{\cal T}$.
Thus we can also write the Floquet Hamiltonian in the enlarged Hilbert space according to
\begin{eqnarray}
\hspace*{-0.7cm}& &{\hat {\cal H}}_{m, \{n^{a}_{l} n^{b}_{l}\};m', \{n^{a}_{l} n^{b}_{l}\}'} = \langle\langle m, \{n^{a}_{l} n^{b}_{l}\}| {\hat {\cal H}} |m, \{n^{a}_{l} n^{b}_{l}\}'\rangle\rangle\nonumber\\
\hspace*{-0.7cm}&& =\delta_{m,m'} \left({\hat H}_{T} + {\hat H}_{\bar U}\right)_{\{n^{a}_{l} n^{b}_{l}\}, \{n^{a}_{l} n^{b}_{l}\}'} + \frac{1}{2} \left(\delta_{m,m' + 1} + \delta_{m,m' - 1}\right) \left[ J^{0}_{\Omega}
\sum_{l}\left(\hat{a}_{l}^{\dagger}\hat{b}_{l}+\rm{h.c.}\right) + \delta U \sum_{l} {\hat n}^{a}_{l}{\hat n}^{b}_{l} \right]_{\{n^{a}_{l} n^{b}_{l}\}, \{n^{a}_{l} n^{b}_{l}\}'}\, .
\end{eqnarray}
In principle, we obtain the full quasi-energy spectrum by exactly diagonalizing the Floquet Hamiltonian. However it is impossible to numerically handle an infinitely large matrix. In practice, because the
spectrum has a repeating structure with respect to the energy axis, we only need to target $D_{\cal P}$ quasi-energy levels in the vicinity of the zero-energy axis with $2N_{D}+1$ cutting frequencies $m = -N_{D}$, $\cdots$,
$N_{D}$. And then we use their translation invariant copies in order to cover the whole spectrum. In this way, we can obtain the full quasi-energy spectrum and find out the positions of the avoided level-crossings,
and thereby the valid parameter regime, which we can reach through the adiabatic switching, by investigating the positions of the avoided level-crossings.
\begin{figure}[t]
\includegraphics[width=0.24 \columnwidth]{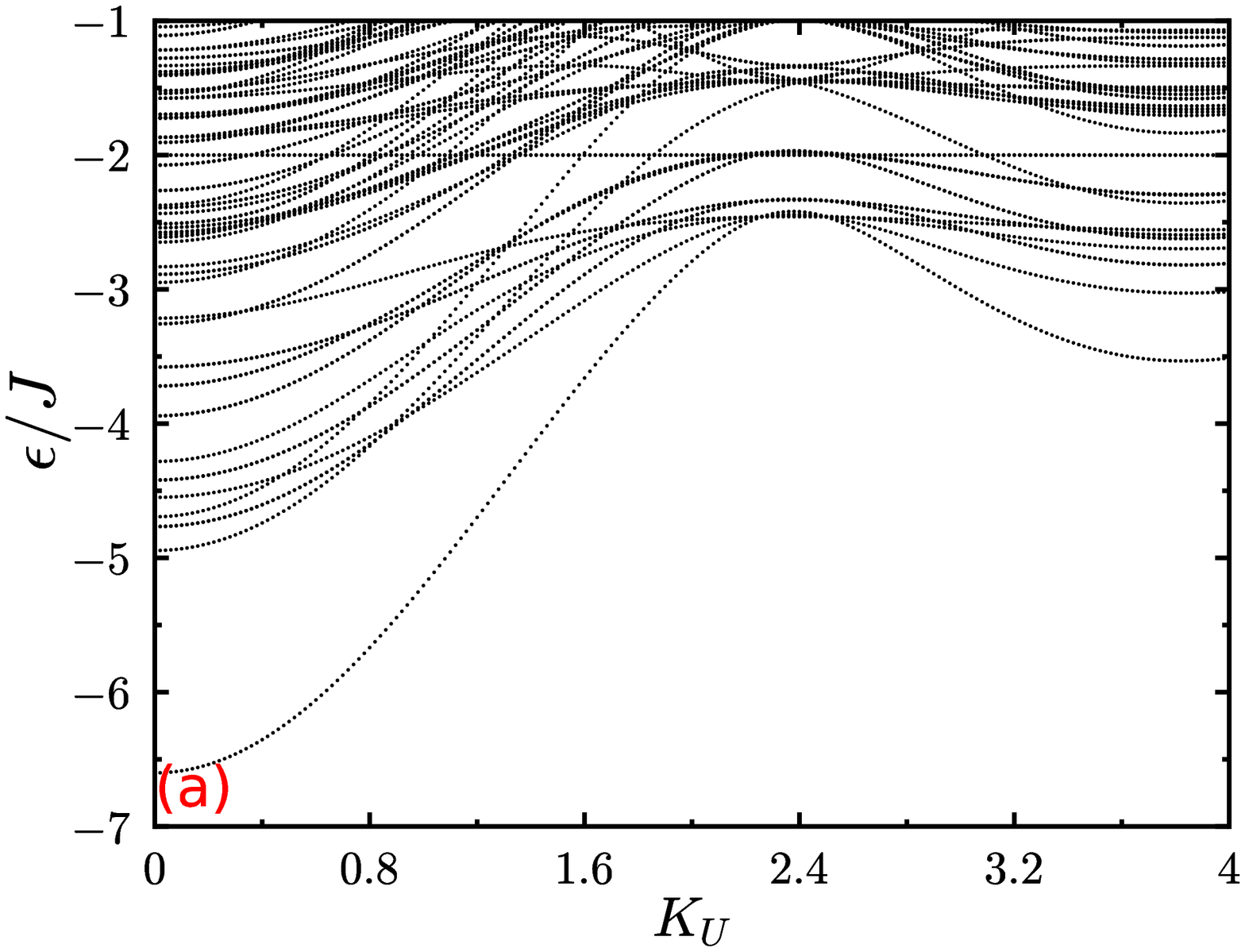}
\includegraphics[width=0.24 \columnwidth]{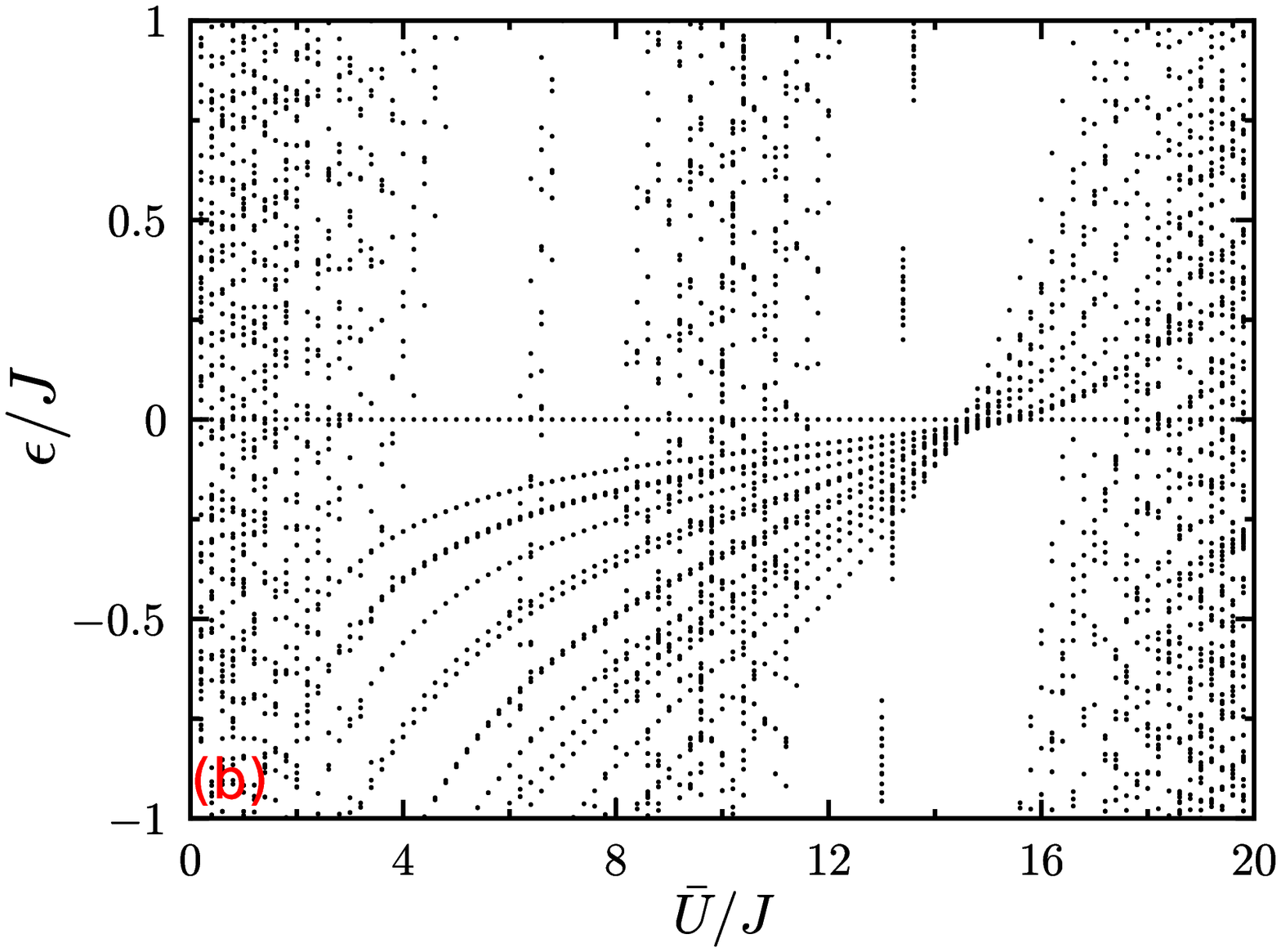}
\includegraphics[width=0.24 \columnwidth]{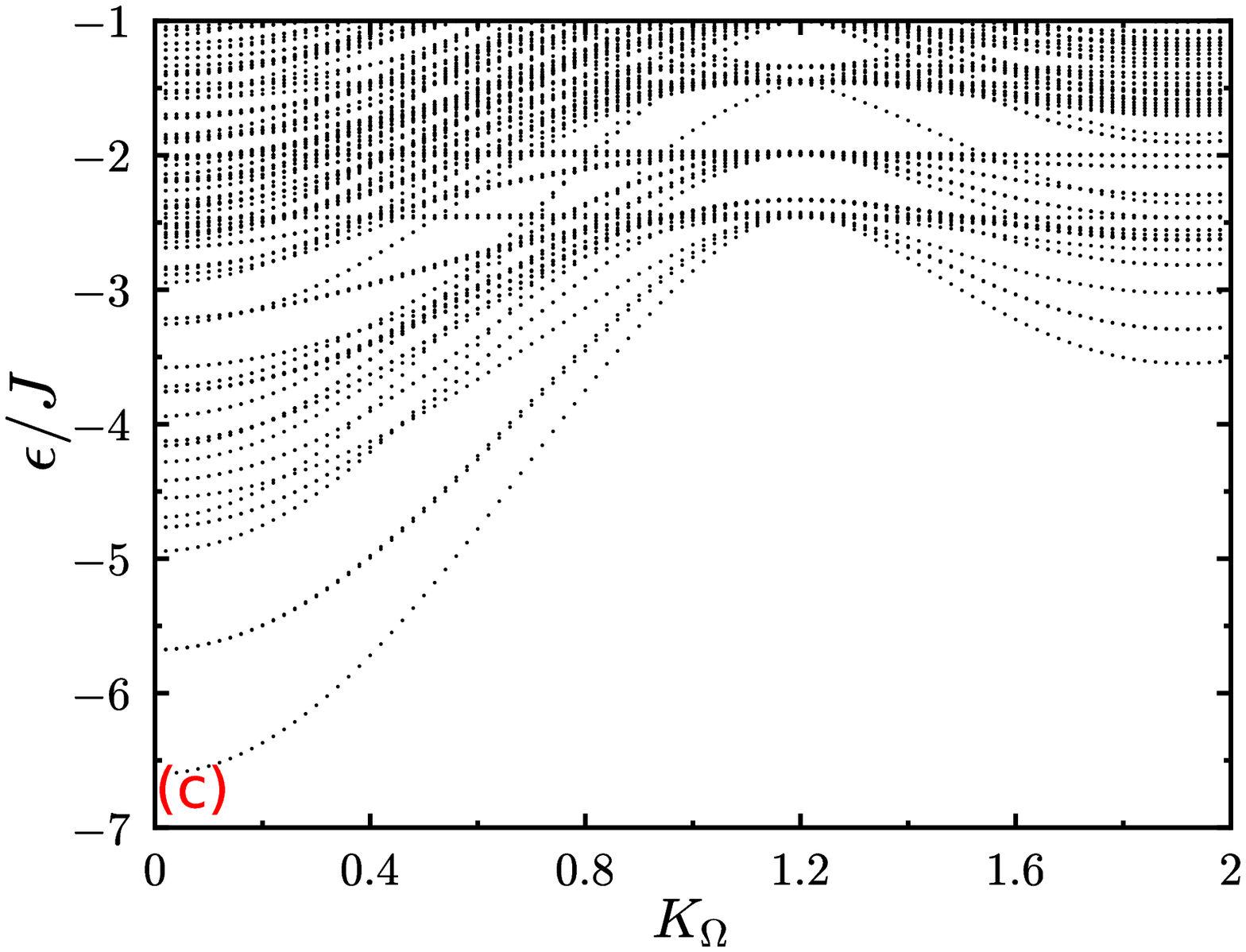}
\includegraphics[width=0.24 \columnwidth]{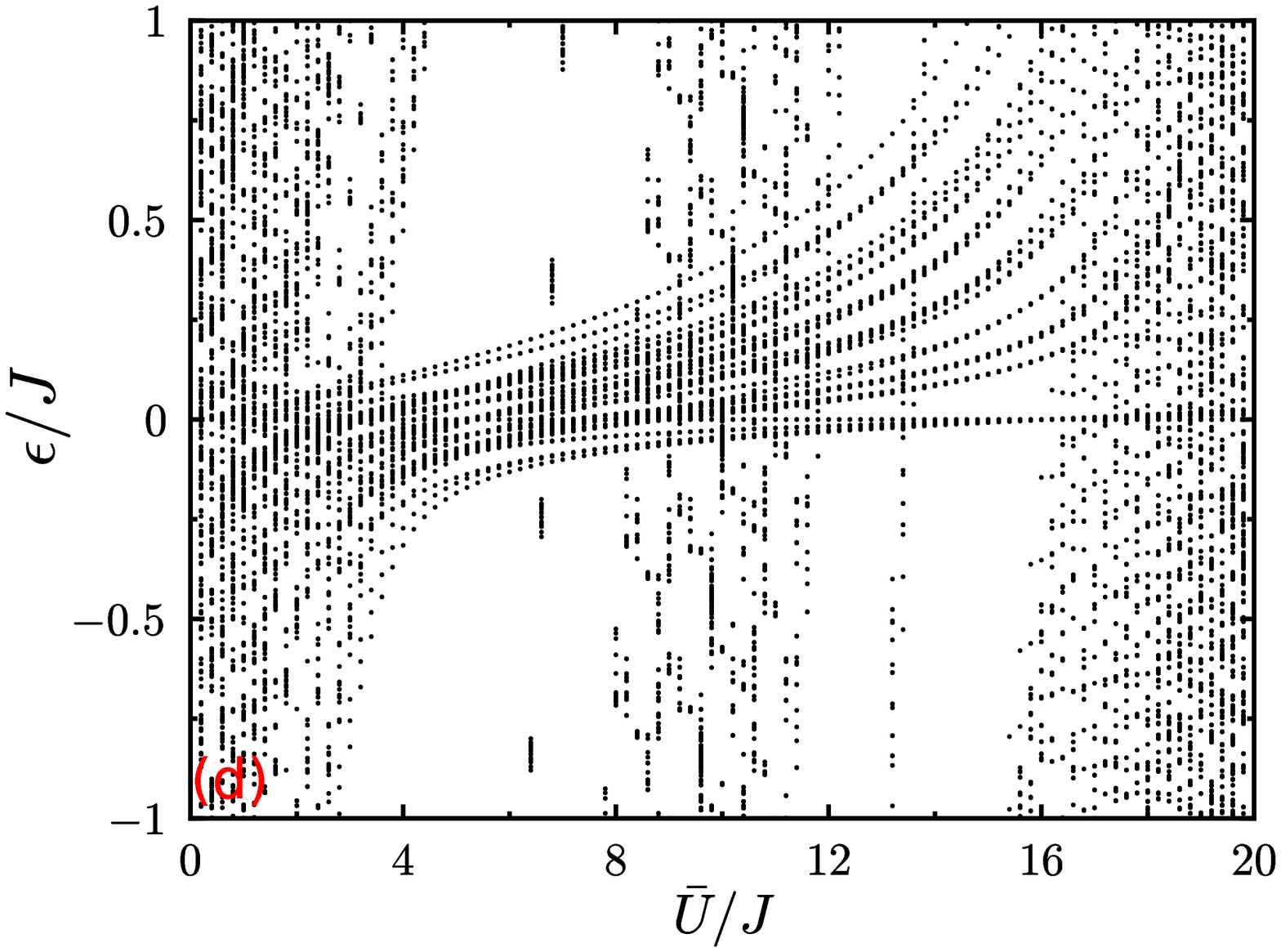}
\caption{
  Quasi-energy spectrum of the Floquet Hamiltonian (\ref{GFH}) as a function of $K_{U}$, $K_{\Omega}$, and $\bar U$, respectively. Here $L=6$, $\omega=20$ and $N^{a}=N^{b}=3$ in panel (a), (b) and $N^{a}+N^{b}=6$
  in panel (c), (d). Specially, (a) $K_{\Omega}=0$ and ${\bar U}/J=1$, (b) $K_{\Omega}=0$ and $K_{U}=1$, (c) $K_{U}=0$ and ${\bar U}/J=1$, (d) $K_{U}=0$ and $K_{\Omega}=1$.
}\label{figs2}
\end{figure}
In Fig.~\ref{figs2}, we depict the quasi-energy spectrum of $6$-sites as a function of $K_{U}$, $K_{\Omega}$, and $U$, respectively. We choose the integer-$1$ filling $N^{a} + N^{b} = 6$ and a relatively high
frequency $\omega=20$. Specially we have chosen $N^{a} = N^{b} = 3$ when $K_{\Omega}=0$. When ${\bar U}/J=1$ is fixed in panel (a) and (c), no problem of generating an adiabatic modulation of $K_{U}$ or $K_{\Omega}$ occurs.
Furthermore, no extremely dense avoided level-crossings are found. When $K_{U}=1$ is fixed in panel (b) and $K_{\Omega}=1$ in panel (d), we find a bunch of dense avoided level-crossings occurring in the vicinity of
${\bar U}/J\approx18$. In comparison with the ground-state phase diagram, where the main interesting phases happen when ${\bar U}/J < 4$, the avoided level-crossings do not appear in this region. Thus we conclude that all
the phases can be achieved by an adiabatic modulation of $\bar U$.
%

\section{Finite-size scaling}
Because of logarithmic corrections, it is challenging to derive the accurate position of the BKT-type transition point from the Mott-insulator to the superfluid phase by the finite-size scaling of the charge gap at
zero temperature or the compressibility $\chi$ at low temperatures. In Fig.~\ref{figs5}, all the curves $L \Delta_c$ collapse for small ${\cal J}_{0}[K]$, which means that the charge gap $\Delta_c$ scales like $1/L$ deep in the 
superfluid region. In the deep Mott-insulator region, when ${\cal J}_{0}[K]$ is large, $\Delta_c$ remains finite in the thermodynamic limit. At the anticipated critical point ${\cal J}_{0}[K]_c=0.624(6)$, we find that the curves
$L \Delta_c$ with different system sizes get slowly close to each other, but reveal no level-crossings. Similarly at the low temperature $\beta / {\bar U} = 2L$, the compressibility converges to a finite value and to zero
in the deep superfluid and the deep Mott-insulator region, respectively. However the turning points for the finite system are slowly approaching ${\cal J}_{0}[K]_c$.
\begin{figure}[t]
\includegraphics[width=0.4 \columnwidth]{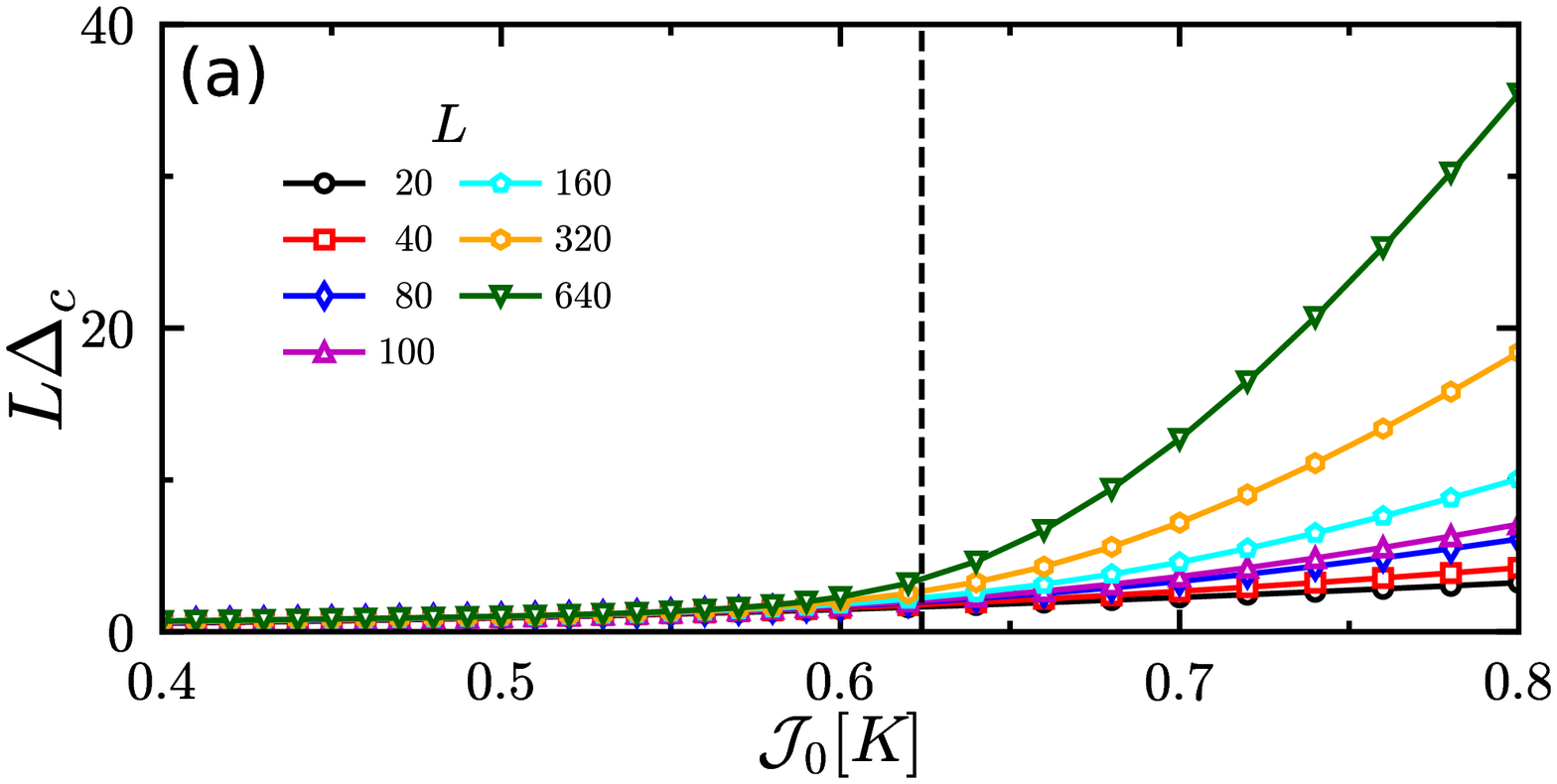}\includegraphics[width=0.4 \columnwidth]{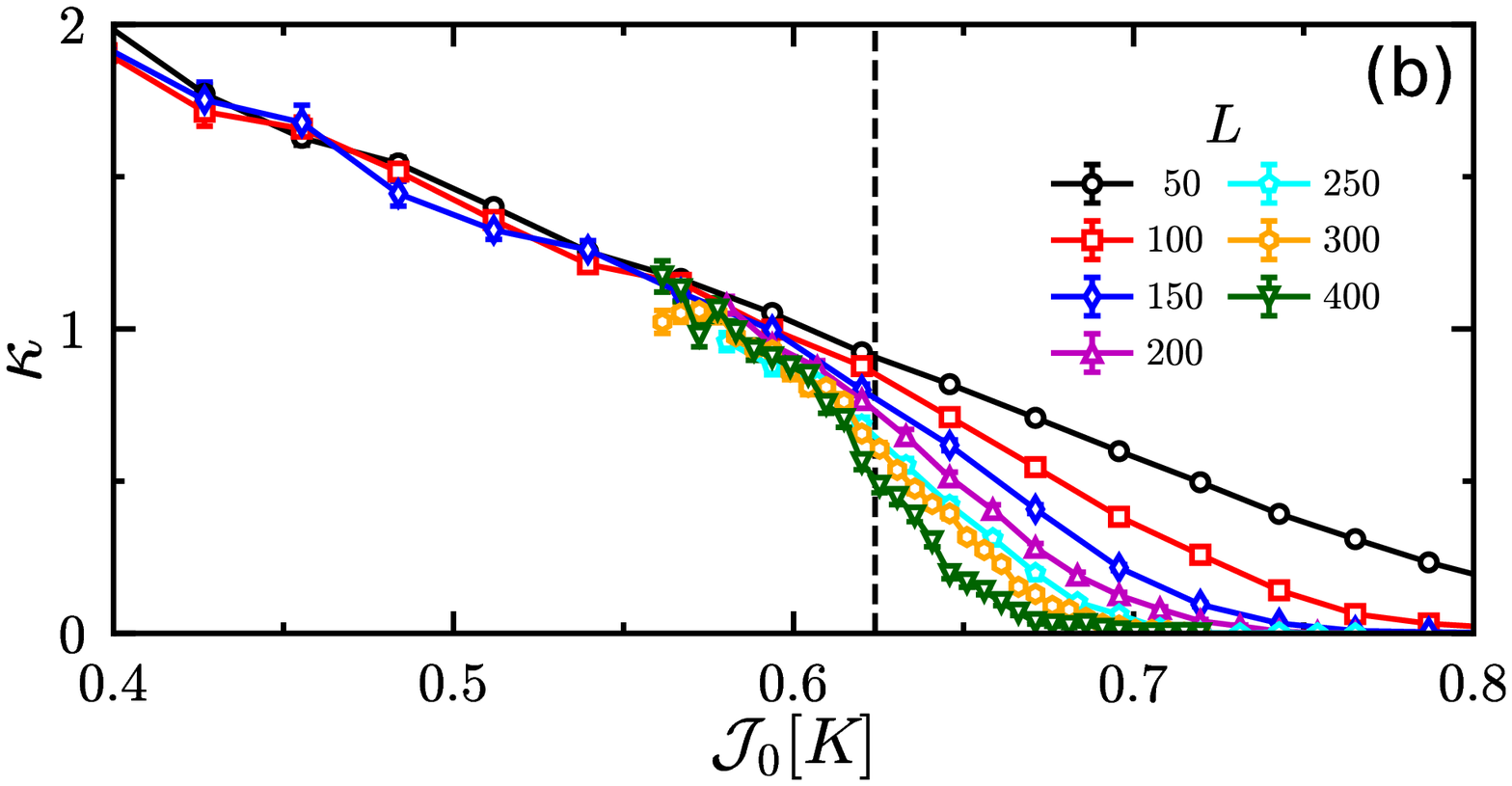}
\caption{Finite-size scaling of charge gap (a) and compressibility (b) for the case of $J=1$ and ${\bar U}/J=0.4$. In DMRG, we choose the maximal truncation dimension $m=4096$ for system sizes $L=20$
  (black $\circ$), $40$ (red $\Square$), $80$ (blue $\diamond$), $100$ (magenta $\Delta$), $160$ (cyan $\pentagon$), $320$ (orange $\hexagon$) and $640$ (green $\nabla$) with open boundary conditions. In QMC we have
  the inverse temperature $\beta/{\bar U}=2L$ with $L=50$ (black $\circ$), $100$ (red $\Square$), $150$ (blue $\diamond$), $200$ (magenta $\Delta$), $250$ (cyan $\pentagon$), $300$ (orange $\hexagon$) and $400$ (green $\nabla$).
}\label{figs5}
\end{figure}

Note that under the Jordan-Wigner transformation our model can be mapped to a density-dependent hopping Fermi-Hubbard model. Thus, with the help of the operator analysis involving the level-spectroscopic technique,\cite{fss}
we can choose the level-crossing of two representative excited states to be the quasi-critical point for the finite system. In the superfluid region, the representative excitation is a particle or
a hole if one adds or removes an atom. Whereas in the Mott-insulator region, the lowest-excitation is a pair of a particle ``$a$" together with a hole ``$b$" or vice versa. The former has a gap
$\Delta^{+}_c=E_{p}-E_0 + {\bar U} / 2$ measured from the energy of system, where we put one more particle $E_{p}$ relative to the ground-state energy $E_{0}$.
The latter has a pseudo-spin gap $\Delta^0_s=E_1 - E_0$ with the first-excitation energy $E_1$ in the same Hilbert space $N^a=N^b=L/2$ for the ground state.

In Fig.~\ref{figs6} (1b) and (2b), the curves of two excitation gaps reveal level-crossings for various finite system sizes. They scale very well as a linear function of $1/L$ in the insets and give us the position of
BKT-type critical points in the thermodynamic limit. The extrapolated values are also consistent with peaks of the fidelity susceptibility, which are obtained from iDMRG calculations.

\section{Gutzwiller mean-field}
Here we exhibit the details of applying the Gutzwiller mean-field (GWMF) method to the problem at hand.
Because of half filling $N_{a}=N_{b}=L/2$ and the hard-core constraint, the probabilities to occupy a site by a particle pair or a hole, as well as that by a single atom $a$
or $b$ are the same. Therefore, taking into account the normalization condition, we perform for the wave function an ansatz in terms of the uniform product matrix state 
\begin{equation}
|\psi_{g}\rangle=\bigotimes^{L}_{l=1}\frac{1}{\sqrt{2}}\left[e^{i\phi_{0,0}}\sin\varphi|0, 0\rangle_l +e^{i\phi_{0,1}}\cos\varphi|0, 1\rangle_l +e^{i\phi_{1,0}}\cos\varphi|1,0\rangle_l +e^{i\phi_{1,1}}\sin\varphi|1,1\rangle_l \right] \, ,
\end{equation}
where $|n^{a}, n^{b}\rangle_l$ denotes the local basis at  site $l$ with $n^{a}$ and $n^{b}$ standing for the numbers of species $a$ and $b$, respectively. Furthermore, $\varphi$ and
$\phi_{n^{a}, n^{b}}$ represent variational parameters, which are determined below. With this the average energy per-site yields
\begin{equation}
e_{g}=\frac{1}{L}\langle \psi_{g}|\hat{H}_{e} |\psi_{g}\rangle=\frac{{\bar U}}{4}(1-\cos2\varphi)-\frac{J}{2}\left(1-\cos^{2}2\varphi)(1+{\cal J}_{0}[K]\cos\delta\phi\right)\,,
\end{equation}
where we have introduced the abbreviation
$\delta\phi=\phi_{0,0}-\phi_{1,0}-\phi_{0,1}+\phi_{1,1}$. Minimizing the energy determines the wave function of the ground state. As $1-\cos^{2}2\varphi$ is always larger than zero, the choice of the
value of $\delta\phi$ in the ground-state wave function depends on the sign of ${\cal J}_{0}[K]$.
In the region ${\cal J}_{0}[K]>0$, we get $\delta\phi=0$ and $\cos2\varphi={\bar U}/[4J(1+|{\cal J}_{0}[K]|)]$, and the condensed density is given by $\rho^{>}_{c}=|\langle {\hat a} \rangle| = |\sin2\varphi(1+e^{i\delta\phi})/4|=|\sin2\varphi|/2>0$
when $J/{\bar U} > 1/4$.
Furthermore, we obtain $\delta\phi=\pi$ in the region ${\cal J}_{0}[K]<0$, $|{\cal J}_{0}[K]|\ll 1$ and $J/{\bar U} > 1/4$, where the condensed density reads $\rho^{>}_{c}=|\langle {\hat a} \rangle| = |\sin2\varphi(1+e^{i\delta\phi})/4|=0$,
while $\rho^{<}_{c}=|\langle {\hat \alpha}\rangle|=|{\hat a} e^{i\pi {\hat n}^{b}_{l}}|=|\sin2\varphi|/2>0$.  That suggests a gauge dressed superfluid phase in the region ${\cal J}_{0}[K]<0$.

\begin{figure}[t]
\includegraphics[width=0.4 \columnwidth]{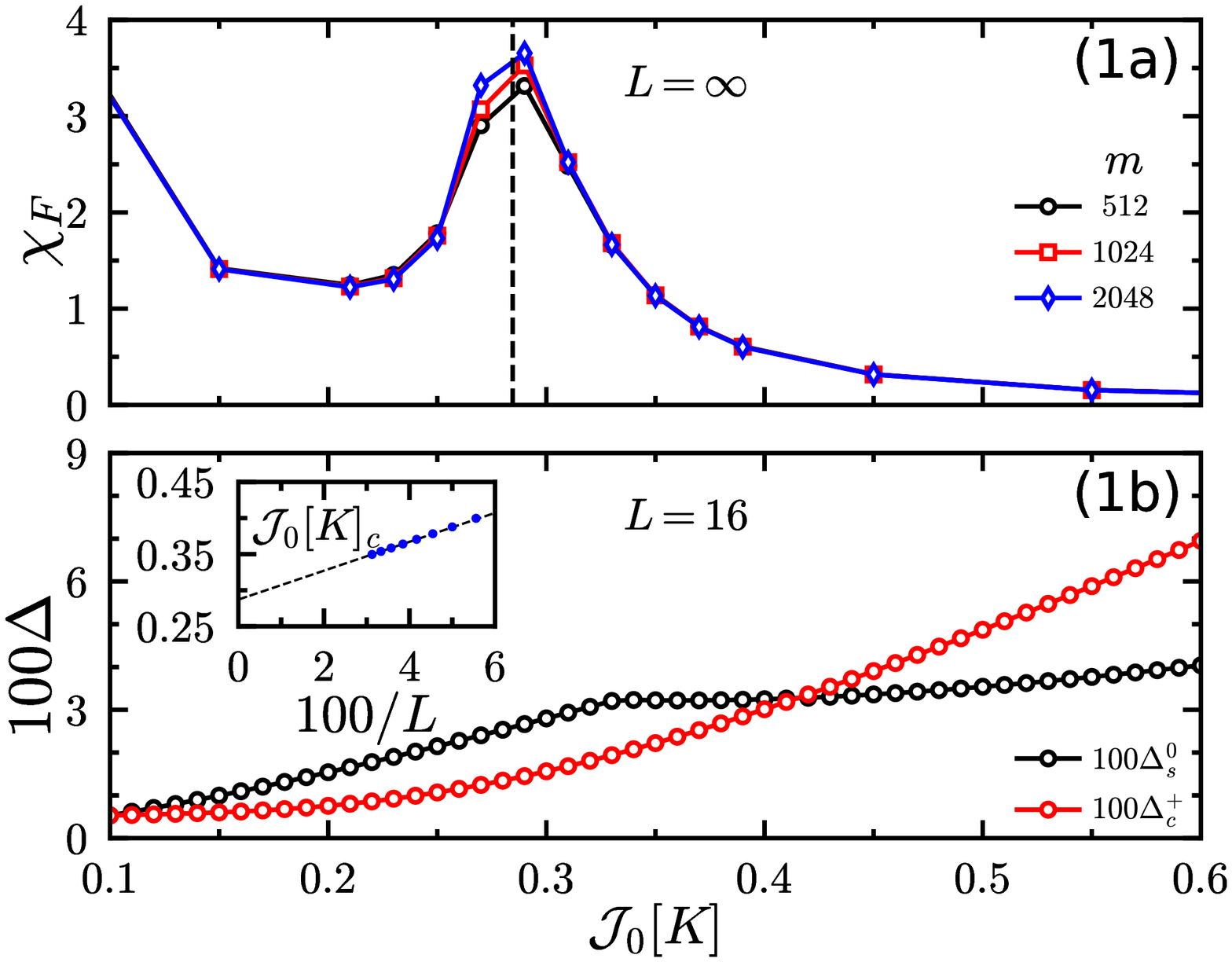}\includegraphics[width=0.4 \columnwidth]{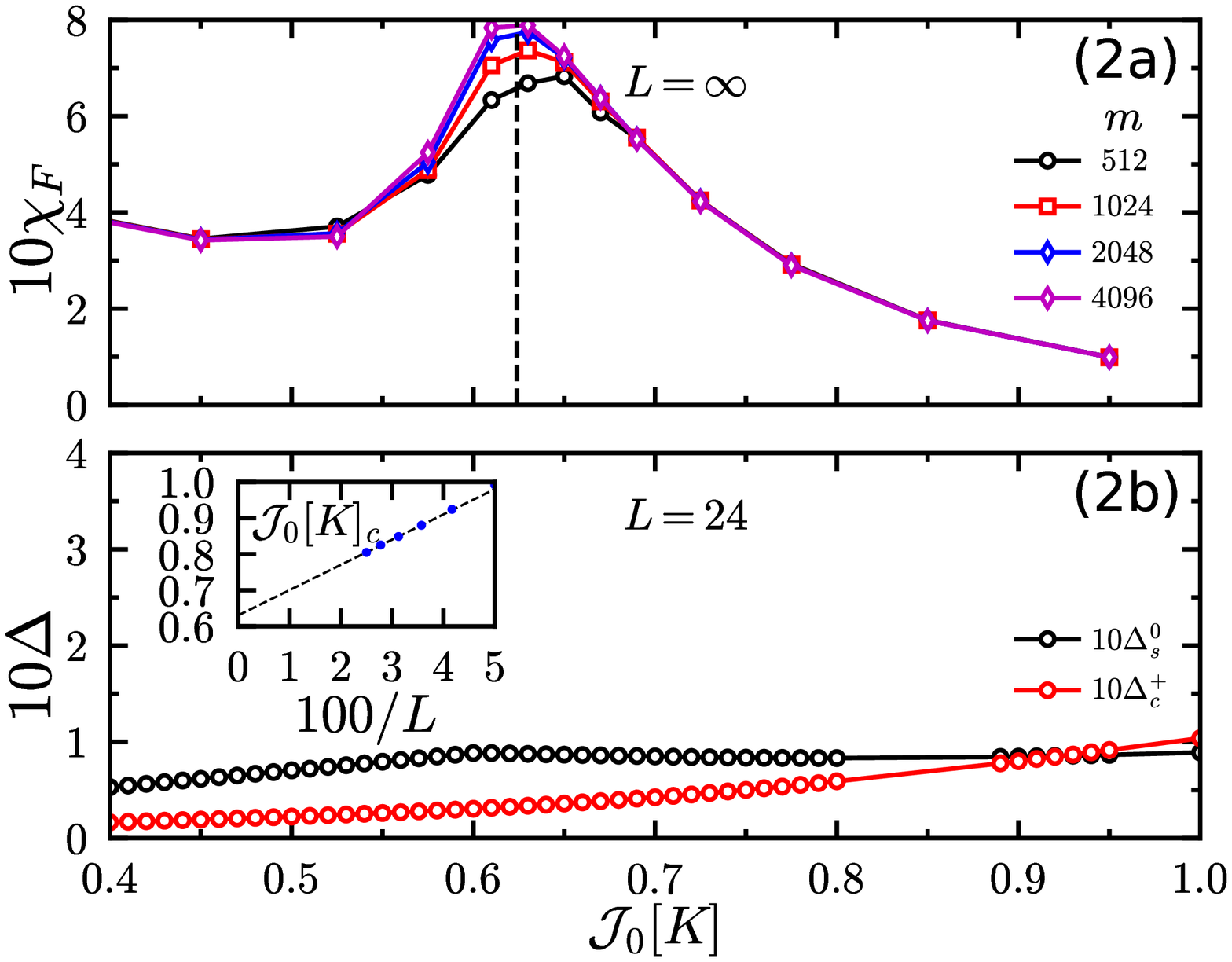}
\caption{
  Determination of BKT-type transition points from Mott-insulator to superfluid phase. In (1a) and (2a), peaks of the fidelity susceptibility measured from iDMRG indicate the transition points ${\cal J}_{0}[K]_c=0.285(3)$ for
  $J/{\bar U}=0.28$ (1a), and ${\cal J}_{0}[K]_c=0.624(6)$ for $J/{\bar U}=0.4$ (2a). In (1b) and (2b), we adopt the level-spectroscopic technique to achieve finite-size scaling. In the first step, we calculate the excitation gaps
  $\Delta^{+}_c=E_{p}-E_0 + {\bar U} / 2$ (red hexagon) and $\Delta^0_s=E_1 - E_0$ (black pentagon), where $E_0$ and $E_1$ are the ground state and the first excited state in the Hilbert space $N^a=N^b=L/2$, respectively, and
  $E_{p}$ is the lowest energy of system, where we add one more particle relative to the ground state. Obviously we find a level-crossing between them called quasi-critical points such as (1b) ${\cal J}_{0}[K]_{qc}=0.415$
  for $L=16$ and $J/{\bar U}=0.28$, (2b) ${\cal J}_{0}[K]_{qc}=0.924$ for $L=24$ and $J/{\bar U}=0.4$. In the second step, we plot these quasi-critical points as a function of $1/L$ in the inset and find that they can be linearly
  extrapolated to the thermodynamic limit quite well. We get the best extrapolation values ${\cal J}_{0}[K]_c=0.287$ for $J/{\bar U}=0.28$ and ${\cal J}_{0}[K]_c=0.630$ for $J/{\bar U}=0.4$, which are consistent with the
  results from iDMRG calculations in (1a) and (2a).
}\label{figs6}
\end{figure}

\section{Integrable Point}

For the effective Hamiltonian in Eq.~(2) of the main text the hopping term of the hardcore species ``$a$" consists of two parts ${\hat {\mathcal H}}_{a}={\hat {\mathcal H}}^{(1)}_{a}+{\hat {\mathcal H}}^{(2)}_{a}$, where
\begin{eqnarray}
{\hat {\mathcal H}}^{(1)}_{a}&=&-J\quad\quad~~\sum^{L}_{l=1} \left\{ {\hat a}^{\dagger}_{l} {\hat a}_{l+1} \left[{\hat n}^{b}_{l} {\hat n}^{b}_{l+1}+(1-{\hat n}^{b}_{l}) (1-{\hat n}^{b}_{l+1})\right] + h.c.\right\}\,,\nonumber\\
{\hat {\mathcal H}}^{(2)}_{a}&=&-J{\cal J}_{0}[K]\sum^{L}_{l=1} \left\{ {\hat a}^{\dagger}_{l} {\hat a}_{l+1} \left[{\hat n}^{b}_{l} (1-{\hat n}^{b}_{l+1})+(1-{\hat n}^{b}_{l}) {\hat n}^{b}_{l+1}\right] + h.c.\right\}\,.
\end{eqnarray}
Here we use the hardcore constraint condition ${\hat a}_{l}{\hat a}^{\dagger}_{l}+{\hat a}^{\dagger}_{l}{\hat a}_{l}=1$.
The second part depends on the ${\cal J}_{0}[K]$, i.e. the normalized driven amplitude $K$, while the first one does not.
Similarly, the hopping term of the species ``$b$" reads ${\hat {\mathcal H}}_{b}={\hat {\mathcal H}}^{(1)}_{b}+{\hat {\mathcal H}}^{(2)}_{b}$,  where
\begin{eqnarray}
{\hat {\mathcal H}}^{(1)}_{b}&=&-J\quad\quad~~\sum^{L}_{l=1} \left\{( {\hat b}^{\dagger}_{l} {\hat b}_{l+1} \left[{\hat n}^{a}_{l} {\hat n}^{a}_{l+1}+(1-{\hat n}^{a}_{l}) (1-{\hat n}^{a}_{l+1})\right] + h.c.\right\}\,,\nonumber\\
{\hat {\mathcal H}}^{(2)}_{b}&=&-J{\cal J}_{0}[K]\sum^{L}_{l=1} \left\{ {\hat b}^{\dagger}_{l} {\hat b}_{l+1} \left[{\hat n}^{a}_{l} (1-{\hat n}^{a}_{l+1})+(1-{\hat n}^{a}_{l}) {\hat n}^{a}_{l+1}\right] + h.c.\right\}\,.
\end{eqnarray}
Furthermore, the onsite interacting term ${\hat {\mathcal H}}_{\bar{U}}=\bar{U} \sum^{L}_{l=1} {\hat n}^{a}_{l} {\hat n}^{b}_{l}$ is the time average of the periodic interaction.

At the zeros of the zeroth-order Bessel function of first kind, namely ${\cal J}_{0}[K]=0$, we have ${\hat {\mathcal H}}_{a}={\hat {\mathcal H}}^{(1)}_{a}$ and ${\hat {\mathcal H}}_{b}={\hat {\mathcal H}}^{(1)}_{b}$.
On the site $l$ we can define number operators of a single-``$a$", a single-``$b$", a hole, and an ``$ab$" pair, respectively, namely
\begin{eqnarray}
{\hat {\mathcal N}}^{a}_{l}={\hat n}^{a}_{l} (1-{\hat n}^{b}_{l}),~
{\hat {\mathcal N}}^{b}_{l}=(1-{\hat n}^{a}_{l}){\hat n}^{b}_{l},~
{\hat {\mathcal N}}^{h}_{l}=(1-{\hat n}^{a}_{l}) (1-{\hat n}^{b}_{l}),~
{\hat {\mathcal N}}^{p}_{l}={\hat n}^{a}_{l} {\hat n}^{b}_{l}.~\nonumber
\end{eqnarray}
Thus, the total number operator naturally reads ${\hat {\mathcal N}}^{a(b,h,p)}_{t}=\sum^{L}_{l=1}{\hat {\mathcal N}}^{a(b,h,p)}_{l}$.
Because of the vanishing commutator
$\left[{\hat {\mathcal H}}_{a(b,\bar{U})}, {\hat {\mathcal N}}^{a(b,h,p)}_{t}\right]=0$, the total numbers of single-``$a$", single-``$b$", hole, and ``$ab$" pair are all conserved in any eigenstate of the Hamiltonian.
Furthermore, hopping terms of the species ``$a$" and ``$b$" can be divided into four individual and equivalent exchange processes:
\begin{eqnarray}
{\hat {\mathcal H}}_{a}&=&-J\sum^{L}_{l=1} \left({\hat S}^{pb,+}_{l} {\hat S}^{pb, -}_{l+1} + {\hat S}^{ha,+}_{l} {\hat S}^{ha, -}_{l+1} + h.c.\right),~\nonumber\\
{\hat {\mathcal H}}_{b}&=&-J\sum^{L}_{l=1} \left({\hat S}^{pa,+}_{l} {\hat S}^{pa, -}_{l+1} + {\hat S}^{hb,+}_{l} {\hat S}^{hb, -}_{l+1} + h.c.\right),
\end{eqnarray}
where we have used the notation
\begin{eqnarray}
{\hat S}^{pa,+}_{l}= \left({\hat a}^{\dagger}_{l} {\hat b}^{\dagger}_{l}\right) {\hat a}_{l},~
{\hat S}^{hb,+}_{l}= \left({\hat a}_{l} {\hat b}_{l}\right) {\hat a}^{\dagger}_{l},~
{\hat S}^{pb,+}_{l}= \left({\hat a}^{\dagger}_{l} {\hat b}^{\dagger}_{l}\right) {\hat b}_{l},~
{\hat S}^{ha,+}_{l}= \left({\hat a}_{l} {\hat b}_{l}\right) {\hat b}^{\dagger}_{l},~\nonumber\\
{\hat S}^{pa, -}_{l}= {\hat a}^{\dagger}_{l} \left({\hat a}_{l} {\hat b}_{l}\right),~
{\hat S}^{hb, -}_{l}= {\hat a}_{l} \left({\hat a}^{\dagger}_{l} {\hat b}^{\dagger}_{l}\right),~
{\hat S}^{pb, -}_{l}= {\hat b}^{\dagger}_{l} \left({\hat a}_{l} {\hat b}_{l}\right),~
{\hat S}^{ha,-}_{l}= {\hat b}_{l} \left({\hat a}^{\dagger}_{l} {\hat b}^{\dagger}_{l}\right)\,.
\end{eqnarray}
The natural basis of a configuration consists of the single-``$a$", the single-``$b$", the hole and the ``$ab$" pair on the respective lattice sites.
From each configuration, we can extract two sub-sequences: the first one $s_{ab}$ is built up by single occupations and the second $s_{ph}$ contains all holes and ``$ab$" pairs.
Supposing that we have an initial configuration with two sub-sequences, hopping processes preserve these two sequences if no exchange happens at edges.
For example, an initial configuration for $L=4$ sites is $|a_{1}p_{2}b_{3}h_{4}\rangle$ with two sub-sequences $|a b\rangle$ and $|p h\rangle$.
We obtain a new configuration $|a_{1}b_{2}p_{3}h_{4}\rangle$ under the exchange process between the ``$ab$" pair on site-$2$ and the single-``$b$" on site-$3$.
However, the new configuration still has two sub-sequences $|a b\rangle$ and $|p h\rangle$.
And thus we consider the two sub-sequences $s_{ab}$ and $s_{ph}$ as two hidden conserved quantities in order to distinguish degenerate states.
The Hilbert space with certain ${\mathcal N}^{a(b,h,p)}_{t}$ can be blocked into $C^{{\mathcal N}^{a}_{t}}_{{\mathcal N}^{a}_{t}+{\mathcal N}^{b}_{t}} C^{{\mathcal N}^{p}_{t}}_{{\mathcal N}^{p}_{t}+{\mathcal N}^{h}_{t}}$ subspaces, where
we do not need to distinguish either ``$a$" from ``$b$" or vacuum from ``$ab$" pair.
Furthermore, we find that the structure of subspaces is invariant, if we replace either ``$a$" (``$b$") by ``$b$" (``$a$") or replace the vacuum (``$ab$" pair) by ``$ab$" pair (vacuum), which leaves ${\mathcal N}^{a}_{t}
+{\mathcal N}^{b}_{t}$ and ${\mathcal N}^{p}_{t}+{\mathcal N}^{h}_{t}$ unchanged.
This means that for $\bar{U}=0$ the Hamiltonian has a larger hidden symmetry $D=Z^{L}_{2}$.
Let us therefore play a trick of preserving the hidden symmetry $D$ as an inner one and regrouping the four states: both ``$a$" and ``$b$" belong to the group "spin-down $\downarrow$", while both vacuum and ``$ab$"
pair belong to the group "spin-up $\uparrow$". With this the Hamiltonian becomes
\begin{eqnarray}
{\hat {\mathcal H}}_{a}+{\hat {\mathcal H}}_{b}&=& \mathbb{I}_{D} \otimes {\hat {\mathcal H}}_{r}\,,
\end{eqnarray}
where ${\hat {\mathcal H}}_{r}=-J\sum^{L}_{l=1} \left({\hat S}^{+}_{l} {\hat S}^{-}_{l+1} + h.c.\right)$ and ${\hat S}^{+(-)}_{l}$ denote the flip-up (down) operator of the normal spin-$1/2$.

Usually, the onsite interacting term with finite $\bar{U}$ breaks the exchange symmetry $D$ such that the vacuum is inequivalent to the ``$ab$" pair.
However the symmetry $D$ can be recovered in the case of the integer-$1$ filling $\sum^{L}_{l=1}\left({\hat n}^{a}_{l} + {\hat n}^{b}_{l}\right)=L$, where we have
\begin{eqnarray}
\hspace*{-0.7cm}{\hat {\mathcal H}}_{U}=\bar{U}\sum^{L}_{l=1} {\hat n}^{a}_{l} {\hat n}^{b}_{l} = \frac{\bar{U}}{2} \sum^{L}_{l=1} \left[2\left({\hat n}^{a}_{l}-\frac{1}{2}\right) \left({\hat n}^{b}_{l}-\frac{1}{2}\right)
+\frac{1}{2}+{\hat n}^{a}_{l}+{\hat n}^{a}_{l}-1\right]=\frac{\bar{U}}{2} \sum^{L}_{l=1} \left[2\left({\hat n}^{a}_{l}-\frac{1}{2}\right) \left({\hat n}^{b}_{l}-\frac{1}{2}\right)+\frac{1}{2}\right]\,.
\end{eqnarray}
Both vacuum and ``$ab$" pair contribute to $\bar{U}/2$, while neither ``$a$" nor ``$b$" have any contribution.
And thus the $\bar{U}$-term can be considered as an effective external magnetic field, which is applied to a redefined spin-$1/2$ and the effective Hamiltonian in the reduced Hilbert space reads
\begin{eqnarray}
{\hat {\mathcal H}}_{r}=-J\sum^{L}_{l=1} \left({\hat S}^{+}_{l} {\hat S}^{-}_{l+1} + h.c.\right) + \frac{\bar{U}}{2} \sum^{L}_{l=1} \left({\hat S}^{z}_{l} + \frac{1}{2}\right)\, ,
\end{eqnarray}
where have introduced ${\hat S}^{z}_{l}=2\left({\hat n}^{a}_{l}-1/2\right) \left({\hat n}^{b}_{l}-1/2\right)$.
By using the common Jordan-Wigner transformation, this becomes an integrable model in the language of spinless fermions.
We know that the ground state energy is $-2J\sum^{{\mathcal N}^{a}_{t}+{\mathcal N}^{b}_{t}}_{l=1} \cos(l\pi/(L+1)) + (\bar{U}/2) ({\mathcal N}^{p}_{t}+{\mathcal N}^{h}_{t})$ with degeneracy  $2^{L}$.
This means that the ground state has a finite residual entropy $\ln2$.

When we consider exchange processes at edges, e.g. with periodic or twisted boundary conditions, the situation becomes a bit more complicated.
From an initial configuration with certain $s_{ab}$ and $s_{ph}$, an exchange process at the edges certainly yields a new configuration with some other $s'_{ab}$ and $s'_{ph}$.
Let us take again $L=4$ as an example: the initial configuration $|a_{1}p_{2}b_{3}h_{4}\rangle$ transits into $|h_{1}p_{2}b_{3}a_{4}\rangle$ under the exchange
process between the single-``$a$" on the site-$1$ and the hole on the site-$4$.
At the same time, the sub-sequences $s_{ab}=|ab\rangle$ and $s_{ph}=|ph\rangle$ change to $s'_{ab}=|ba\rangle$ and $s'_{ph}=|hp\rangle$.
Therefore we have groups of relevant Hilbert subspaces with periodic boundary conditions.
In one group consisting of ${\mathcal N}_{s}$ subspaces, the hopping process between two Hilbert subspaces only happens at edges and provides a phase shift $Q_{q}=2q\pi/{\mathcal N}_{s}$ after a renormalization
group manipulation, where $q=0, 1, 2,\cdots,{\mathcal N}_{s}-1$.
And thus the single particle spectrum is equal to $e_{m,q}=-2J \cos[2m\pi/L+({\mathcal N}^{a}_{t}+{\mathcal N}^{b}_{t})\pi/L + Q_{q}/L]$.

In the following we will determine the physical properties of the integrable point.
Let us have at first a look at the single-particle correlation function of the species ``$a$", namely
\begin{eqnarray}
  \left\langle {\hat a}^{\dagger}_{l} {\hat a}_{l'}  \right\rangle = \left\langle {\hat a}^{\dagger}_{l} \left( {\hat b}_{l} {\hat b}^{\dagger}_{l} + {\hat b}^{\dagger}_{l} {\hat b}_{l} \right) {\hat a}_{l'}
  \left( {\hat b}_{l'} {\hat b}^{\dagger}_{l'} + {\hat b}^{\dagger}_{l'} {\hat b}_{l'} \right)  \right\rangle=\left\langle \left({\hat S}^{va,-}_{l} + {\hat S}^{pb,+}_{l} \right) \left({\hat S}^{va,+}_{l'}
  + {\hat S}^{pb,-}_{l'} \right) \right\rangle=\left\langle {\hat S}^{va,-}_{l} {\hat S}^{va,+}_{l'} + {\hat S}^{pb,+}_{l} {\hat S}^{pb,-}_{l'} \right\rangle\nonumber
\end{eqnarray}
where both mixing terms $\left\langle{\hat S}^{va,-}_{l} {\hat S}^{pb,-}_{l'}\right\rangle$ and $\left\langle{\hat S}^{pb,+}_{l} {\hat S}^{va,+}_{l'}\right\rangle$ are missing because none of them holds
the total number of the single-``$a$", single-``$b$", vacuum and ``$ab$" pair at the integrable point.
When we choose the balanced filling $\sum^{L}_{l=1} {\hat n}^{a}_{l} = \sum^{L}_{l=1} {\hat n}^{b}_{l}$, the probabilities of exchange processes between the single ``$a$" (``$b$") and vacuum (``$ab$" pair) are
always equal and thus the above single-particle correlation function becomes
\begin{eqnarray}
  \left\langle {\hat a}^{\dagger}_{l} {\hat a}_{l'}  \right\rangle =\frac{1}{4}\left\langle {\hat S}^{-}_{l} {\hat S}^{+}_{l'} + {\hat S}^{+}_{l} {\hat S}^{-}_{l'} \right\rangle_r=
  \frac{1}{2}\left\langle {\hat S}^{+}_{l} {\hat S}^{-}_{l'} \right\rangle_r\nonumber\, ,
\end{eqnarray}
where we have used the relation $\left\langle {\hat S}^{+}_{l} {\hat S}^{-}_{l'}\right\rangle_r=\left\langle {\hat S}^{-}_{l} {\hat S}^{+}_{l'}\right\rangle_r$ because the effective Hamiltonian represents a real matrix.

Next we investigate the superfluid density at the integrable point. To this end
we use the original definition of the superfluid density (or "spin stiffness") in terms of the second-order response to the twisted phase at the edge bond.
Supposing that the twisted angle is $\theta$, the hopping terms in the Hamiltonian becomes
\begin{eqnarray}
  {\hat {\mathcal H}}_{a}(\theta)&=&-J\sum^{L-1}_{l=1} \left({\hat S}^{pb,+}_{l} {\hat S}^{pb, -}_{l+1} + {\hat S}^{va,+}_{l} {\hat S}^{va, -}_{l+1} + h.c.\right) -Je^{i\theta}\left({\hat S}^{pb,+}_{L} {\hat S}^{pb, -}_{1}
  + {\hat S}^{va,-}_{L} {\hat S}^{va,+}_{1} + h.c.\right)\,,~\nonumber\\
  {\hat {\mathcal H}}_{b}(\theta)&=&-J\sum^{L-1}_{l=1} \left({\hat S}^{pa,+}_{l} {\hat S}^{pa, -}_{l+1} + {\hat S}^{vb,+}_{l} {\hat S}^{vb, -}_{l+1} + h.c.\right)-Je^{i\theta}\left({\hat S}^{pa,+}_{L} {\hat S}^{pa, -}_{1}
  + {\hat S}^{vb,-}_{L} {\hat S}^{vb,+}_{1} + h.c.\right)\,,
\end{eqnarray}
where the exchange process between the single ``$a$" (``$b$") and the ``$ab$" pair carries a positive phase, while the one between the single ``$a$" (``$b$") and the vacuum carries a negative phase.
The hidden inner symmetry $D$ disappears and, thus, the model can not be mapped to the effective spin-$1/2$ XY model.
Again the onsite interacting term is invariant.
The ground-state energy with infinitesimal twisted angle $\theta\ll 1$ can be expanded in the vicinity of $\theta=0$, namely
\begin{eqnarray}
E_g(\theta)=E_g(0) + \frac{1}{2} \rho_{s} \theta^{2} + o(\theta^3)\, ,
\end{eqnarray}
where the first-order term disappears because for $\theta=0$ the system holds the time-reversal symmetry and has no residual "current".
In fact we can prove that the energy response vanishes in case of integer-$1$ filling.
From the effective model, the ground-state occurs when ${\mathcal N}^{a}_{t}+{\mathcal N}^{b}_{t}={\mathcal N}^{p}_{t}+{\mathcal N}^{h}_{t}=L/2$.
Therefore we conclude ${\mathcal N}^{p}_{t}={\mathcal N}^{h}_{t}=L/4$.
We have $L/2$ relevant Hilbert subspaces: $L/4$ subspaces are connected by the exchange processes between single occupations on the site-$1$ and holes on the site-$L$ carrying a twisted phase $\exp(i\theta)$,
while the other $L/4$ subspaces are connected by ones between single occupations on the site-$1$ and pairs on the site-$L$ carrying a twisted phase $\exp(-i\theta)$.
As a result, the residual twist phase in this group vanishes under the gauge transformation.
Thus, the ground state has no energy response to the twisted phase on the boundary, which means
that their superfluid density vanishes.

\end{widetext}


\begin{thebibliography}{100}

\bibitem{Anderson_1995}
M.H.~Anderson, J.R.~Ensher, M.R.~Matthews, C.E.~Wieman, and E.A.~Cornell,
  Observation of Bose-Einstein Condensation in a Dilute Atomic Vapor, Science
  {\bf 269}, 198 (1995).

\bibitem{Davis_1995}
K.~B. Davis, M.~O. Mewes, M.~R. Andrews, N.~J. van Druten, D.~S. Durfee, D.~M.
  Kurn, and W.~Ketterle, Bose-Einstein Condensation in a Gas of Sodium Atoms,
  \prl~{\bf 75}, 3969 (1995).

\bibitem{Fisher_1989}
M.~P.~A. Fisher, P.~B. Weichman, G.~Grinstein, and D.~S. Fisher, Boson
  localization and the superfluid-insulator transition, Phys.~Rev.~B {\bf
  40}, 546 (1989).

\bibitem{Jaksch_1998}
D.~Jaksch, C.~Bruder, J.~I. Cirac, C.~W. Gardiner, and P.~Zoller, Cold Bosonic
  Atoms in Optical Lattices, \prl~{\bf 81}, 3108 (1998).

\bibitem{Greiner_2002}
M.~Greiner, O.~Mandel, T.~Esslinger, T.~W. H{\"a}nsch, and I.~Bloch, Quantum
  phase transition from a superfluid to a Mott insulator in a gas of ultracold
  atoms, Nature {\bf 415}, 39 (2002).

\bibitem{Hoffmann_2009}
A.~Hoffmann and A.~Pelster, Visibility of cold atomic gases in optical lattices
  for finite temperatures, Phys.~Rev.~A {\bf 79}, 053623 (2009).

\bibitem{Chin_2010}
C.~Chin, R.~Grimm, P.~Julienne, and E.~Tiesinga, Feshbach resonances in
  ultracold gases, Rev.~Mod. Phys.~{\bf 82}, 1225 (2010).

\bibitem{Gross_2002}
K.~Gross, C.~P. Search, H.~Pu, W.~Zhang, and P.~Meystre, Magnetism in a lattice
  of spinor Bose-Einstein condensates, Phys.~Rev.~A {\bf 66}, 033603
  (2002).

\bibitem{Demler_2002}
E.~Demler and F.~Zhou, Spinor Bosonic Atoms in Optical Lattices: Symmetry
  Breaking and Fractionalization, \prl~{\bf 88}, 163001
  (2002).

\bibitem{Mobarak}
M. Mobarak and A. Pelster,
Superfluid Phases of Spin-1 Bosons in Cubic Optical Lattice,
Laser Phys. Lett. {\bf 10}, 115501 (2013).

\bibitem{Duan_2003}
L.-M. Duan, E.~Demler, and M.~D. Lukin, Controlling Spin Exchange Interactions
  of Ultracold Atoms in Optical Lattices, \prl~{\bf 91},
  090402 (2003).

\bibitem{Hofstetter_2002}
W.~Hofstetter, J.~I. Cirac, P.~Zoller, E.~Demler, and M.~D. Lukin,
High-Temperature Superfluidity of Fermionic Atoms in Optical Lattices,
\prl~{\bf 89}, 220407 (2002).

\bibitem{Kuklov_2003}
A.~B. Kuklov and B.~V. Svistunov, Counterflow Superfluidity of Two-Species
  Ultracold Atoms in a Commensurate Optical Lattice, \prl~{\bf 90}, 100401 (2003).

\bibitem{Altman_2003}
E.~Altman, W.~Hofstetter, E.~Demler, and M.~D. Lukin, Phase diagram of
  two-component bosons on an optical lattice, New J.~Phys.~{\bf 5},
 113 (2003).

\bibitem{Kuno_2013}
Y.~Kuno, K.~Kataoka, and I.~Ichinose, Effective field theories for
 two-component repulsive bosons on lattice and their phase diagrams, Phys.~Rev.~B {\bf 87}, 014518 (2013).

\bibitem{ring}
H.-N. Dai, B.~Yang, A.~Reingruber, H.~Sun, X.-F. Xu, Y.-A. Chen, Z.-S. Yuan,
  and J.-W. Pan, Four-body ring-exchange interactions and anyonic statistics
  within a minimal toric-code Hamiltonian, Nat. Phys. {\bf 13}, 1195 (2017).

\bibitem{M_lmer_1998}
K.~M{\o}lmer, Bose Condensates and Fermi Gases at Zero Temperature, Phys.~Rev.~Lett.~{\bf 80}, 1804 (1998).

\bibitem{Viverit_2000}
L.~Viverit, C.~J. Pethick, and H.~Smith, Zero-temperature phase diagram of
  binary boson-fermion mixtures, Phys.~Rev.~A {\bf 61}, 053605 (2000).

\bibitem{Bhaseen_2009}
M.~J. Bhaseen, M.~Hohenadler, A.~O. Silver, and B.~D. Simons, Polaritons and
  Pairing Phenomena in Bose-Hubbard Mixtures, \prl~{\bf
  102}, 135301 (2009).

\bibitem{Bretin_2004}
V.~Bretin, S.~Stock, Y.~Seurin, and J.~Dalibard, Fast Rotation of a
  Bose-Einstein Condensate, \prl~{\bf 92}, 050403 (2004).

\bibitem{Schweikhard_2004}
V.~Schweikhard, I.~Coddington, P.~Engels, S.~Tung, and E.~A. Cornell,
  Vortex-Lattice Dynamics in Rotating Spinor Bose-Einstein Condensates,
  \prl~{\bf 93}, 210403 (2004).

\bibitem{Lin_2009}
Y.-J. Lin, R.~L. Compton, K.~Jim{\'e}nez-Garc{\'\i}a, J.~V. Porto, and I.~B.
  Spielman, Synthetic magnetic fields for ultracold neutral atoms, Nature {\bf
  462}, 628 (2009).

\bibitem{Lin_2011}
Y.-J. Lin, R.~L. Compton, K.~Jim{\'e}nez-Garc{\'\i}a, W.~D. Phillips, J.~V.
  Porto, and I.~B. Spielman, A synthetic electric force acting on neutral
  atoms, Nature Phys.~{\bf 7}, 531 (2011).

\bibitem{Vidanovic}
I. Vidanovic, A. Balaz, H. Al-Jibbouri, and A. Pelster,
Nonlinear BEC Dynamics Induced by a Harmonic Modulation of the s-wave Scattering Length,
Phys. Rev. A {\bf 84}, 013618 (2011).

\bibitem{Liberto_2014}
M.~Di~Liberto, C.~E.~Creffield, G.~I.~Japaridze, and C.~Morais~Smith,
Quantum simulation of correlated-hopping models with fermions in optical lattices,
Phys.~Rev.~A {\bf 89}, 013624 (2014).

\bibitem{Greschner_2015}
S.~Greschner and L.~Santos, Anyon Hubbard Model in One-Dimensional Optical
 Lattices, \prl~{\bf 115}, 053002 (2015).

\bibitem{Tang_2015}
G.~Tang, S.~Eggert, and A.~Pelster, Ground-state properties of anyons in a
  one-dimensional lattice, New J.~Phys.~{\bf 17}, 123016 (2015).

\bibitem{Str_ter_2016}
C.~Str{\"a}ter, S.~C. Srivastava, and A.~Eckardt, Floquet Realization and
  Signatures of One-Dimensional Anyons in an Optical Lattice, Phys.~Rev.~Lett.~{\bf 117}, 205303 (2016).

\bibitem{Keilmann_2011}
T.~Keilmann, S.~Lanzmich, I.~McCulloch, and M.~Roncaglia, Statistically induced
  phase transitions and anyons in 1D optical lattices, Nat.~Comm.~{\bf 2}, 361 (2011).

\bibitem{Ramos_2008}
E.~R.~F. Ramos, E.~A.~L. Henn, J.~A. Seman, M.~A. Caracanhas, K.~M.~F.
  Magalh{\~a}es, K.~Helmerson, V.~I. Yukalov, and V.~S. Bagnato, Generation of
  nonground-state Bose-Einstein condensates by modulating atomic interactions,
  Phys.~Rev.~A {\bf 78}, 063412 (2008).

\bibitem{Pollack_2010}
S.~E. Pollack, D.~Dries, R.~G. Hulet, K.~M.~F. Magalh{\~a}es, E.~A.~L. Henn,
  E.~R.~F. Ramos, M.~A. Caracanhas, and V.~S. Bagnato, Collective excitation of
  a Bose-Einstein condensate by modulation of the atomic scattering length,
  Phys.~Rev.~A {\bf 81}, 053627 (2010).

\bibitem{Eckardt_2017}
A.~Eckardt, Colloquium: Atomic quantum gases in periodically driven optical
  lattices, Rev.~Mod. Phys.~{\bf 89}, 011004 (2017).

\bibitem{Arimondo_2012}
E.~Arimondo, D.~Ciampini, E.~A., M.~Holthaus, and O.~Morsch, Kilohertz-Driven Bose
Einstein Condensates in Optical Lattices,
Adv. At. Mol. Opt. Phy. {\bf 61}, 515
(2012).

\bibitem{Rapp_2012}
A.~Rapp, X.~Deng, and L.~Santos, Ultracold Lattice Gases with Periodically
  Modulated Interactions, Phys. Rev. Lett. {\bf 109}, 203005 (2012).
  
\bibitem{Wang_2014}
T.~Wang, X.-F. Zhang, F.~E. A.~d. Santos, S.~Eggert, and A.~Pelster, Tuning the
  quantum phase transition of bosons in optical lattices via periodic
  modulation of thes-wave scattering length, Phys.~Rev.~A {\bf 90}, 013633
  (2014).

\bibitem{Greschner_2014}
S.~Greschner, L.~Santos, and D.~Poletti, Exploring Unconventional Hubbard
  Models with Doubly Modulated Lattice Gases, Phys. Rev. Lett. {\bf 113},
  183002 (2014).

\bibitem{Meinert_2016}
F.~Meinert, M.~Mark, K.~Lauber, A.~Daley, and H.-C. N{\"a}gerl, Floquet
  Engineering of Correlated Tunneling in the Bose-Hubbard Model with Ultracold
  Atoms, \prl~{\bf 116}, 205301 (2016).

\bibitem{Struck_2011}
J.~Struck, C.~Olschlager, R.~Le~Targat, P.~Soltan-Panahi, A.~Eckardt,
  M.~Lewenstein, P.~Windpassinger, and K.~Sengstock, Quantum Simulation of
  Frustrated Classical Magnetism in Triangular Optical Lattices, Science {\bf
  333}, 996 (2011).

\bibitem{Struck_2012}
J.~Struck, C.~{\"O}lschl{\"a}ger, M.~Weinberg, P.~Hauke, J.~Simonet,
  A.~Eckardt, M.~Lewenstein, K.~Sengstock, and P.~Windpassinger, Tunable Gauge
  Potential for Neutral and Spinless Particles in Driven Optical Lattices,
  \prl~{\bf 108}, 225304 (2012).

\bibitem{Hauke_2012}
P.~Hauke, O.~Tieleman, A.~Celi, C.~{\"O}lschl{\"a}ger, J.~Simonet, J.~Struck,
  M.~Weinberg, P.~Windpassinger, K.~Sengstock, M.~Lewenstein, and A. Eckardt,
  Non-Abelian Gauge Fields and Topological Insulators in Shaken Optical
  Lattices, \prl~{\bf 109}, 145301 (2012).

\bibitem{Struck_2013}
J.~Struck, M.~Weinberg, C.~{\"O}lschl{\"a}ger, P.~Windpassinger, J.~Simonet,
  K.~Sengstock, R.~H{\"o}ppner, P.~Hauke, A.~Eckardt, M.~Lewenstein, and
  et~al., Engineering Ising-XY spin-models in a triangular lattice using
  tunable artificial gauge fields, Nature Phys.~{\bf 9}, 738 (2013).

\bibitem{Aidelsburger_2011}
M.~Aidelsburger, M.~Atala, S.~Nascimb{\`e}ne, S.~Trotzky, Y.-A. Chen, and
  I.~Bloch, Experimental Realization of Strong Effective Magnetic Fields in an
  Optical Lattice, \prl~{\bf 107}, 255301 (2011).

\bibitem{Lignier_2007}
H.~Lignier, C . Sias, D. Ciampini, Y. Singh, A. Zenesini, O. Morsch, and E.
Arimondo, Dynamical Control of Matter-Wave Tunneling in Periodic Potentials,
Phys. Rev. Lett. {\bf 99}, 220403 (2007).

\bibitem{exp2}
G. Jotzu, M. Messer, F. G\"org, D. Greif, R. Desbuquois,
and T. Esslinger,
Creating State-Dependent Lattices for Ultracold Fermions by Magnetic Gradient
Modulation,
Phys. Rev. Lett. 115, 073002 (2015)

\bibitem{exp3}
F. G\"org, M. Messer, K. Sandholzer, G. Jotzu, R. Desbuquois, and T. Esslinger,
Enhancement and sign change of magnetic correlations in a driven quantum
many-body system,
Nature {\bf 553}, 481 (2018).
  
\bibitem{Zenesini_2009}
A. Zenesini, H. Lignier, D. Ciampini, O. Morsch, and 
E. Arimondo, Coherent Control of Dressed Matter Waves, Phys. Rev. Lett.
 102, 100403 (2009).

\bibitem{SM}
See Supplemental Material at http://link.aps.org/
supplemental/10.1103/PhysRevLett.xxx.xxxxx for discussions on two proposals of the experimental realization, higher orders in the effective Hamiltonian, the role of the Kick operator, avoided level crossing of the quasi-energy spectrum, real-time dynamics of the original time-dependent Hamiltonian, finite size scaling, and the Gutzwiller mean field method.

\bibitem{Lieb_1968}
E.~H. Lieb and F.~Y. Wu, Absence of Mott Transition in an Exact Solution of the
  Short-Range, One-Band Model in One Dimension, Phys. Rev. Lett. {\bf 20},
  1445 (1968).

\bibitem{White_1992}
S.~R. White, Density matrix formulation for quantum renormalization groups,
  \prl~{\bf 69}, 2863 (1992).

\bibitem{White_1993}
S.~R. White, Density-matrix algorithms for quantum renormalization groups,
  Phys.~Rev.~B {\bf 48}, 10345 (1993).

\bibitem{Peschel_1999}
I.~Peschel, X.~Q. Wang, M.~Kaulke, and K.~Hallberg, eds., {\em Density-Matrix
  Renormalization}.
\newblock Springer Berlin Heidelberg, 1999.

\bibitem{Schollwoeck_2005}
U.~Schollw{\"o}ck, The density-matrix renormalization group, Rev.~Mod.~Phys.~{\bf 77}, 259 (2005).

\bibitem{McCulloch_2008}
I.~P. McCulloch, Infinite size density matrix renormalization group,
  arXiv:0804.2509 (2008).

\bibitem{Hu_2011}
S.~Hu, B.~Normand, X.~Wang, and L.~Yu, Accurate determination of the Gaussian
  transition in spin-1 chains with single-ion anisotropy, Phys.~Rev.~B
  {\bf 84}, 220402(R) (2011).

\bibitem{Hu_2014}
S.~Hu, A.~M. Turner, K.~Penc, and F.~Pollmann, Berry-Phase-Induced Dimerization
  in One-Dimensional Quadrupolar Systems, \prl~{\bf 113},
  027202 (2014).

\bibitem{sse1}
A.~W. Sandvik, Stochastic series expansion method with operator-loop update,
  Phys. Rev. B {\bf 59}, R14157 (1999).

\bibitem{sse2}
O.~F. Sylju\aa{}sen and A.~W. Sandvik, Quantum Monte Carlo with directed loops,
  Phys. Rev. E {\bf 66}, 046701 (2002).

\bibitem{sse3}
K.~Louis and C.~Gros, Stochastic cluster series expansion for quantum spin
  systems, Phys. Rev. B {\bf 70}, 100410(R) (2004).

\bibitem{Kuhner}
T.~D. K\"uhner and H.~Monien, Phases of the one-dimensional Bose-Hubbard model,
  Phys. Rev. B {\bf 58}, R14741 (1998).

\bibitem{Roth_2003}
R.~Roth and K.~Burnett, Superfluidity and interference pattern of ultracold
  bosons in optical lattices, Phys.~Rev.~A {\bf 67}, 031602(R) (2003).

\bibitem{book}
F.~Essler, H.~Frahm, F.~G{\"o}hmann, A.~Kl{\"u}mper, and V.~Korepin, {\em The
  One-Dimensional Hubbard Model}.
\newblock Cambridge University Press, 2005.

\bibitem{You_2007}
W.-L. You, Y.-W. Li, and S.-J. Gu, Fidelity, dynamic structure factor, and
  susceptibility in critical phenomena, Phys.~Rev.~E {\bf 76}, 022101
  (2007).

\bibitem{Campos_Venuti_2007}
L.C.~Venuti and P.~Zanardi, Quantum Critical Scaling of the Geometric
  Tensors, \prl~{\bf 99}, 095701 (2007).

\bibitem{Osterloh_2002}
A.~Osterloh, L.~Amico, G.~Falci, and R.~Fazio, Scaling of entanglement close to
  a quantum phase transition, Nature {\bf 416}, 608 (2002).

\bibitem{Wu_2004}
L.-A. Wu, M.~S. Sarandy, and D.~A. Lidar, Quantum Phase Transitions and
  Bipartite Entanglement, \prl~{\bf 93}, 250404 (2004).

\bibitem{Laflorencie_2016}
N.~Laflorencie, Quantum entanglement in condensed matter systems, 
Phys.~Rep.~{\bf 646}, 1 (2016).

\bibitem{fss}
M.~Nakamura, Tricritical behavior in the extended Hubbard chains, Phys.~Rev.~B {\bf 61}, 16377 (2000). 

\bibitem{wannier}
R.~Walters, G.~Cotugno, T.~H. Johnson, S.~R. Clark, and D.~Jaksch, Ab initio
  derivation of Hubbard models for cold atoms in optical lattices, Phys. Rev. A
  {\bf 87}, 043613 (2013).

\bibitem{Schmidt_2006}
K.~P. Schmidt, J.~Dorier, A.~L\"auchli, and F.~Mila, Single-particle versus
  pair condensation of hard-core bosons with correlated hopping, Phys. Rev. B
  {\bf 74}, 174508 (2006).
\bibitem{Verhaar_2002}
A. Marte, T. Volz, J. Schuster, S. D\"{u}rr, G. Rempe, E. G. M. van Kempen, and B. J. Verhaar,
Phys. Rev. Lett. {\bf 89}, 283202 (2002).


\bibitem{zwerger} W. Zwerger, J. Opt. B {\bf 5}, S9 (2003).

\bibitem{Naber_2016}
J. B. Naber, L. Torralbo-Campo, T. Hubert, and R. J. C. Spreeuw,
Phys. Rev. A {\bf 94}, 013427 (2016).

\bibitem{Sapriel_1979}
J. Sapriel, S. Francis, and B. Kelly,
{\it Acousto-Optics}
(Wiley, 1979).

\bibitem{Eklund_1975}
H. Eklund, A. Roos, and S. T. Eng,
Opt. Quantum Elect. {\bf 7}, 73 (1975).

\bibitem{Floquet}
G. Floquet,
Ann. Ecole Norm. Sup. {\bf 12}, 47 (1883).

\bibitem{Eckardt_2015}
A. Eckardt and E. Anisimovas,
New J. Phys. {\bf 17}, 093039 (2015).
%
\bibitem{Bukov_2015}
M. Bukov, L. D'Alessio, and A. Polkovnikov,
Adv. Phys. {\bf 64}, 139 (2015).
%
\bibitem{Eckardt_2008}
A. Eckardt and M. Holthaus,
Phys. Rev. Lett. {\bf 101}, 245302 (2008).

\end{thebibliography}
\end{document}